\documentclass[journal]{IEEEtran}
\usepackage{amssymb}
\usepackage{times}
\usepackage{epsfig}
\usepackage{graphicx}
\usepackage{amsmath}
\usepackage{amssymb}

\usepackage{color}
\usepackage{bm}
\usepackage{algorithm}
\usepackage{algorithmic}
\usepackage{multirow}
\usepackage{subfigure}
\usepackage{amsthm}
\usepackage{booktabs}
\usepackage{textcomp}
\usepackage{mathrsfs}
\usepackage{makecell}
\usepackage{diagbox}
\usepackage{setspace}

\usepackage{balance}
\usepackage{float}
\usepackage{subeqnarray}
\usepackage{cases}
\usepackage{tabularx}
\usepackage{lscape}

\usepackage{bbding}
\usepackage{pifont}
\usepackage{wasysym}

\newcommand{\tabincell}[2]{\begin{tabular}{@{}#1@{}}#2\end{tabular}}

\usepackage[colorlinks,linkcolor=blue]{hyperref}

\begin{document}
%

\title{Triple-level Model Inferred Collaborative Network Architecture for Video Deraining}

%

\author{Pan Mu, Zhu Liu, Yaohua Liu, 
	   Risheng Liu,~\IEEEmembership{Member,~IEEE}
       Xin Fan,~\IEEEmembership{Member,~IEEE}
\thanks{This work was supported by the National Key R\&D Program of China (2020YFB1313503), the National Natural Science Foundation of China (Nos. 61922019) and the Fundamental Research Funds for the Central Universities.} 
 	\thanks{Pan Mu is with the College of Computer Science and Technology, Zhejiang University of Technology, Hangzhou, China, and also with the School of Mathematical Sciences, Dalian University of Technology, Dalian, China. (e-mail: panmu@zjut.edu.cn)
}
\thanks{Zhu Liu and Yaohua Liu are with the School of Software, Dalian University of Technology, Dalian 116024, China. (email:liuzhu\_dlut@mail.dlut.edu.cn, liuyaohua\_918@163.com)}
\thanks{Risheng Liu is with the DUT-RU International School of Information 	Science \& Engineering, Dalian University of Technology, Dalian 116024, 	China, and also with the Pazhou Lab, Guangzhou, China. (Corresponding author, e-mail: rsliu@dlut.edu.cn)}
\thanks{Xin Fan is with the DUT-RU International School of Information Science 	\& Engineering, Dalian University of Technology, Dalian 116024, China. (email:xin.fan@dlut.edu.cn)}
\thanks{Manuscript received April 19, 2005; revised August 26, 2015.}

}

%
%

\markboth{Journal of \LaTeX\ Class Files,~Vol.~14, No.~8, August~2015}%
{Shell \MakeLowercase{\textit{et al.}}: Bare Demo of IEEEtran.cls for IEEE Journals}
%



\maketitle

\begin{abstract}
	Video deraining is an important issue for outdoor vision systems and has been investigated extensively. However, designing optimal architectures by the aggregating model formation and data distribution is a challenging task for video deraining. In this paper, we develop a model-guided triple-level optimization framework to deduce network architecture with cooperating optimization and auto-searching mechanism, named Triple-level Model Inferred Cooperating Searching (TMICS), for dealing with various video rain circumstances. In particular, to mitigate the problem that existing methods cannot cover various rain streaks distribution, we first design a hyper-parameter optimization model about task variable and hyper-parameter. Based on the proposed optimization model, we design a collaborative structure for video deraining. This structure includes Dominant Network Architecture (DNA) and Companionate Network Architecture (CNA) that is cooperated by introducing an Attention-based Averaging Scheme (AAS). To better explore inter-frame information from videos, we introduce a macroscopic structure searching scheme that searches from Optical Flow Module (OFM) and Temporal Grouping Module (TGM) to help restore latent frame. In addition, we apply the differentiable neural architecture searching from a compact candidate set of task-specific operations to discover desirable rain streaks removal architectures automatically. Extensive experiments on various datasets demonstrate that our model shows significant improvements in fidelity and temporal consistency over the state-of-the-art works. Source code is available at \href{https://github.com/vis-opt-group/TMICS}{https://github.com/vis-opt-group/TMICS}. 
\end{abstract}


%
\IEEEpeerreviewmaketitle

\section{Introduction}
%
%
%
%
\IEEEPARstart{W}{ith} the flourishing development of computer technology, outdoor vision system plays a critical role in many real-world applications. For instance, the advent of low-cost technology in the field of video capture systems has made it easier for various organizations to adopt surveillance technology. However, bad weather harms perceptual performance and degrades video quality. This degradation could prejudice outdoor multimedia systems and influence the visibility of real-world images captured by camera drones. Thus, developing efficient rain streaks removal method is imperative for a wide range of computer vision tasks, such as video surveillance, intelligent vehicles, object detection, tracking, and remote sensing monitoring, etc.

\begin{figure}[t]
	\centering 
	\begin{tabular}{c@{\extracolsep{0.2em}}c}
		\includegraphics[width=0.235\textwidth]{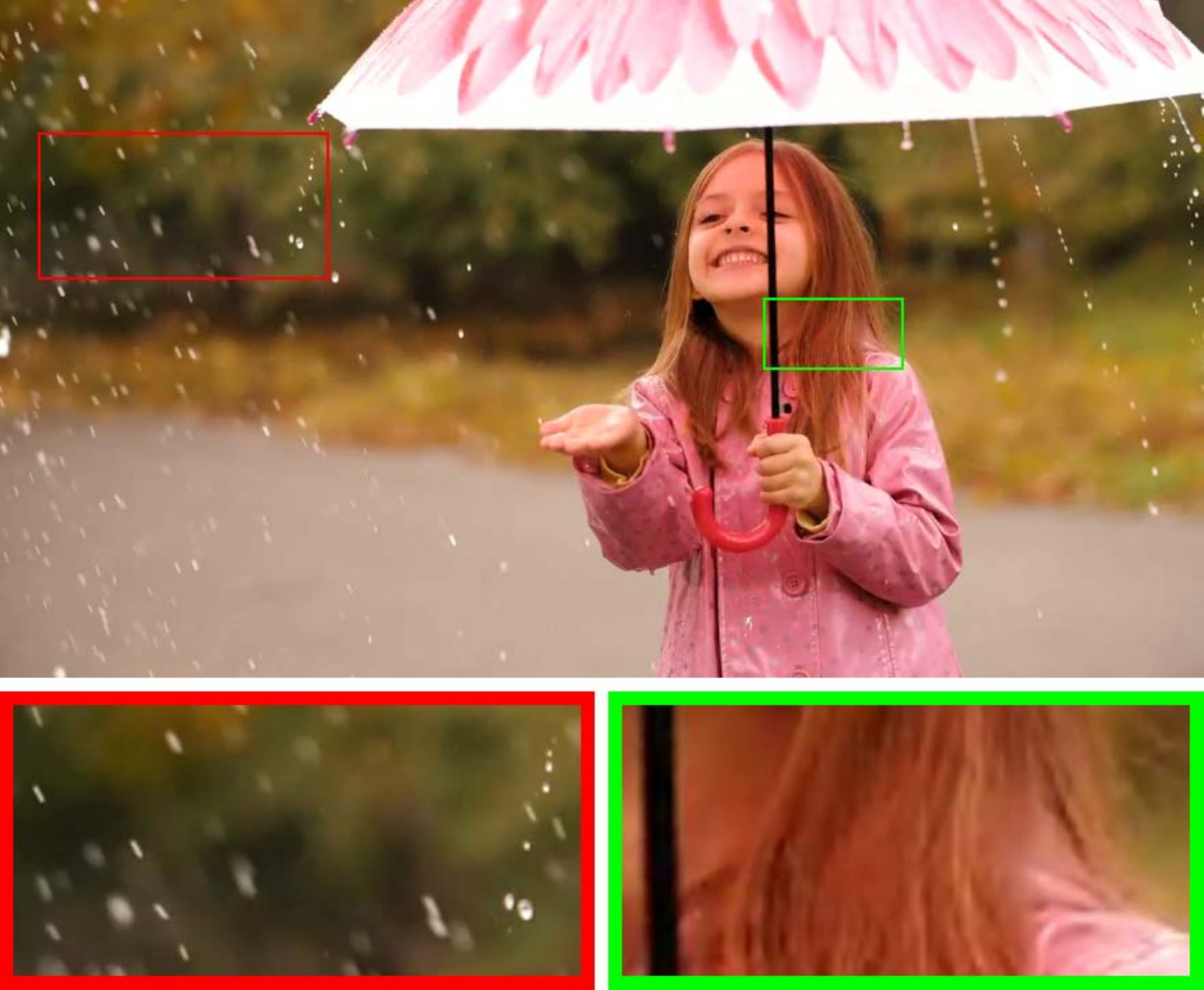}
		&\includegraphics[width=0.235\textwidth]{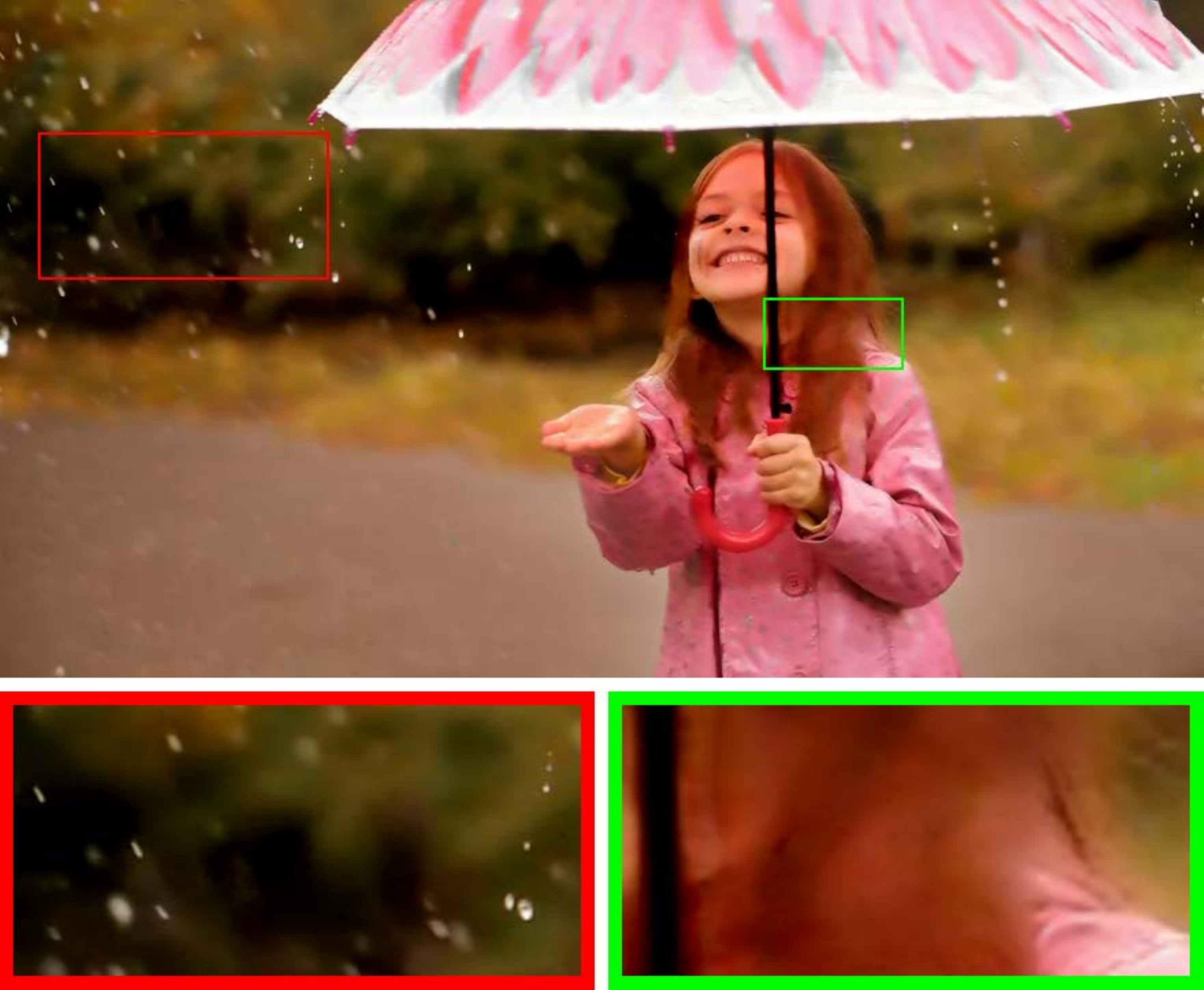}\\
		\footnotesize Input & \footnotesize JORDER \\
		\includegraphics[width=0.235\textwidth]{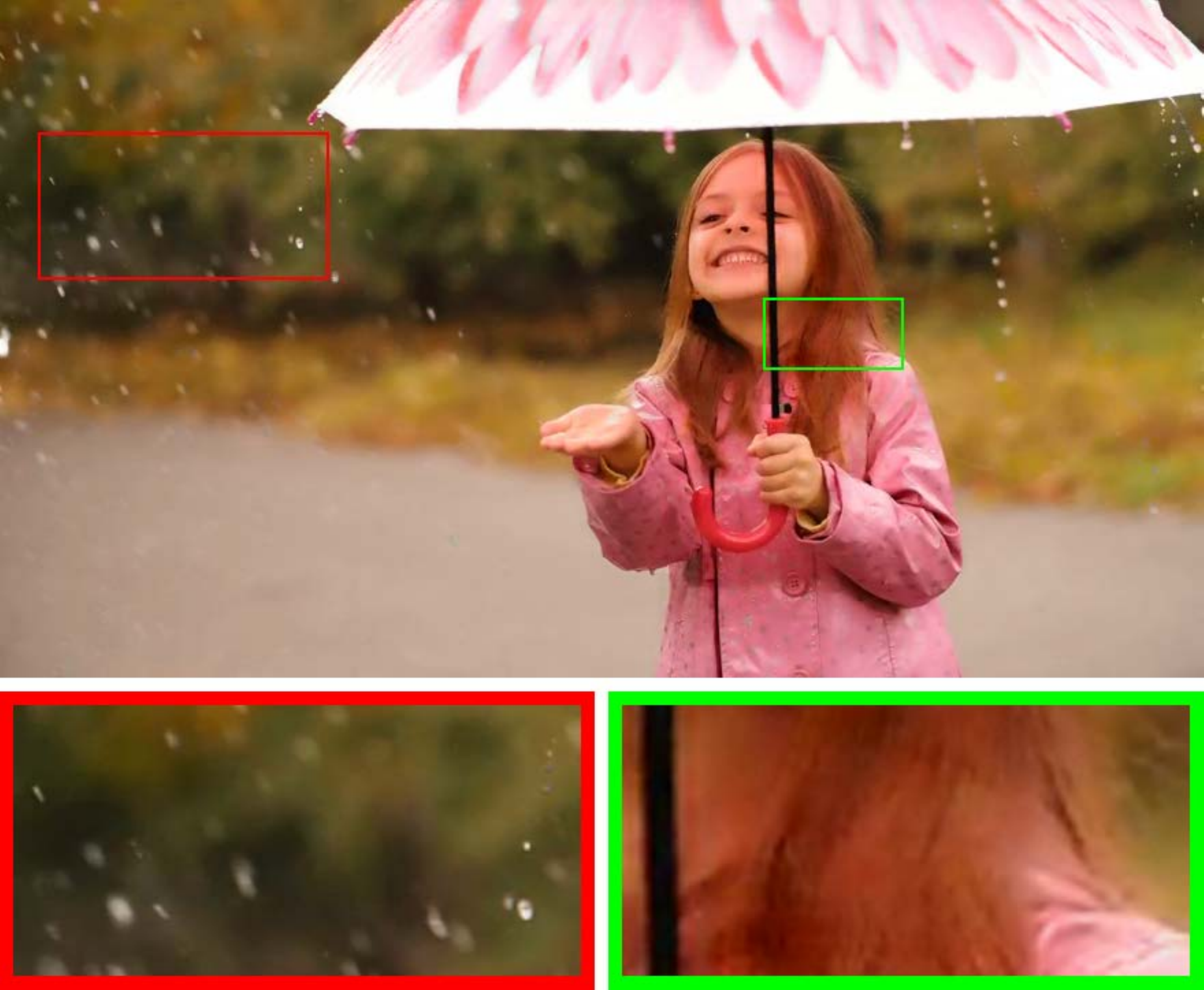}
		&\includegraphics[width=0.235\textwidth]{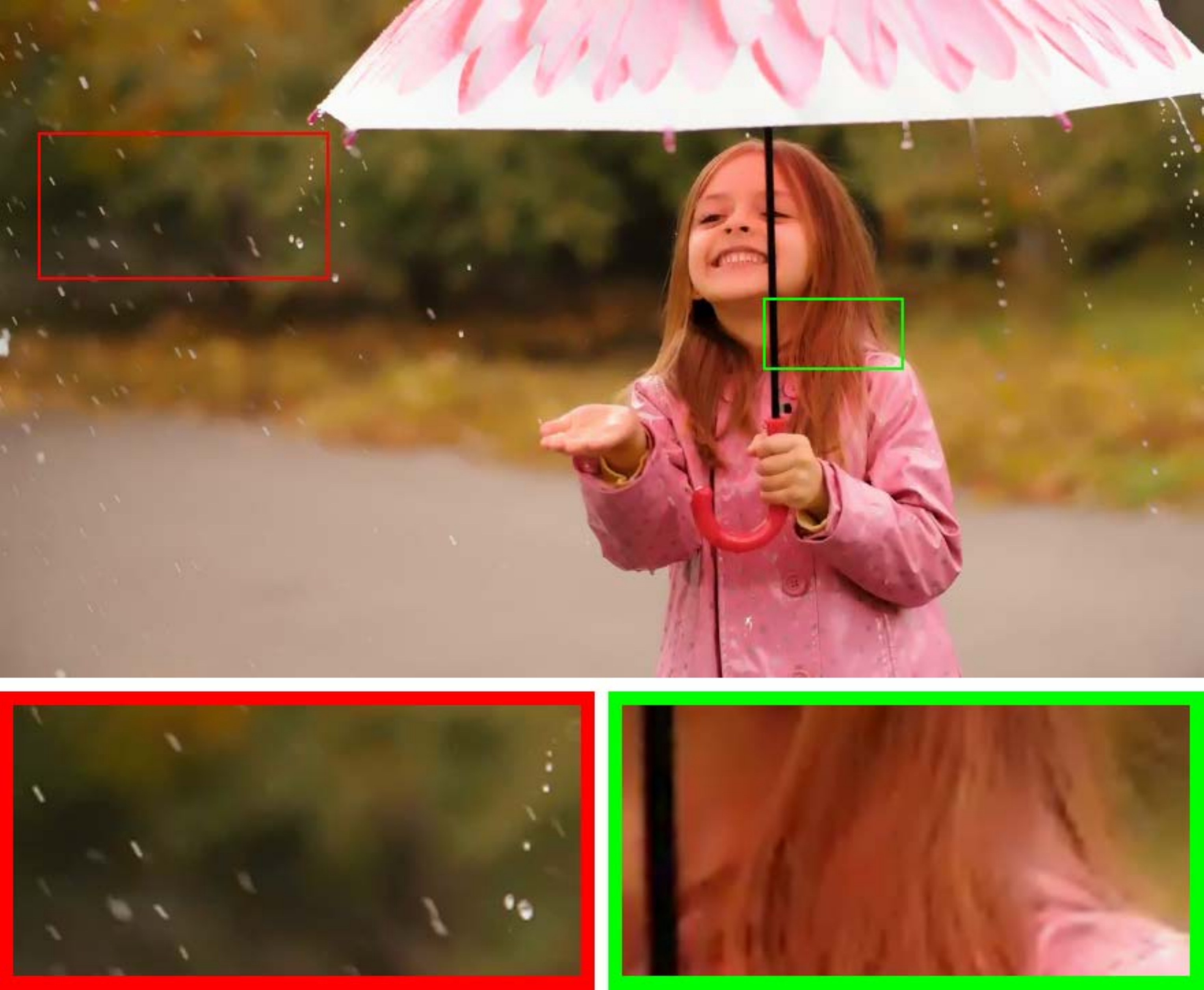}\\
		\footnotesize J4RNet & \footnotesize  FastDerain\\
		\includegraphics[width=0.235\textwidth]{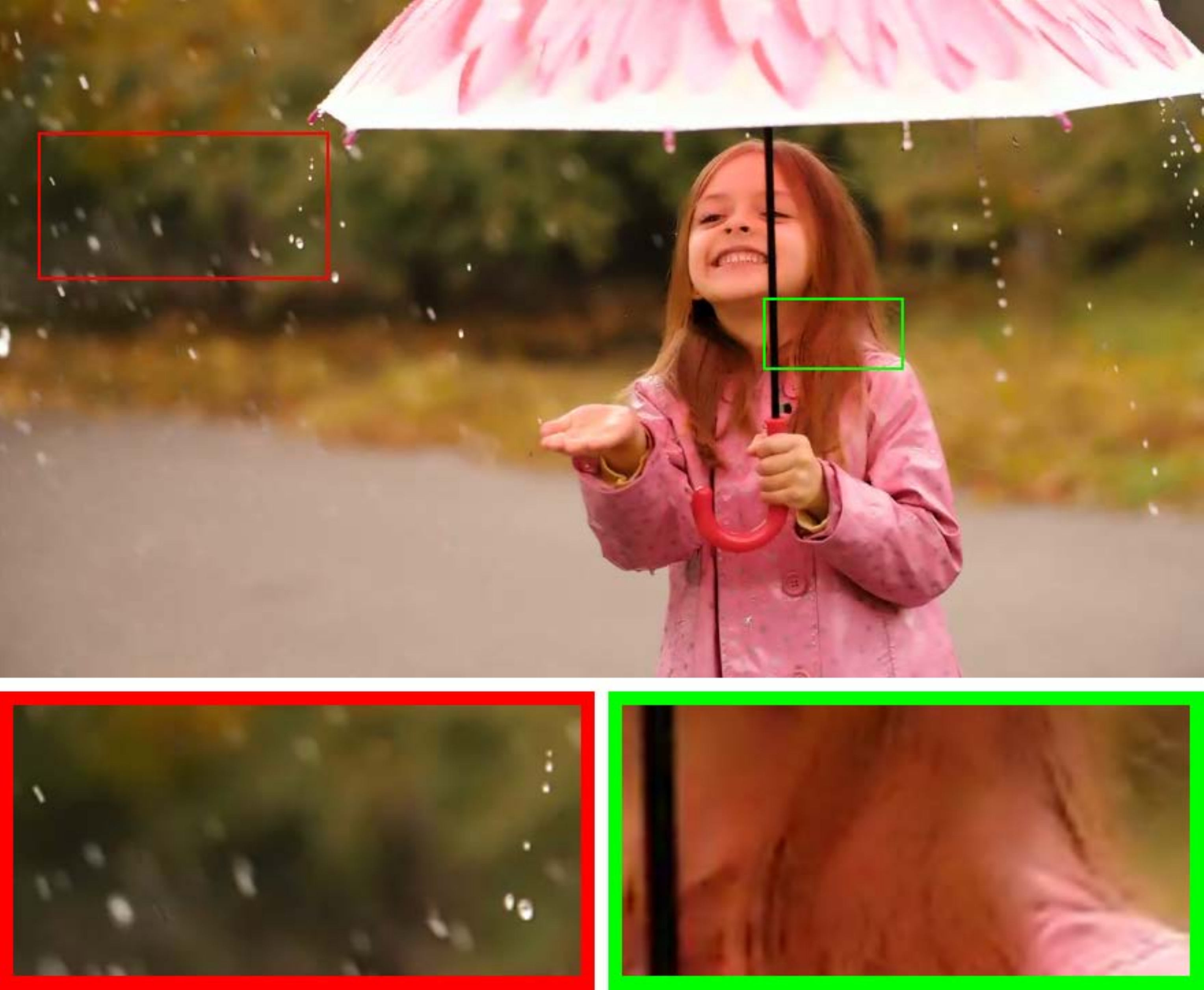}
		&\includegraphics[width=0.235\textwidth]{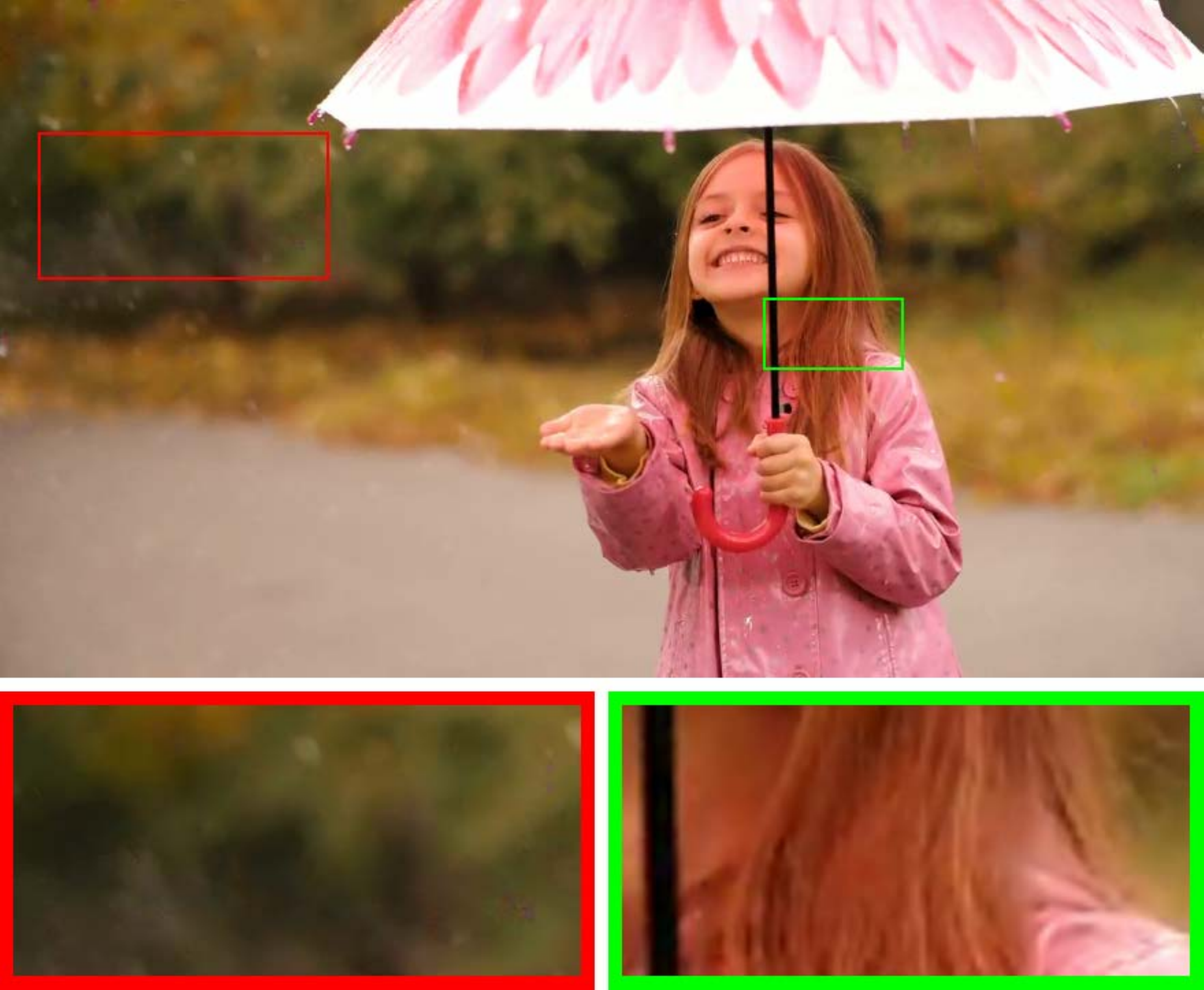}\\
		\footnotesize SPANet  &  \footnotesize TMICS (Ours) \\ 
	\end{tabular}
	\caption{Visual comparison with different video deraining methods on real-world rainy  frames. Comparing with JORDER \cite{yang2017deep}, J4RNet~\cite{liu2018erase}, FastDeRain~\cite{jiang2019fastderain} and SPANet~\cite{wang2019spatial}, our developed method (i.e., TMICS) performs the best visual quality under these various types of complex rain streaks obviously.}\label{fig:first_figure} 
\end{figure}

An early study for video deraining introduced temporal median filters~\cite{starik2003simulation} to deal with each pixel. In~\cite{garg2004detection}, by developing two separate models that capture the dynamics and the photometry of rain through its physical properties, a comprehensive model for the visual appearance of rain was proposed. Subsequently, a large number of research efforts have been dedicated to remove rain streaks from different rain scenes. The existing methods mainly include two categories: frequency domain-based methods and time domain-based schemes. 

Frequency domain-based methods rely on the characteristic in frequency space. For example, in~\cite{barnum2007spatio}, by applying physical and statistical characteristics in the frequency domain, researchers derived a simple rain model to determine the general properties of a single streak. However, as for estimating rain streaks, frequency domain-based methods tend to roughly, and often show errors for complicated rain streaks.

To further improve the performance of video derain, some schemes focus on utilizing temporal dynamics information of consecutive frames to remove rain streaks from rain videos~\cite{chen2013rain,jiang2017novel,ren2017video,liu2018erase,jiang2019fastderain,wang2020video}. Specifically, by dividing rain streaks into two types (i.e., sparse and dense scenes), frequency domain-based matrix decomposition~\cite{ren2017video} methodology was applied for removing rain streaks. Based on the discriminative characteristics of rain streaks, tensor-based video derain methods were developed~\cite{jiang2017novel,li2018video,jiang2019fastderain,wang2020video}. These schemes are not effective for heavy rain due to lacking of data information. Recently, data-driven based deep learning methodologies are popularly used in rain streak removal applications, such as CNN-based reconstruction network~\cite{liu2018erase}, recurrent dual-level flow network~\cite{yang2019frame} and self-learned network~\cite{yang2020self}, etc. However, the performance of CNN-based methods depends on well-designed architectures and subtle adjustments that are hard to be constructed. 

\begin{figure*}[t]
	\centering 
	\begin{tabular}{c}
		\includegraphics[width=1\textwidth]{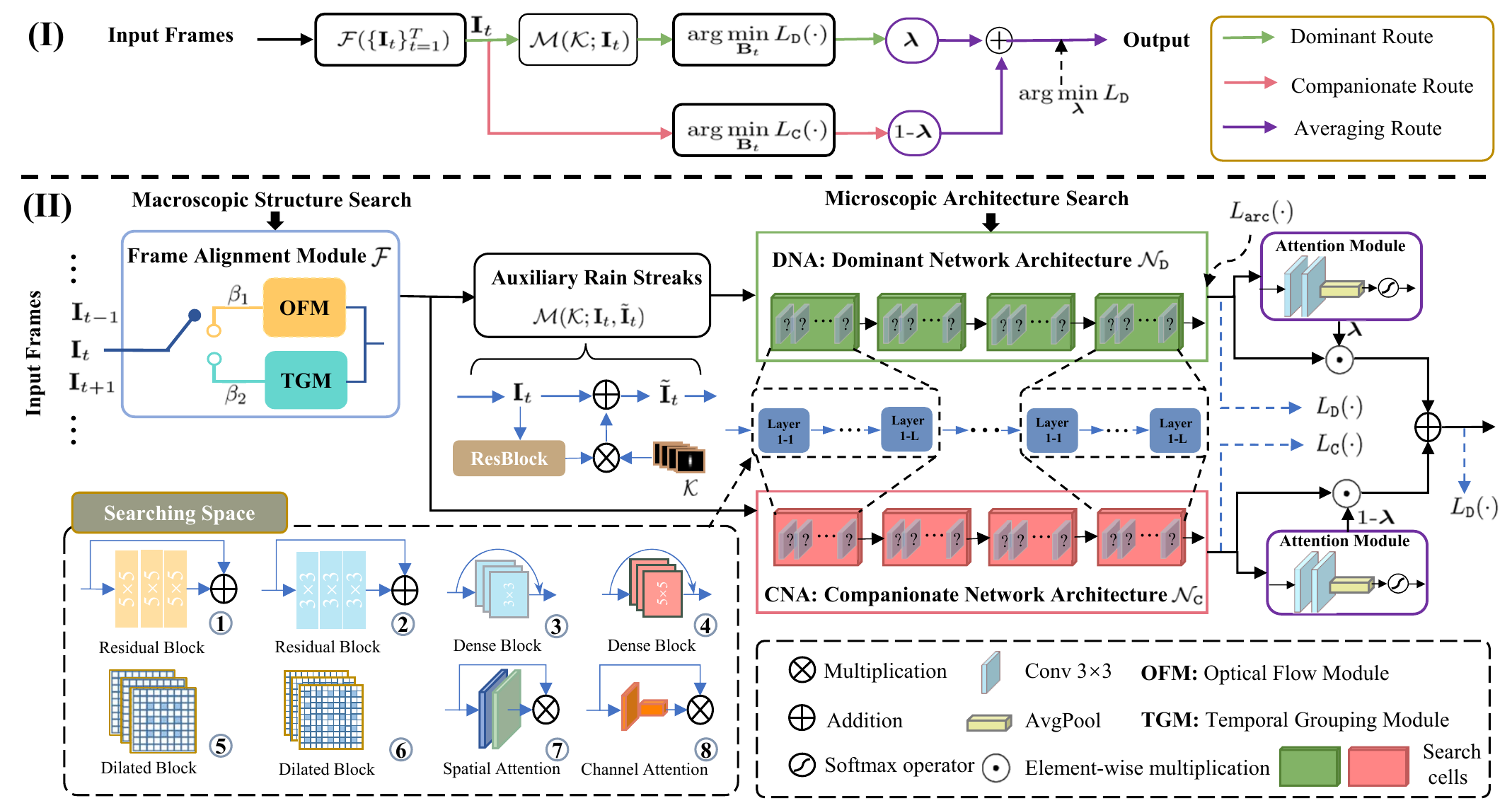}\\
	\end{tabular}
	\caption{Schematic of proposed framework TMICS for video deraining. (I): A rough framework based on the designed model in Eq. \eqref{eq:bi-model}. (II): The detailed procedure of TMICS with deep network architecture.} \label{fig:pipeline}
\end{figure*}

To mitigate the above issues, we develop a model-guided auto-searching method for removing different video rain streaks. Specifically, based on the formulated triple-level video deraining optimization framework, a collaborative learning scheme is deduced via introducing cooperating optimization and network architecture searching mechanism, named as Triple-level Model Inferred Cooperating Searching (TMICS). The pipeline is shown in Figure~\ref{fig:pipeline}. 
Firstly, different from existing video deraining works that only concern motion case, we introduce a macroscopic structure searching scheme from Optical Flow-estimation Module (OFM) and Temporal Grouping Module (TGM) for inter-frames information extraction. Secondly, the existing learning-based video de-raining methods rely on training data and cannot cover rain streaks distribution information that differs significantly from training data. To mitigate this issue, we design a hyper-parameter optimization model about task variable and hyper-parameter. For the task variable propagation, we develop a collaborative structure, i.e., Dominant Network Architecture (DNA) and Companionate Network Architecture (CNA), to deduce the cooperating optimization procedure. In addition, we introduce an Attention-based Averaging Scheme (AAS) to effectively fuse features from collaborative structures by the guidance of hyper-parameter. 
Thirdly, the performance of CNN-based methods depends on well-designed architectures and subtle adjustments that are hard to be constructed. To deal with this problem, based on the above hyper-parameter optimization model, we apply the microscopic network architectures searching~\cite{liu2018darts} from a compact task-specific search space to discover desirable video deraining architectures. Finally, the ablation study demonstrates the effectiveness of the developed modules. Extensive evaluations illustrate that the proposed framework performs favorably against state-of-the-art video deraining methods. Figure~\ref{fig:first_figure} demonstrates the superiority of our proposed approach over existing methods in a challenging video sequence. 
The contributions are summarized as follows: 
\begin{itemize}
	\item We develop a triple-level model-driven framework with macroscopic structure searching and microscopic collaborative architecture searching schemes for video deraining. The designed model not only optimizes task variables (or network parameters), but also optimizes hyper-parameters and architecture weights. 
	\item To mitigate the problem that existing methods cannot cover various rain streaks distribution, we design a hyper-parameter optimization model which has the ability to estimate various rain streaks. We design a collaborative structure, i.e., DNA and CNA, via AAS to preserve details and structure. This structure helps improve the generalization capability of network. 
	\item We apply automatically microscopic search strategy from task-specific search space to discover desirable network. To find a suitable video frame estimation module, we design a macroscopic structure searching scheme via combining OFM and temporal TGM. The above search strategies provide a trade-off between automated search and experience-driven. 
	\item Extensive experiments demonstrate the effectiveness of our proposed method and show comparable results against state-of-the-art video deraining methods on different video rainy benchmarks. 
\end{itemize}

\section{Related Work}

\textbf{Single Image Derain:} 
Traditional single image deraining methods usually apply inherent physical features to characterize rain streaks. For example, in~\cite{kang2011automatic}, sparse coding was applied to divide rain streaks from high frequency layer. Prior-based strategies explore prior knowledge to recover clear image from the rainy one, such as morphological component analysis~\cite{fu2011single}, low-rank assumptions~\cite{chang2017transformed}, guided filter~\cite{zheng2013single,ding2016single}, dictionary learning~\cite{luo2015removing}, Gaussian mixture model~\cite{li2016rain} and joint convolutional analysis and synthesis sparse representation~\cite{gu2017joint}, etc. 
In recent years, deep learning-based methods govern rain streaks removal literature, such as shallow CNN-based schemes~\cite{fu2017removing,ren2019progressive}, dilated convolution~\cite{yang2017deep}, dense blocks~\cite{zhang2018density-aware} and self-supervised method~\cite{yang2020towards}. Besides, model-driven methodologies have also witnessed the rapid progress of deep learning in image deraining field~\cite{wang2020structural,mu2019learning,chen2020flexible,yang2020single,liu2021investigating}. Additionally, by introducing directional gradient operator of arbitrary direction, an efficient and robust constraints-based model was proposed in~\cite{ran2020single}.

\textbf{Video Derain:} 
Different from single image rain streaks removal, video deraining can additionally make use of the temporal correlation and dynamics to explore the intrinsic properties. An early attempt method for video deraining was developed in~\cite{starik2003simulation,garg2004detection,garg2007vision} that utilized a space-time correlation model to analyze the visual effects of rain streaks. Thereafter, a variety of methods are proposed for video deraining. One branch of these methods is to explore the inherent rain streaks priors and general background signals, such as morphological component analysis~\cite{kang2012self}, spatio-temporal correlation of patch groups~\cite{chen2013generalized}, directional prior of rain streaks~\cite{jiang2017novel}, Gaussian mixture model~\cite{wei2017should}, low-rank regularization~\cite{kim2015video,wang2020video}, matrix decomposition~\cite{ren2017video} and tensor model~\cite{jiang2017novel,li2018video,wang2020video}. Deep learning based schemes have been investigated in rain streaks removal application~\cite{liu2018erase,yang2019frame,yang2020self,sharma2021high,yang2021recurrent}. For example, in~\cite{liu2018erase}, a CNN based reconstruction network was developed for video rain streaks removal by integrating rain degradation classification, spatial texture appearances based rain removal. In~\cite{yang2019frame}, a recurrent network was designed for synthesizing visually authentic rain videos to predict the rain-related variables and perform an inverse scheme to estimate rain-free frame.

\textbf{Rain Model:} In general, video rain model can be formulated as $\mathbf{I}_t = \mathbf{B}_t + \mathbf{R}_t$ where $t = 1,\cdots,T$ means the time-steps of video frames, $\mathbf{I}_t\in\mathbb{R}^{m\times n}$, $\mathbf{B}_t^{m\times n}$ and $\mathbf{R}_t^{m\times n}$ denote the captured image with rain streaks, rain-free background and rain streaks respectively. In~\cite{ren2017video}, a reconstructed rain model was developed by separating rain streaks into two types, i.e. sparse and dense rain streaks. This model can be written as $$\mathbf{I}_t = \mathbf{B}_t + \mathbf{F}_t + \mathbf{R}_t^s + \mathbf{R}_t^d,$$ where $\mathbf{F}_t$, $\mathbf{R}_t^s$ and $\mathbf{R}_t^d$ represent the intensity fluctuations caused by foregrounds, sparse and dense rain streaks respectively. 
In~\cite{yang2019frame}, rain accumulation and accumulation flow were considered in the following form $$\mathbf{I}_t = \beta_t\mathbf{B}_t + \mathbf{R}_t + (1-\beta_t)\mathbf{A}_t + \mathbf{U}_t,$$ where $\mathbf{A}_t$ is the global atmospheric light, $\beta_t$ means the atmospheric transmission which is correlated with the scene depth and $\mathbf{U}_t$ denotes the rain accumulation flow layer.

\textbf{Neural Architecture Search:} Neural Architecture Search (NAS)~\cite{elsken2019neural} aims to discover high-performance task-specific architectures to replace heavy manual design automatically. Early search strategies for NAS apply evolutionary algorithm~\cite{real2019regularized,real2017large} and reinforcement learning~\cite{zoph2016neural}, which achieve remarkable performance along with much inefficient computation, spent on architecture evaluations at the same time. In light of these computation issues, various of gradient-based differentiable approaches have been proposed. 

By performing gradient-descent form for the bi-level search scheme, a continuous relaxation method is developed~\cite{liu2018darts} to optimize both the model weights and architecture parameters. Several differentiable architecture search algorithms also show comparable performance on low-level vision tasks. As for image restoration (e.g., image deraining), HiNAS~\cite{zhang2020memory} searches for both inner cell architectures and outer layer widths to be memory and computation efficient. CLEARER~\cite{gou2020clearer} combines multi-scale search space with task-flexible modules for image denoising and deraining.

\section{Our Approach}\label{sec:alg}

In this section, we first construct a triple-level optimization model to jointly optimize network architectures, task variables and hyper-parameters in Section~\ref{sec:model}. Subsequently, we provide a detailed procedure to deduce the optimization model in Section~\ref{sec:alg_procedure}. Finally, the training details are illustrated in Section~\ref{sec:train_details}. The overall procedure is presented in Algorithm~\ref{alg:procedure} and the detailed pipeline is shown in Figure~\ref{fig:pipeline}.

\subsection{Model Formulation for Video Deraining}\label{sec:model}
In this work, we take into account more complicated rain synthesis model and exploring the temporal information of rainy videos. With this goal, we first re-construct rainy video frame as the following model 
\begin{equation}\label{eq:rain_model}
\mathbf{I}_t = \bm{\lambda}\circ(\mathbf{B}_t + \mathbf{R}_t)+(1-\bm{\lambda})\circ(\mathbf{B}_t + \mathcal{K}\otimes\mathbf{R}_t),\ t = 1,\cdots,T,
\end{equation}
where $\mathbf{I}_t\in\mathbb{R}^{m\times n}$, $\mathbf{B}_t\in\mathbb{R}^{m\times n}$ and $\mathbf{R}_t\in\mathbb{R}^{m\times n}$. Here $\circ$ and $\otimes$ represent the element-wise multiplication and convolution operation respectively. $\mathcal{K}$ denotes the constructed model aiming to generate auxiliary rain streaks. $\bm{\lambda}\in\mathbb{R}^{m\times n}$ means the weight between different rain streaks types. Indeed, $\mathbf{R}_t$ is usually considered as independent and identically distributed sample.

Note that the main aim of video deraining is to find a clear background frame, i.e., $\mathbf{B}_t$. Under the above model~\eqref{eq:rain_model}, the introduced hyper-parameter $\bm{\lambda}$ aims to help us find a better task variable. Thus, a hyper-parameter optimization model (see~\cite{liu2021generic,liu2021investigatingsurvy,liu2021towards}) is introduced to jointly optimize task variable $\mathbf{B}_t$ and hyper-parameter $\bm{\lambda}$ that can be formulated as the following scheme 
\begin{equation}\label{eq:bi-model}
\begin{array}{l}
\min\limits_{\bm{\lambda}} L_{\mathtt{D}}(\bm{\lambda},\mathbf{B}_t(\bm{\lambda});\mathcal{K},\mathbf{I}_t)\\ 
s.t., \mathbf{B}_t(\bm{\lambda})\in\arg\min L_{\mathtt{C}}\left(\mathbf{B}_t(\bm{\lambda});\mathbf{I}_t\right),\\
\quad\quad\quad\ \ \mathbf{I}_{t} = \mathcal{F}\left(\{\mathbf{I}_p\}_{p=1}^T\right),
\end{array}
\end{equation}
where $L_{\mathtt{D}}$ and $L_{\mathtt{C}}$ denote the loss function in the upper-level and lower-level respectively, $\mathcal{F}(\cdot)$ represents the temporal alignment module, and $T$ is the input frame number. Indeed, module $\mathcal{F}(\cdot)$ is designed for introducing frame information which is usually used through different forms in video derain methods~\cite{yang2019frame,yang2020self}.

As for optimizing $\mathbf{B}_t$ in Eq.~\eqref{eq:bi-model}, it can be written as an averaging scheme based on~\cite{liu2020generic,sabach2017first,liu2020investigating}. Thus, for any fixed $\bm{\lambda}$, we have the following averaging-based scheme
\begin{equation}\label{eq:SAM}
\mathbf{B}_t\! \gets\! \bm{\lambda}\circ \underbrace{\arg\min_{\mathbf{B}_t} L_{\mathtt{D}}(\bm{\lambda},\mathbf{B}_t;\mathcal{K},\mathbf{I}_t)}_{\mathcal{N}_{\mathtt{D}}(\bm{\omega}^*_\mathtt{D};\mathcal{K},\mathcal{D}_{tr}(\mathbf{I}_t,\mathbf{I}_{gt}))}\! + (1-\bm{\lambda})\circ\underbrace{\arg\min\limits_{\mathbf{B}_t} L_{\mathtt{C}}(\mathbf{B}_t;\mathbf{I}_t)}_{\mathcal{N}_{\mathtt{C}}(\bm{\omega}^*_\mathtt{C};\mathcal{D}_{tr}(\mathbf{I}_t,\mathbf{I}_{gt}))},
\end{equation}
where $\mathcal{N}_{\mathtt{D}}$ and $\mathcal{N}_{\mathtt{C}}$ respectively denote the constructed Dominant Network Architecture (DNA) and Companionate Network Architecture (CNA). $\bm{\omega}^*_{\mathtt{D}}$ and $\bm{\omega}^*_{\mathtt{C}}$ represent the learned optimal network parameters for $\mathcal{N}_{\mathtt{D}}$ and $\mathcal{N}_{\mathtt{C}}$ respectively. $\mathbf{I}_t$ and $\mathbf{I}_{gt}$ represent the input rainy image of the current frame and the corresponding ground truth respectively. As for $\bm{\lambda}$-subproblem, a single numerical parameter cannot well cooperate the constructed dominant and companionate network. Thus, with the above obtained $\mathbf{B}_t$, we introduce an Attention-based Averaging Scheme (AAS) to learn $\bm{\lambda}$-subproblem.

Moreover, designing network by experience in video deraining requires substantial efforts. How to trade-off between automated search and experience-driven is a crucial task for discovering structure $\mathcal{F}$ and task-specific architectures (i.e., $\mathcal{N}:=\{\mathcal{N}_{\mathtt{D}}, \mathcal{N}_{\mathtt{C}}\}$). Motivated by the success of NAS in high-level vision field~\cite{liu2018darts}, under the bi-level optimization scheme in Eq.~\eqref{eq:bi-model}, we construct a triple-level optimization model to search network architectures for video rain streaks removal. The above model can be reformulated as the following form 
\begin{equation}
\begin{aligned}\label{eq:nas_obj}
&\min\limits_{\bm{\alpha},\bm{\beta}} L_{\mathtt{arc}}\left(\bm{\beta},\mathcal{N}_{\mathtt{D}}(\bm{\omega}^*_{\mathtt{D}}(\bm{\alpha})),\mathcal{N}_{\mathtt{C}}(\bm{\omega}^*_{\mathtt{C}}(\bm{\alpha}));\mathcal{D}_{val}\right),\\
&s.t.,\left\{
\begin{array}{l}
\bm{\omega}^*_{\mathtt{D}}(\bm{\alpha})\in\arg\min\limits_{\bm{\omega}_{\mathtt{D}}} L_{\mathtt{D}}(\bm{\lambda},\mathcal{N}_{\mathtt{D}}(\bm{\omega}_{\mathtt{D}}(\bm{\alpha});\mathcal{K},\mathcal{D}_{tr}(\mathbf{I}_t,\mathbf{I}_{gt}))),\\
\bm{\omega}^*_{\mathtt{C}}(\bm{\alpha})\in\arg\min\limits_{\bm{\omega}_{\mathtt{C}}} L_{\mathtt{C}}(\mathcal{N}_{\mathtt{C}}(\bm{\omega}_{\mathtt{C}}(\bm{\alpha});\mathcal{D}_{tr}(\mathbf{I}_t,\mathbf{I}_{gt}))),\\
\quad\quad \mathbf{I}_{t} = \mathcal{F}\left(\bm{\beta};\{\mathbf{I}_t\}_{t=1}^T\right),
\end{array}
\right.
\end{aligned}
\end{equation}
where $\bm{\alpha}$ and $\bm{\beta}$ are the architecture relaxation weights, $\mathcal{D}_{tr}$ and $\mathcal{D}_{val}$ denote training and validation datasets. Indeed, Eq.~\eqref{eq:bi-model} and Eq.~\eqref{eq:nas_obj} imply a triple-level optimization model with network parameters $\bm{\omega}=(\bm{\omega}_{\mathtt{D}}$, $\bm{\omega}_{\mathtt{C}})$ (the first level), architecture parameter $\bm{\alpha}$, $\bm{\beta}$ (the second level) and hyper-parameter $\bm{\lambda}$ (the third level). Then, we provide detailed procedure for finding the optimal solution of the above optimization formulation in the following section~\ref{sec:alg_procedure}. We summarize some necessary definitions in Table~\ref{tab:definition}.

\subsection{Optimization Procedure}\label{sec:alg_procedure}

With the developed optimization models in Eqs.~\eqref{eq:bi-model}-\eqref{eq:SAM}, this part introduces microscopic architecture searching for $\mathcal{N}=\{\mathcal{N}_{\mathtt{D}},\mathcal{N}_{\mathtt{C}}\}$ and macroscopic structure searching for frame alignment module $\mathcal{F}$. Specifically, we construct a collaborative rain streaks removal structure, i.e., $\mathcal{N}_{\mathtt{D}}$ and $\mathcal{N}_{\mathtt{C}}$, to jointly characterize different rain streaks distribution. Then, with the above structures, we design an attention-based module to simulate $\bm{\lambda}$. 
\subsubsection{Optimizing Hyper-parameter}
As for $\bm{\lambda}$, we design an attention mechanism to obtain an adaptive map that can directly learn inter-channel information from global context and contribute to our performance. Indeed, the attention-based module auto-cooperates different video rain streaks distribution and helps improve the generalization capability of the network. In detail, given two restored frames, we perform two shared $3\times3$ convolutions to extract features and then introduce the global average pooling to encode weights of each spatial channels. Soft-max operator is adopted to generate $\bm{\lambda}$ for each estimated frame. 

\subsubsection{Macroscopic Structure Search for Frame Alignment Scheme $\mathcal{F}$} 
\begin{figure}[t]
	\centering 
	\begin{tabular}{c}
		\includegraphics[width=0.48\textwidth]{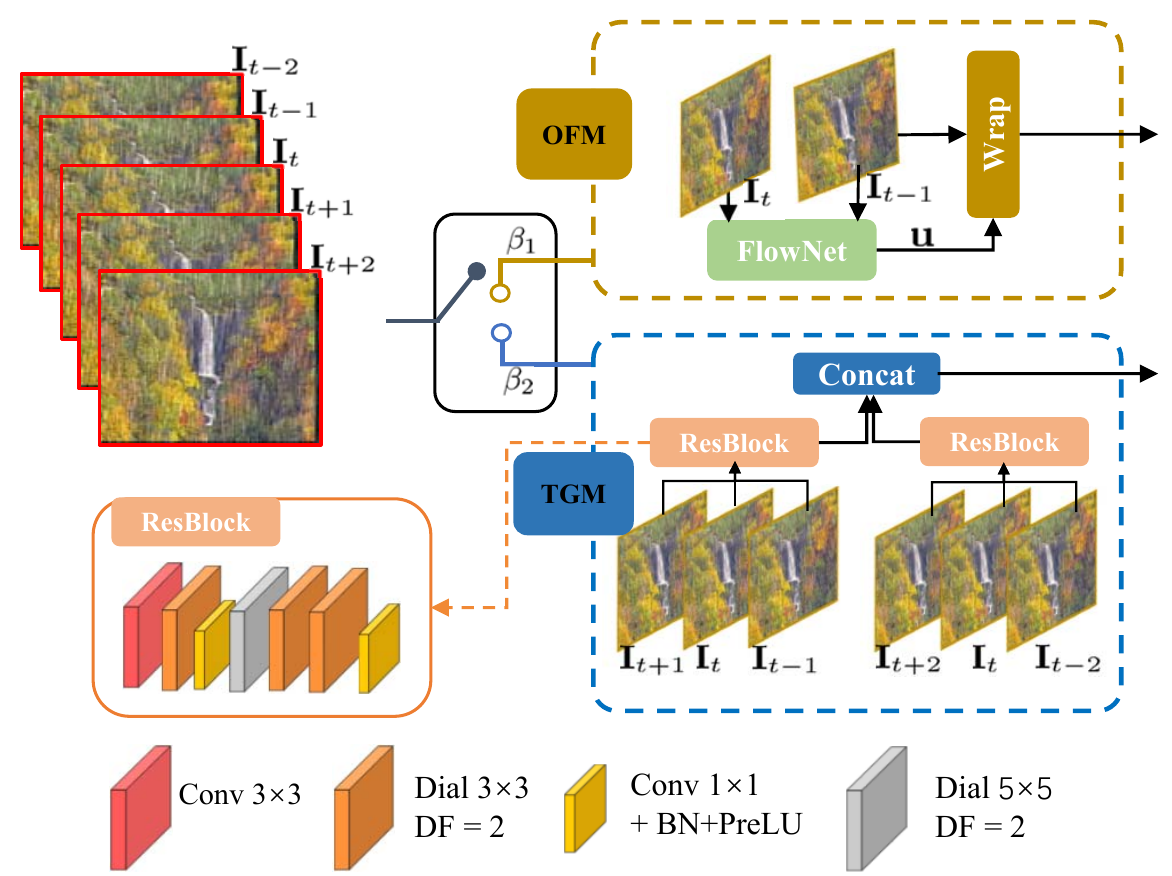}\\
	\end{tabular}
	\caption{A diagram of the architecture to illustrate OFM and TGM. Five consecutive frames are used to derain the middle frame. ``Concat'' means the concatenation operation.} \label{fig:frame_structure}
\end{figure}

Choosing suitable alignment modules to estimate spatial-temporal information from task-specific data distribution is important for video deraining. In this work, we introduce a macroscopic structure search mechanism to select a suitable alignment module. In other words, we choose either TGM or OFM for frame alignment module $\mathcal{F}$ by NAS. Indeed, we can formulate this search strategy using relaxation weights,
\begin{equation*}
\mathcal{F}\left(\bm{\beta};\{\mathbf{I}_t \}_{t=1}^T\right)= \bm{\beta}_{1} \mathcal{F}_\mathtt{OFM}(\{\mathbf{I}_t \}_{t=1}^T) +  \bm{\beta}_{2} \mathcal{F}_\mathtt{TGM}(\{\mathbf{I}_t \}_{t=1}^T),
\end{equation*}
where $\sum_i \bm{\beta}_{i}=1$. We leverage these hybrid aligned features to feed $\mathcal{N}$ in the search phase. The detailed structure is shown in Figure~\ref{fig:frame_structure}.

For the motion estimation (i.e., OFM), by applying recurrent all-pairs field transforms as stated in~\cite{teed2020raft}, we introduce an optical flow prediction to obtain the spatial-temporal information from a sequence video frames $\{\mathbf{I}_{t}\}_{t = 1,\dots,T} $. With this pre-trained network, it produces a series of aligned frames for subsequent spatial-temporal information extraction,
\begin{equation*}
\mathcal{F}_{\mathtt{OFM}}\left(\{\mathbf{I}_t \}_{t=1}^T\right)=\mathbf{I}_{t} + \mathbf{u},
\end{equation*}
where $\mathbf{u}$ means the updated flow.

For TGM, the input sequence is divided into two groups~\cite{isobe2020video}. Specifically, given a consecutive rainy video frame sequence, we select $2M+1$ neighboring frames to calculate the reference frame $\mathbf{I}_t$. The two groups are 
\begin{equation*}
\begin{array}{l}
	G_1:=\{\mathbf{I}_{t-2M},\cdots,\mathbf{I}_{t-2}, \mathbf{I}_t, \mathbf{I}_{t+2},\cdots,\mathbf{I}_{t+2M}\}, \\
	G_2:=\{\mathbf{I}_{t-2M+1},\cdots,\mathbf{I}_{t-1}, \mathbf{I}_t, \mathbf{I}_{t+1},\cdots,\mathbf{I}_{t+2M-1}\}.
\end{array}
\end{equation*}
Note that each group represents a certain type of frame rate. Then, the residual block and fusion block with shared weights are applied to extract and fuse spatial-temporal information within the above two groups, 
\begin{equation*}
	\mathcal{F}_{\mathtt{TGM}}\left(\{\mathbf{I}_p \}_{p=1}^T\right)= Concat(\mathcal{N}_{\mathtt{Res}}(G_1),\mathcal{N}_{\mathtt{Res}}(G_2)),
\end{equation*}
where $\mathcal{N}_{\mathtt{Res}}$ is the residual block as stated in Figure \ref{fig:frame_structure}.

\subsubsection{Constructing Auxiliary Rain Streaks} 
After the above calculation, we roughly estimate rain streaks by using a simple residual network structure $\mathcal{N}_{\mathtt{Res}}$ which can be written as $\widetilde{\mathbf{R}}_t = \mathcal{N}_{\mathtt{Res}}(\mathbf{I}_t)$. We apply the estimated rain streaks $\widetilde{\mathbf{R}}_t$ to encode the physical structural factors underlying rain streaks which can be written as 
\begin{equation*}
\widehat{\mathbf{R}}_t\gets\sum_{i=1}^N W_i * (\mathcal{K}_i\otimes\widetilde{\mathbf{R}}_t) + \widetilde{\mathbf{R}}_t.
\end{equation*}
This model enables the capability of characterizing a wide range of rain streaks, such as small/large, blur rain streaks etc. Here $\mathcal{K}_i$ represents the constructed filter, $W_i$ denotes the corresponding weight, and $*$ means multiply operation. In this model, we apply two groups of filters to simulate rain types. Consequently, the re-constructed frames $\tilde{\mathbf{I}}_t$ can be 
\begin{equation*}
\tilde{\mathbf{I}}_t\gets\mathcal{M}(\mathcal{K};\mathbf{I}_t):=\mathbf{I}_t + \widehat{\mathbf{R}}_t. 
\end{equation*}
As real-world rainy scene usually contains many rain streaks cases, thus it is troublesome to cover intricate rain streaks distribution (such as heavy or combined rain streaks) with simple rain model. The constructed model provides us an opportunity to enable the capability of characterizing a wide range of rain streaks, such as small/large, blur rain streaks etc. 

\subsubsection{Microscopic Architecture Search for $\mathcal{N}$}\label{sec:nas} 

\begin{figure}[t]
	\centering 
	\begin{tabular}{c}
		\includegraphics[width=0.45\textwidth]{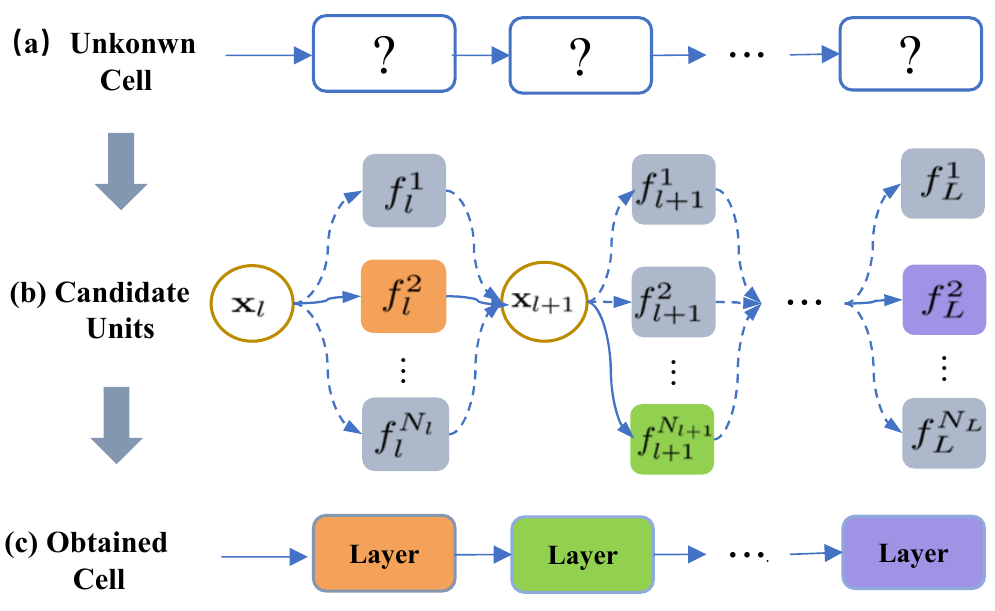}\\
	\end{tabular}
	\caption{A diagram of the auto-searching architecture to illustrate the optimization procedure of one cell. (a) The unknown units in initial cell. (b) The searching procedure on candidate units with continuous relaxation. (c) The final obtained cell through the learned relaxation weights.
	} \label{fig:nas_structure}
\end{figure}

Generally, designing an efficient task-specific neural network structure (i.e., $\mathcal{N}$) requires substantial architecture engineering. In other words, it is significant but difficult to decide which module and how many convolutions, dilation, residual blocks are applied in a network. Focus on these key points, we apply the differentiable architecture search strategy~\cite{liu2018darts} from a discrete set of candidate operation cells to discover the task-specific DNA and CNA network simultaneously. These cells consist of an ordered sequence of $L$ nodes to automatically discover desirable rain streaks removal architectures, and the pipeline is shown in Figure~\ref{fig:nas_structure}. For each intermediate, it relaxes the discrete architecture parameters into a continuous distribution to perform a differentiable search.

\begin{algorithm}[t]
	\caption{Procedure for solving the model in Eq.~\eqref{eq:nas_obj}}
	\label{alg:procedure}
	\begin{algorithmic}[1]
		\REQUIRE Input necessary parameters.
		\ENSURE $\mathbf{x}^{\ast}$.
		\WHILE{not converged}
		\STATE Estimating frame content with $\mathbf{I}_t \gets\mathcal{F}\left(\bm{\beta};\{\mathbf{I}_p\}_{p=1}^T\right)$.\label{step:flow}
		\STATE Obtaining complex rain streaks by $\tilde{\mathbf{I}}_t\gets \mathcal{M}(\mathcal{K};\mathbf{I}_t)$.\label{step:generator_rain}
		\STATE Learning microscopic architecture $\mathcal{N}$ through:
		\WHILE{not converged}
		\STATE 
		Updating $\bm{\omega_{\mathtt{D}}}^*(\bm{\alpha})$, $\bm{\omega_{\mathtt{D}}}^*(\bm{\alpha})$ with weighted shared $\bm{\alpha}$ by \\
		$\bm{\omega}^*_{\mathtt{D}}(\bm{\alpha})\in\arg\min\limits_{\bm{\omega}_{\mathtt{D}}} L_{\mathtt{D}}(\bm{\lambda},\mathcal{N}_{\mathtt{D}}(\bm{\omega}_{\mathtt{D}}(\bm{\alpha});\mathcal{K},\mathcal{D}_{tr}(\mathbf{I}_t,\mathbf{I}_{gt}))),$
		$\bm{\omega}^*_{\mathtt{C}}(\bm{\alpha})\in\arg\min\limits_{\bm{\omega}_{\mathtt{C}}} L_{\mathtt{C}}(\mathcal{N}_{\mathtt{C}}(\bm{\omega}_{\mathtt{L}}(\bm{\alpha});\mathcal{D}_{tr}(\mathbf{I}_t,\mathbf{I}_{gt}))).$
		\ENDWHILE
		\STATE Updating architecture $\bm{\alpha}$ and $\bm{\beta}$ by gradient descent:\\
		$\bm{\beta} \in\arg\min\limits_{\bm{\beta}} L_{\mathtt{arc}}(\bm{\beta},\mathcal{N}_{\mathtt{D}}(\bm{\omega}^*_{\mathtt{D}}(\bm{\alpha})),\mathcal{N}_{\mathtt{C}}(\bm{\omega}^*_{\mathtt{C}}(\bm{\alpha}));\mathcal{D}_{val}).$\\
		$\bm{\alpha}\in\arg\min\limits_{\bm{\alpha}} L_{\mathtt{arc}}(\bm{\beta},\mathcal{N}_{\mathtt{D}}(\bm{\omega}^*_{\mathtt{D}}(\bm{\alpha})),\mathcal{N}_{\mathtt{C}}(\bm{\omega}^*_{\mathtt{C}}(\bm{\alpha}));\mathcal{D}_{val}).$
		\STATE With the obtained $\bm{\omega}^*_{\mathtt{D}}$ and $\bm{\omega}^*_{\mathtt{C}}$, updating $\bm{\lambda}$ by
		\STATE $\bm{\lambda} \in\arg\min L_{\mathtt{D}}(\bm{\lambda},\mathcal{N}_{\mathtt{D}}(\bm{\omega}^*_{\mathtt{D}}),\mathcal{N}_{\mathtt{D}}(\bm{\omega}^*_{\mathtt{C}});\mathcal{K},\mathcal{D}_{tr}).$
		\STATE Output $\bm{\omega}^*_{\mathtt{D}}$, $\bm{\omega}^*_{\mathtt{C}}$, $\bm{\beta}$, $\bm{\alpha}$ and $\bm{\lambda}$.
		\ENDWHILE
	\end{algorithmic}
\end{algorithm}

\begin{table*}[htb!]
	\centering
	\caption{Summary of some important notations. }
	\setlength{\tabcolsep}{0.8mm}{
		\begin{tabular}{c|c||c|c||c|c}
			\hline
			Notation & Description & Notation & Description & Notation & Description\\
			\hline
			$\circ$  & Element-wise multiplication & $\mathbf{I}_t$ & Input frame &  $\mathcal{F}$ & Frame alignment scheme\\
			$\otimes$ & Convolution operation & $\mathbf{B}_t$ & Background frame &$\mathcal{F}_\mathtt{OFM}$ & Optical flow-estimation module  \\
			$\ast$    &  Multiply & $\mathbf{R}_t$ & Rain streaks &   $\mathcal{F}_\mathtt{TGM}$ & Temporal grouping module \\
			\hline
			$\mathcal{K}$ &  Constructed filter & $\bm{\alpha}$ & Architecture weights &  $\mathcal{N}\!:=(\mathcal{N}_{\mathtt{D}},\mathcal{N}_{\mathtt{C}})$ & Task-specific architectures \\
			$\bm{\lambda}$ & Hyper-parameter & $\bm{\beta}$ &  Architecture weights &  $\mathcal{N}_{\mathtt{Res}}$ & Residual Block \\
			$\mathcal{M}$  & Generating complex input frame &$\bm{\omega}$ &  Network parameters & &\\
			\hline
		\end{tabular}
	}
	\label{tab:definition}
\end{table*}

\begin{table}[t]
	\small
	\centering
	\caption{Summary of the training / test image number and rain streaks types of four different datasets used in this work. }
	\setlength{\tabcolsep}{0.3mm}{
		\begin{tabular}{c|c|c}
			\Xhline{0.8pt}
			Dataset & {\tabincell{c}{Image Num.\\ training / test}} & Rain Types  \\		
			\hline
			RainSynLight25 &  1900 / 775 &   sythetic rain streaks~\cite{liu2018erase}  	\\
			RainSynComplex25 & 1900 / 775  &   sythetic rain streaks~\cite{liu2018erase} 	\\
			NTURain &  2400 / 1682  &  sythetic and real rain streaks \cite{chen2018robust}	\\
			LasVR &  -- / 200 & \tabincell{c}{sythetic rain streaks~\cite{liu2019removing}}	\\
			\Xhline{0.8pt}
		\end{tabular}
	}
	\label{tab:datasets}
\end{table}

\textbf{Task-oriented Search Space.}  
As proper search space is crucial for finding the architecture backbone, we introduce a series of basic operators as the compact candidate search space, including residual blocks, dense blocks, dilated convolutions and attention mechanisms, which have been corroborated their effectiveness for rain removal tasks~\cite{yang2017deep,wang2019spatial}. The pre-defined operators are listed in the following: 
\begin{itemize}
	\item Residual:  $3\times 3$ and $5\times 5$ residual blocks;
	\item Dense:  $3\times 3$  and  $5\times 5$ dense blocks;
	\item Attention:  spatial attention and  channel attention;
	\item Dilated:  $3\times 3$ and  $5\times 5$ dilated convolution with DF=2,
\end{itemize}
where DF denotes the dilated rate. Note that each convolution is followed by a batch normalization layer and a ReLU activation layer, and all operators include three layers of convolutions. We take $l$-th layer as an example and calculate the choice of a particular operation by relaxing networks with a soft-max operation. Let $\mathbf{x}_l$ be the input of the $l$-th layer and $ l\in\{1,\cdots,L \} $, then we have 
\begin{equation}
	\mathbf{x}_{l+1} = \sum_{i=1}^{N_l} {\alpha}^{i}_l f^i_{l}(\mathbf{x}_l),
\end{equation}
where $f^i_{l}$ denotes the operation realized by the $i$-th operation at $l$-th layer and ${\alpha}^{i}_l$ means the corresponding weight. Here ${\alpha}^{i}_l\in [0,1]$ and $\sum_i{\alpha}^{i}_l=1$.

\textbf{The Final Derived Architectures.} 
Subsequently, we can leverage the differentiable search to perform Eq.~\eqref{eq:nas_obj}. More specific, we set four layers of intermediates to construct the basic blocks (i.e., cells). Then we cascade four cells to establish the concrete architectures (i.e., DNA and CNA) at the searching and training phase. More searching and training configurations are shown in the following Section~\ref{sec:train_details}. It is worth emphasizing that the relaxed weights are shared by both networks to simplify the searching procedure. Derived from the searching phase, we can obtain the final architectures for light and heavy rain scenarios respectively. Specifically, as for the micro search, the cell for light rain consists of $3\times3$ residual block, spatial attention, $3\times 3$ dilated convolutions and channel attention orderly. Moreover, the cell for heavy rain includes $5\times5$ residual block, spatial attention, channel attention and  $3\times 3$ residual block, which indicates cells for heavy rain have the wider respective fields and residual information to capture long-range rain streaks reasonably. The attention mechanisms and dilated convolutions are also effective and verified in previous works~\cite{wang2019spatial,yang2017deep}. As for temporal alignment module $\mathcal{F}$, TGM has been searched for heavy rain and OFM has been searched for light rain. That may be caused by the serious warping error of OFM for heavy rain.

In summary, designing an appropriate deep network architecture for each task and data set is tedious and time-consuming. The neural architecture search strategy attempts to find suitable architecture automatically and provides a trade-off between automated search and experience-driven. Indeed, the designed microscopic automatically searched architectures from a task-specific search space to discover desirable video deraining architectures. Further, the macroscopic structure searching scheme combines optical flow-estimation and temporal grouping module to jointly concern motion and rain streaks occlusion circumstance. Besides, the designed collaborative scheme can preserve detail and structure of video when estimating various rain streaks distribution. Thus, the derived model are efficient compared with some existing handcrafted networks.

\subsection{Implementation Details}\label{sec:train_details}
In this part, we report more details about the loss functions, searching configurations and training configurations.

\textbf{Loss Functions.} As for $L_{\mathtt{D}}$ and $L_{\mathtt{C}}$ described in Section~\ref{sec:alg}, we apply the negative SSIM loss (i.e., $L_{\mathtt{SSIM}}$, see~\cite{yang2020single}) and $\ell_1$ loss between the restored video frames $\mathbf{x}$ (i.e., $\mathbf{x}=\mathcal{N}_{\mathtt{D}}(\bm{\omega}_{\mathtt{D}}(\bm{\alpha}))$) and their ground truths $\mathbf{y}$ as 
\begin{equation}
L_{\mathtt{D}}=-\frac{1}{M}\sum_{i=1}^M L_{\mathtt{SSIM}}(\mathbf{x},\mathbf{y}) + \rho \|\mathbf{x}-\mathbf{y}\|_1, \label{eq:loss2}
\end{equation}
where $\rho$ is a parameter. The same loss function is applied for $L_{\mathtt{C}}$ with $\mathbf{x}=\mathcal{N}_{\mathtt{C}}(\bm{\omega}_{\mathtt{C}}(\bm{\omega}))$. As for $L_{\mathtt{arc}}$, it is constructed as
\begin{equation}\label{eq:nas}
L_{\mathtt{arc}} = L_{\mathtt{D}} + \eta L_{\mathtt{reg}},
\end{equation}
where $\eta$ is a positive parameter and $L_{\mathtt{reg}}$ is the regularization about architecture parameter
\begin{equation*}
L_{\mathtt{reg}}=-\sum_{l=1}^{L} \sum_{i=1}^{N} \alpha^{i}_{l}\log(\alpha^{i}_{l}).
\end{equation*}
Actually, the above $L_{\mathtt{reg}}$ will impose the distribution of $\bm{\alpha}_{l}$ to ignore the weakest candidates gradually for discovering the most suitable operation of $i$-th layer effectively.

\textbf{Search Configurations.} 
Considering the different distribution of rain streaks, we choose the RainSynLight25 and RainSynComplex25  as the  datasets to search the diverse optimal architectures respectively. During the searching phase, we only sample twenty groups of rainy frames to search 80 epochs with batch size of 4. Furthermore we define the train and validation loss using Eq.~\eqref{eq:nas} with $\rho$ = 0.75 and $\eta$ = 0.01. The SGD optimizer is employed for the search phase. The learning rate of searching architecture is $1e^{-4}$. On the other hand, the initial learning rate of updating parameters is  $3e^{-4}$ and decays to $10^{-6} $ with cosine annealing strategy. In order to provide a warm start for network parameters, we only update  network weights before the first 30 epochs. 

\textbf{Training Configurations.} 
Derived from the searching phase, we perform two-stage training strategy. First, we train the cooperate architecture (i.e., DNA and CNA) end-to-end with $\ell_{1}$ loss and negative SSIM loss (i.e., Eq.~\eqref{eq:loss2}  with $\rho$ = 0.75). We exploit the cosine annealing strategy to decay the initial learning rate $3\times 10^{-4}$ to $10^{-6}$. Employing the Adam optimizer and cropping the patches of size $240 \times 240$ randomly, we train the whole networks for 160 epochs. Data augmentations, including the horizontal and vertical flipping are implemented. After obtaining the pre-trained part of above networks, we train the fused attention mechanism leveraging a small learning rate of $1e^{-6}$ for 50 epochs.

\begin{table}[t]
	\small
	\renewcommand\arraystretch{1.02}
	\centering
	\caption{Performance (PSNR and SSIM) of an ablation study about different frames on RainSynComplex25 and LasVR data sets.}
	\setlength{\tabcolsep}{0.5mm}{
	\begin{tabular}{c | c  c  c}
		\Xhline{0.8pt}
		Frames & 3 & 5 & 7 \\
		\hline
		RainSynComplex25 & 25.69 / 0.8276 & \textbf{28.90} / \textbf{0.8743} & 23.91 / 0.7731 \\
		LasVR & 32.36 / 0.9063 & \textbf{33.41} / \textbf{0.9102} & 32.16 / 0.9070 \\
		\Xhline{0.8pt}
	\end{tabular}\label{tab:diff_frames}
}
\end{table}

\begin{figure}[t]
	\centering \begin{tabular}{c}
		\includegraphics[width=0.48\textwidth]{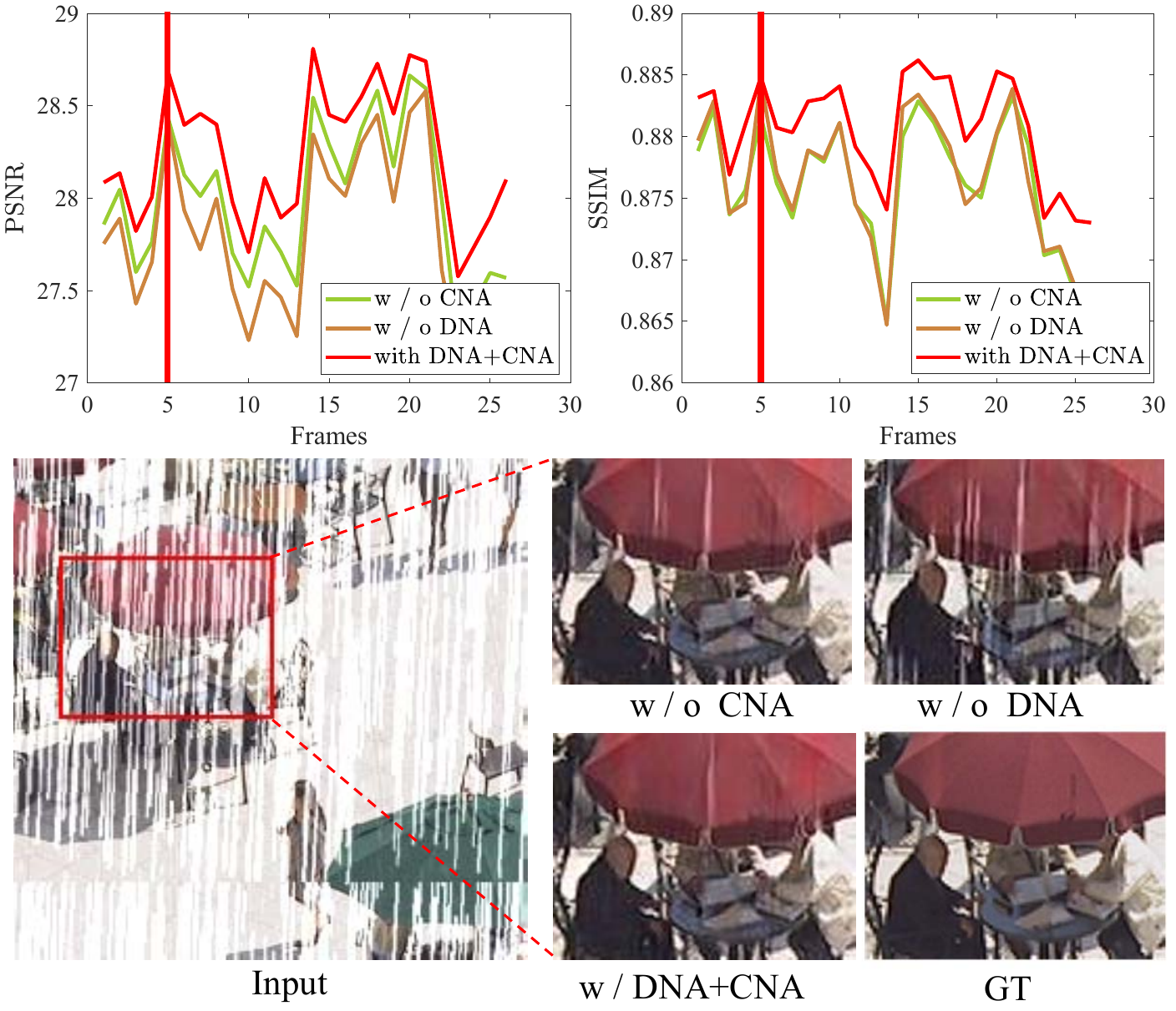}
	\end{tabular}
	\caption{Ablation study about the triple level models. PSNR and SSIM scores of a video sequence are plotted in the top row with three different settings, i.e., w / o CNA, w / o DNA, w / DNA + CNA. The bottom row shows their visual performances.} \label{fig:diff_model}
\end{figure}

\begin{figure}[t]
	\centering \begin{tabular}{c}
		\includegraphics[width=0.48\textwidth]{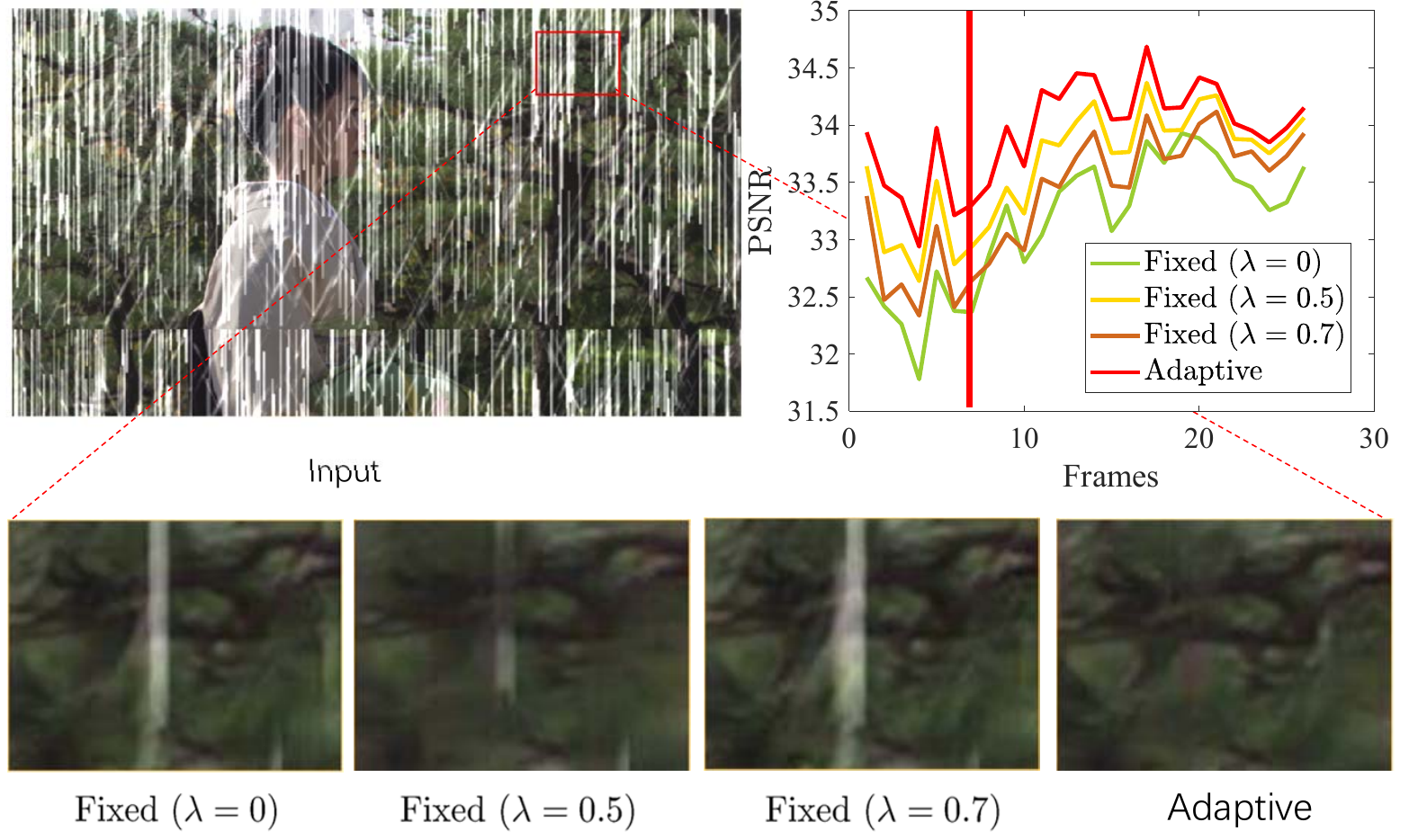}
	\end{tabular}
	\caption{Ablation study about the hyper-parameter $\bm{\lambda}$.  In the top row, the left subfigure is 7-th frame of input video and the right hand is the PSNR scores of a video with different hyper-parameter settings. The bottom row is the performance cropped from the visual results of different settings. The ``Adaptive'' setting means attention mechanism.} \label{fig:diff_lambda}
\end{figure}

\begin{table}[t]
	\small
	\centering
	\caption{Performance (PSNR and SSIM) of estimating different strategies, including MDA (Manually Designed Architecture), ASA (Auto-Searching Architecture), GARS (Generating Auxiliary Rain Streaks), on SynHybrid and LasVR data sets.}
	\setlength{\tabcolsep}{1.5mm}{
		\begin{tabular}{c c | c c c c c}
			\Xhline{0.8pt}
			\multicolumn{2}{c|}{Settings}& (a) & (b) & (c) & (d) & (e)\\
			\hline
			\multicolumn{2}{c|}{OFM} & \ding{51}&\ding{51} &\ding{55} &\ding{51}   &\ding{51}\\
			\multicolumn{2}{c|}{TGM} &\ding{55} & \ding{55}& \ding{51}& \ding{51} & \ding{51}\\
			\multicolumn{2}{c|}{MDA} & \ding{51}&\ding{55} &\ding{55} &\ding{55}  &\ding{55}\\
			\multicolumn{2}{c|}{ASA} &\ding{55} & \ding{51}& \ding{51}& \ding{51} & \ding{51} \\
			\multicolumn{2}{c|}{GARS} &\ding{55} & \ding{55}& \ding{55}& \ding{55} & \ding{51}\\
			\hline
			\multirow{2}{*}{SynHybrid} & & 30.35 & 32.7369 &  32.8993 & 33.9986 & \textbf{34.1054} \\
			& & 0.9058 & 0.9347 & 0.9376 & 0.9411 & \textbf{0.9424} \\
			\hline
			\multirow{2}{*}{LasVR}  &  & 31.3549 & 32.6170 &  32.6902 & 33.1070 & \textbf{33.4157} \\
  			& & 0.9042 &  0.9074 & 0.9088  & 0.9120 & \textbf{0.9144} \\
			\Xhline{0.8pt}	
		\end{tabular}
	}
	\label{tab:diff_modules}
\end{table}

\begin{table*}[t]
	\centering
	\caption{Averaged PSNR and SSIM results among different rain streaks removal methods on three synthesized video datasets. TMICS\_S denotes the result from single DNA. Red and blue colors are used to indicate top $\textcolor{red}{1^{st}}$ and $\textcolor{blue}{2^{nd}}$ rank, respectively.}
	\setlength{\tabcolsep}{1.8mm}{
		\begin{tabular}{c|cccccccc|cc}
			\hline
			Methods  &  MS-CSC & FastDeRain &DID-MDN & JORDER&SPANet & J4R-Net \\
			\hline
			RainSynLight25 &24.43 / 0.7312 &31.57 / 0.9508  &23.78 / 0.8140 &30.37 / 0.9235 &27.46 / 0.8844 &32.96 / 0.9434  \\
			RainSynComplex25 & 16.57 / 0.5833  & 26.92 / 0.8011 & 17.51 / 0.5888  & 20.20 / 0.6335  & 18.25 / 0.5824  & 24.13 / 0.7163  \\
			NTURain & 25.63 / 0.7600 & 30.54 / 0.9255 &25.67 / 0.8819 &32.69 / 0.9451& 30.58 / 0.9413& 30.73 / 0.9407  \\
			\hline\hline
			Methods & SpacCNN  & CLEARER & DualFlow & SLDNet & TMICS\_S & TMICS \\
			\hline
			RainSynLight25   &  32.78 / 0.9239 &  27.08 / 0.8858  &35.80 / 0.9622 &- / - &\textcolor{blue}{\textbf{36.10}} / \textcolor{blue}{\textbf{0.9674 }}&\textcolor{red}{\textbf{36.65}} / \textcolor{red}{\textbf{0.9689}} \\
			RainSynComplex25   & 21.21 / 0.5854   &  17.51 / 0.7108  & 27.72 / 0.8239 & - / - & \textcolor{blue}{\textbf{28.90}} / \textcolor{blue}{\textbf{0.8743}}  & \textcolor{red}{\textbf{29.49}} / \textcolor{red}{\textbf{0.8933}} \\
			NTURain & 33.11 / 0.9475  & 29.02 / 0.9158 & 36.05 / 0.9676 & 34.89 / 0.9540& \textcolor{blue}{\textbf{36.64}} / \textcolor{blue}{\textbf{0.9702}}& \textcolor{red}{\textbf{37.38}} / \textcolor{red}{\textbf{0.9704}}\\
			\hline
		\end{tabular}
	}
	\label{tab:datasets_results}
\end{table*}

\begin{figure*}[t]
	\centering \begin{tabular}{c@{\extracolsep{0.2em}}c@{\extracolsep{0.2em}}c@{\extracolsep{0.2em}}c@{\extracolsep{0.2em}}c}
		\includegraphics[width=0.192\textwidth]{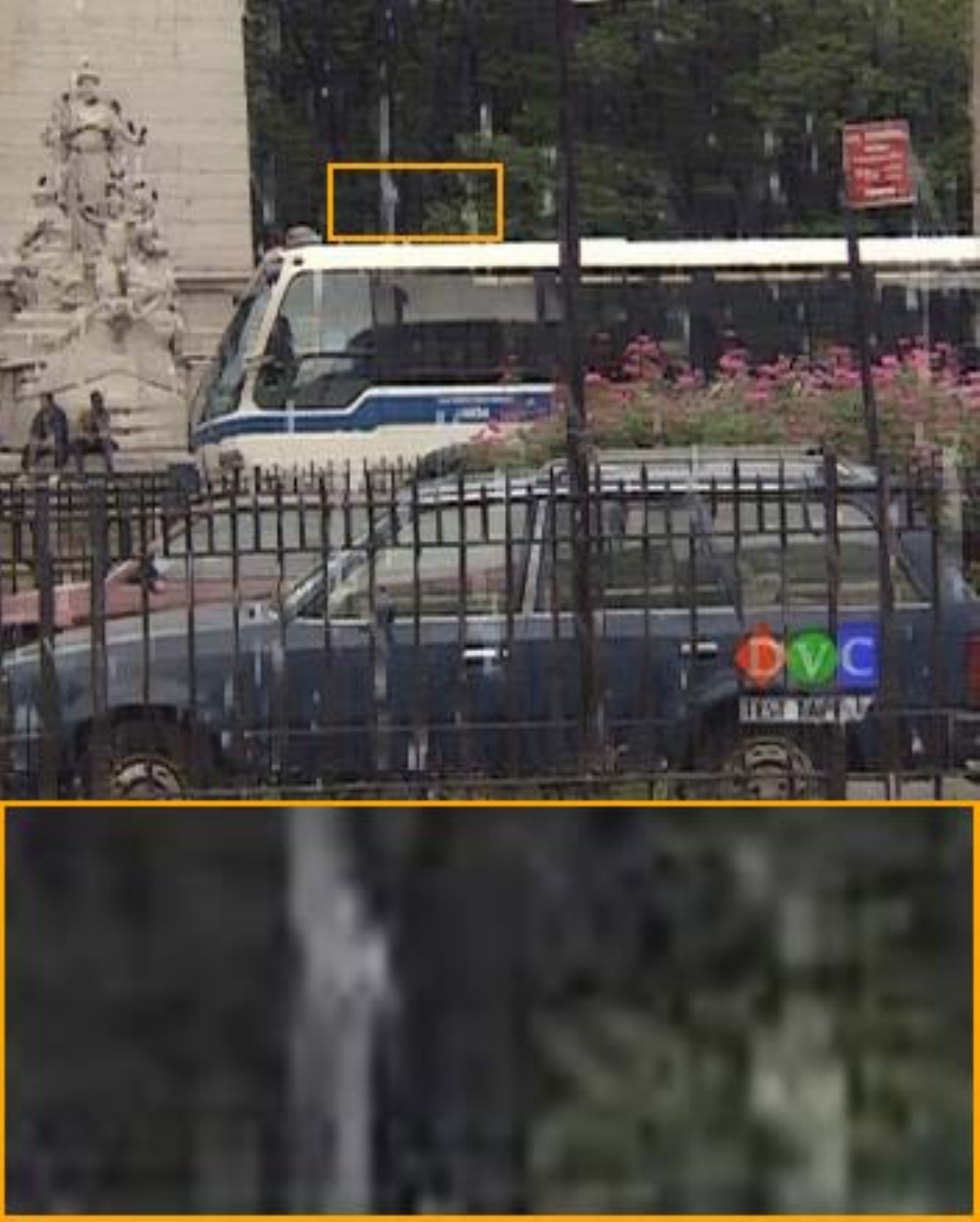}
		&\includegraphics[width=0.192\textwidth]{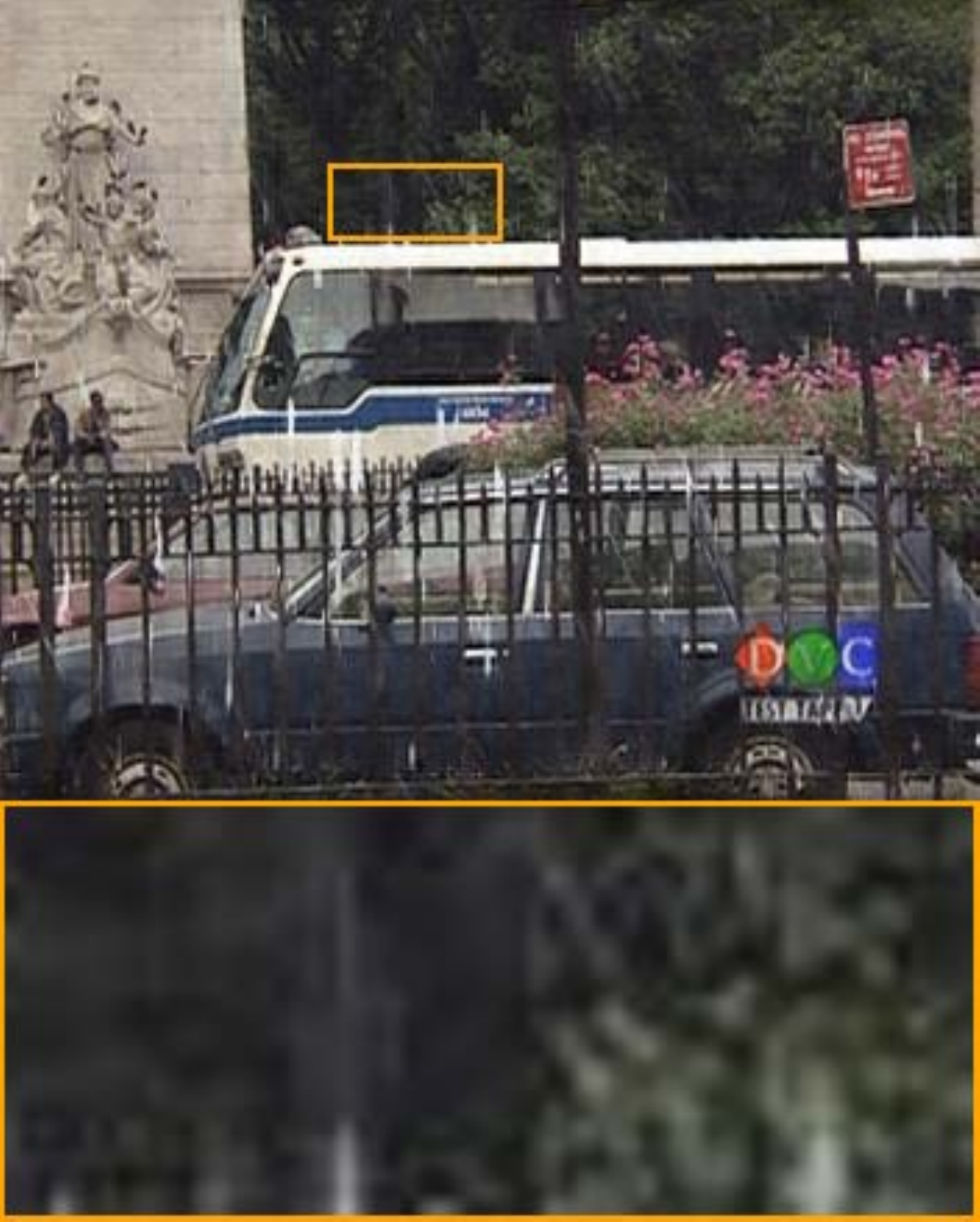}
		&\includegraphics[width=0.192\textwidth]{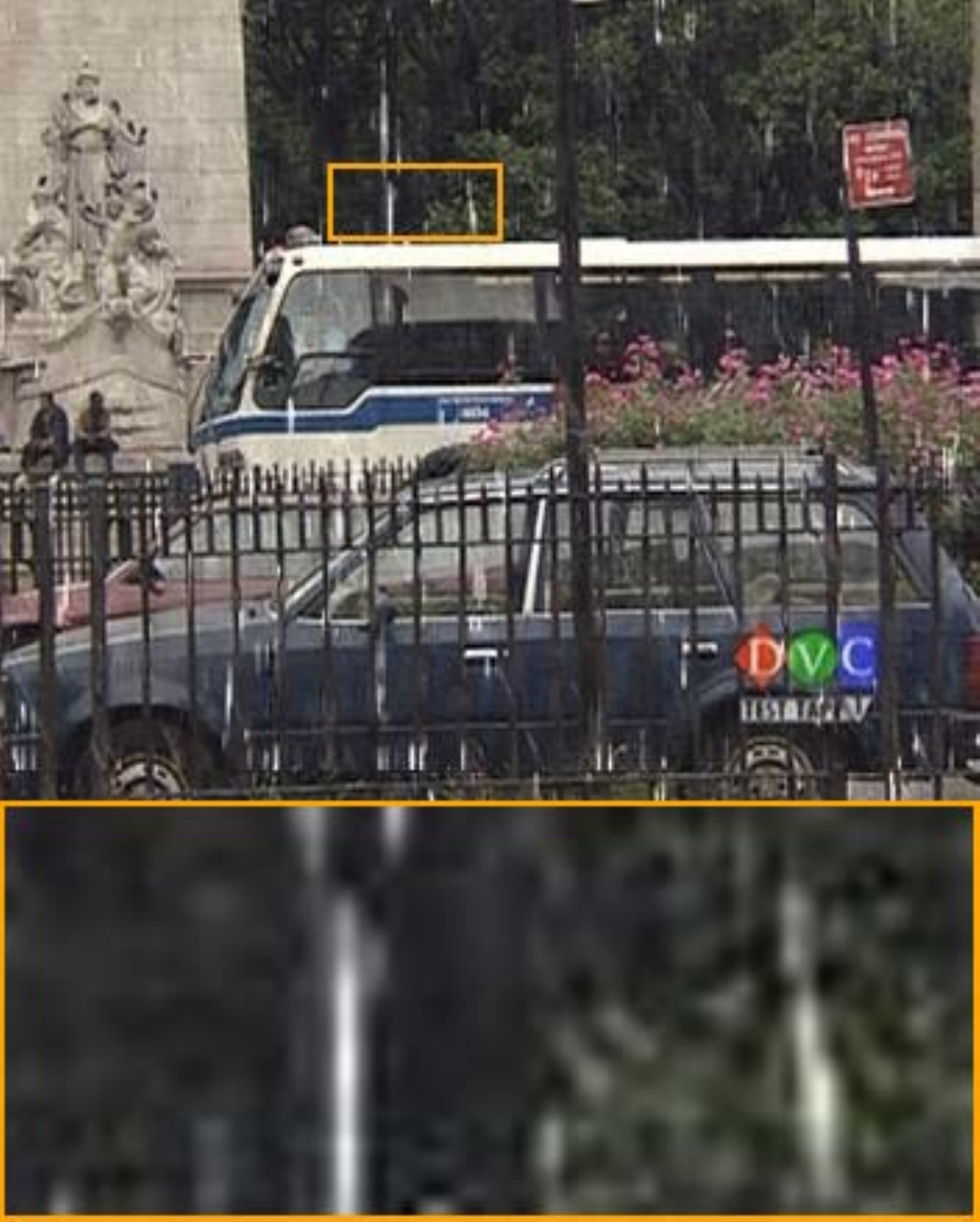}	
		&\includegraphics[width=0.192\textwidth]{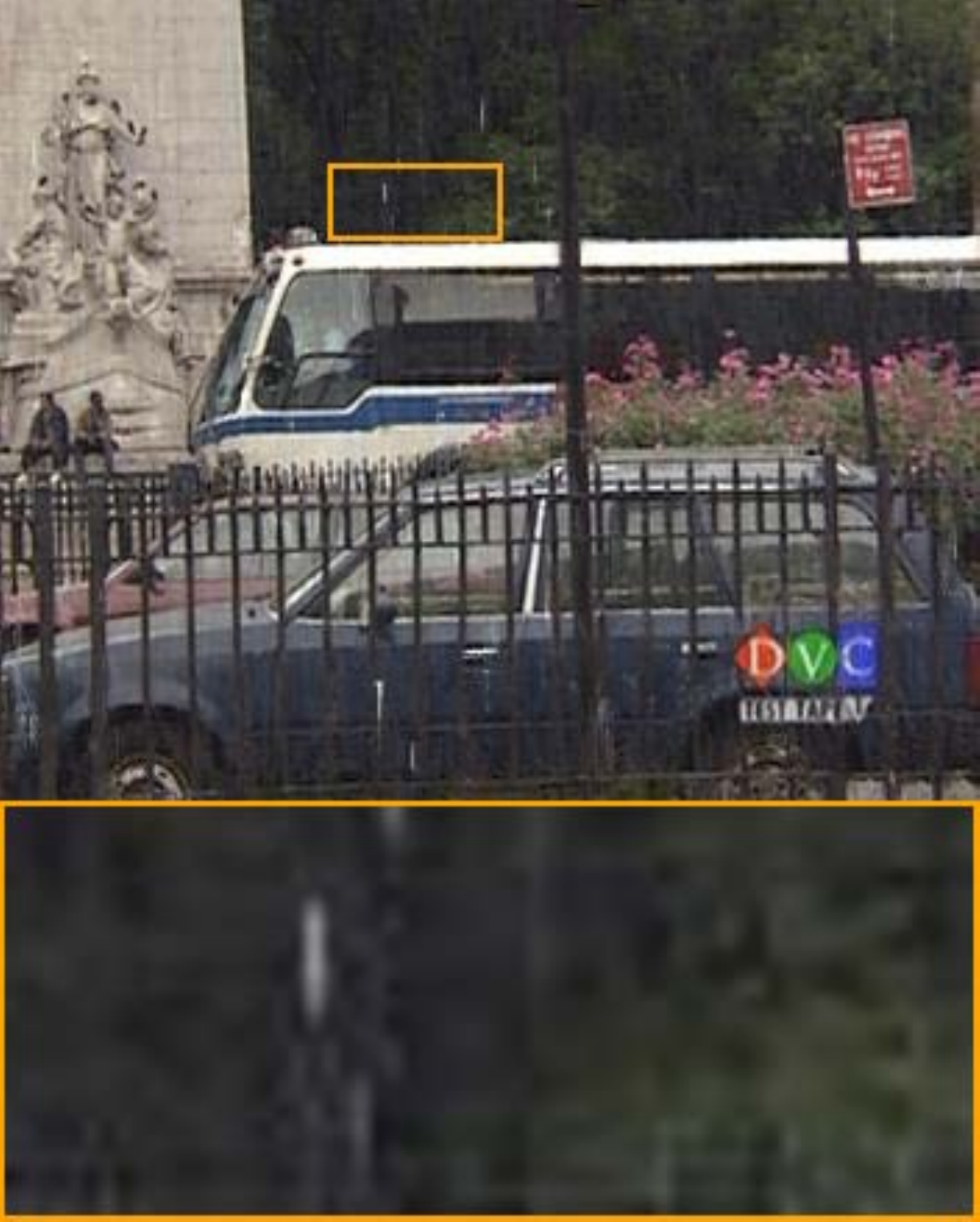}
		&\includegraphics[width=0.192\textwidth]{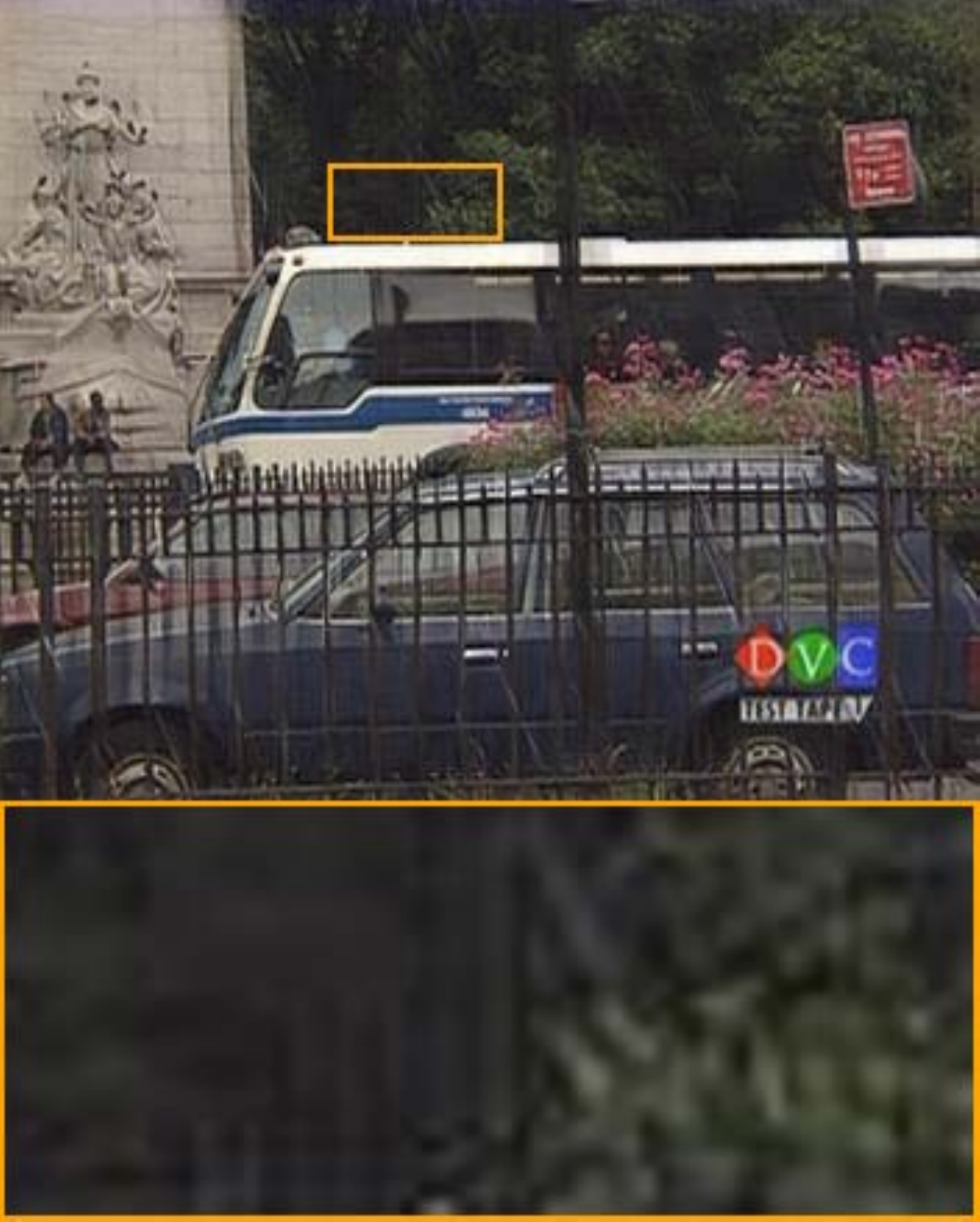}\\
		
		\includegraphics[width=0.192\textwidth]{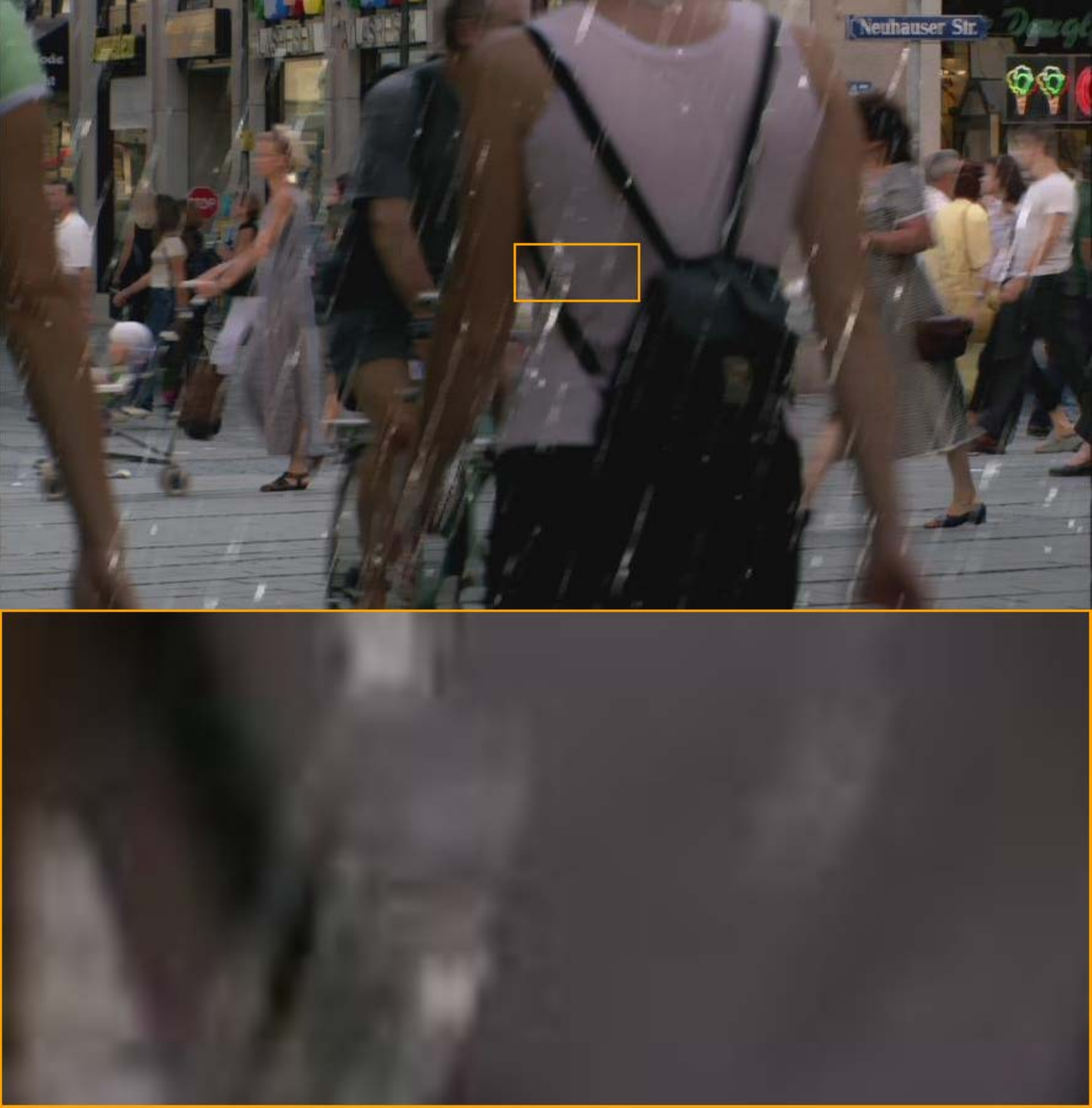}
		&\includegraphics[width=0.192\textwidth]{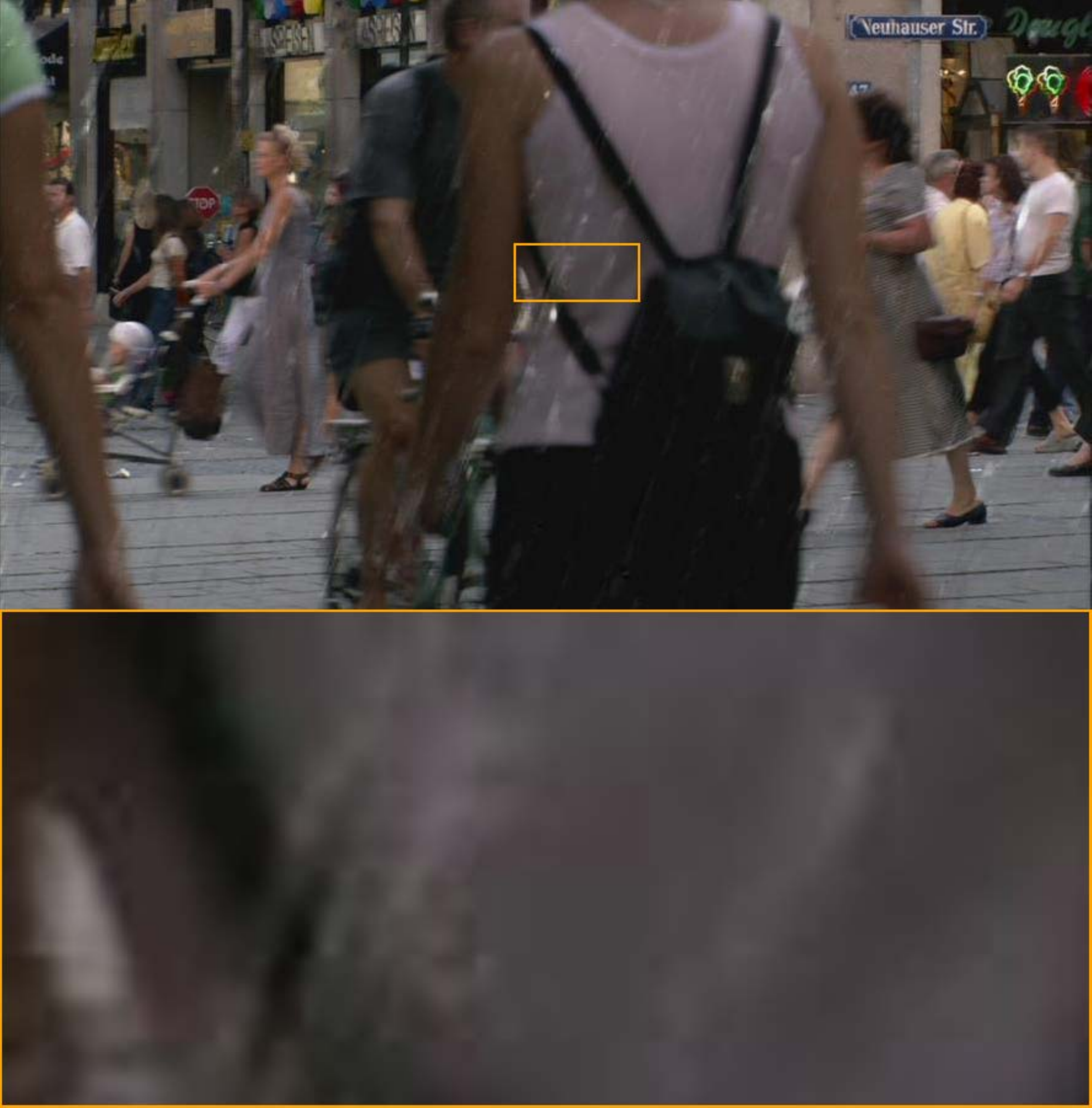}
		&\includegraphics[width=0.192\textwidth]{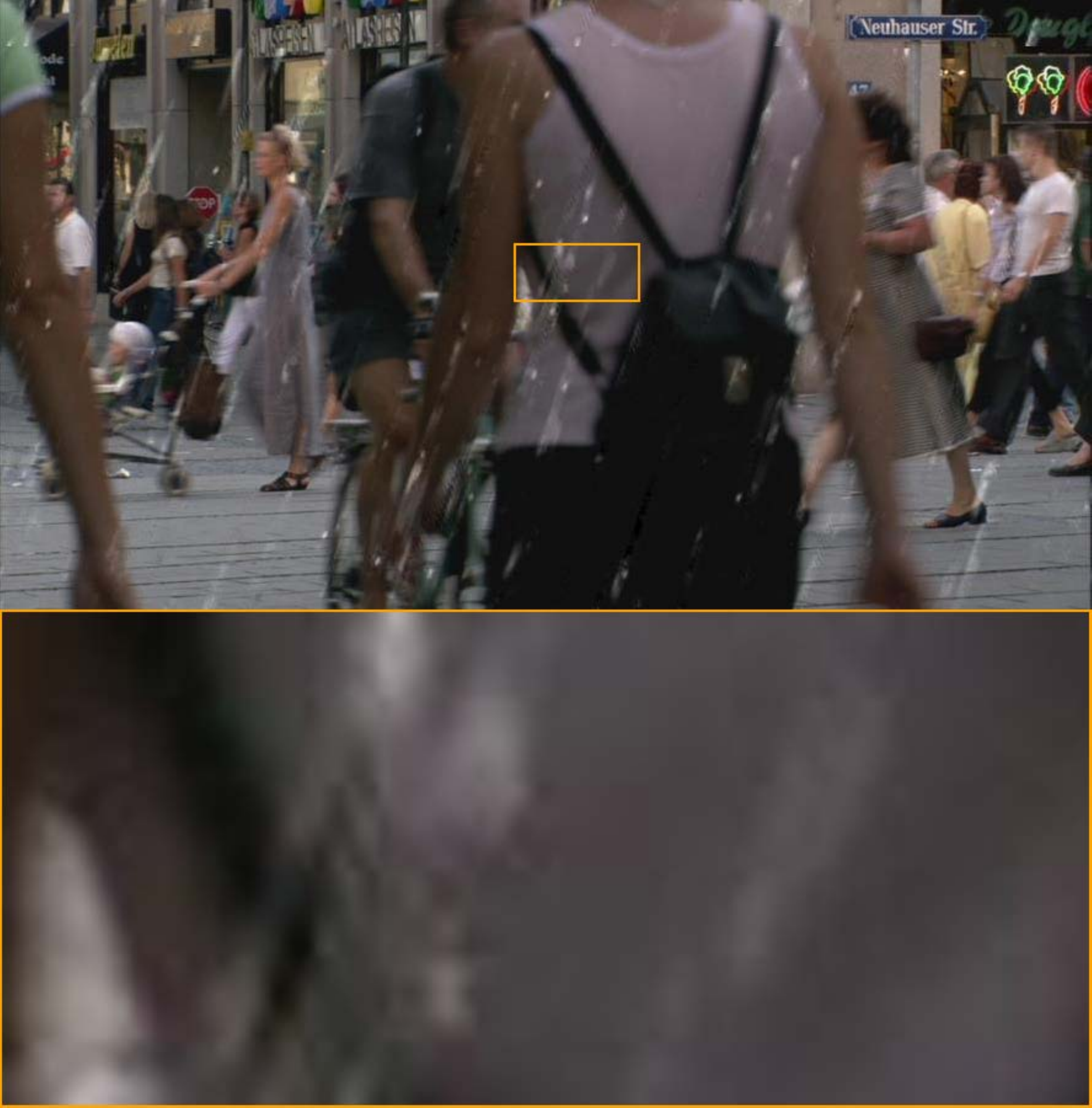}	
		&\includegraphics[width=0.192\textwidth]{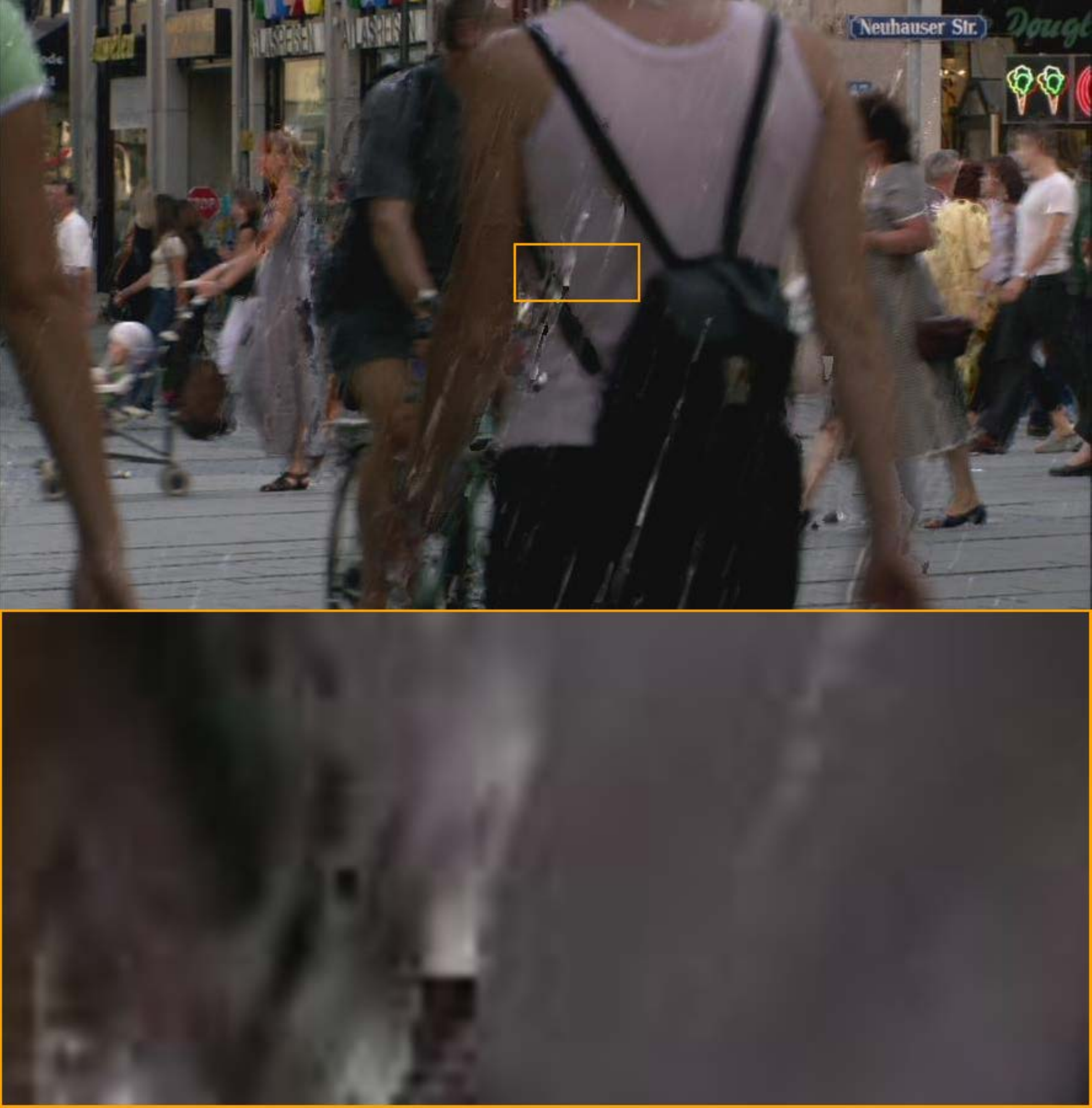}
		&\includegraphics[width=0.192\textwidth]{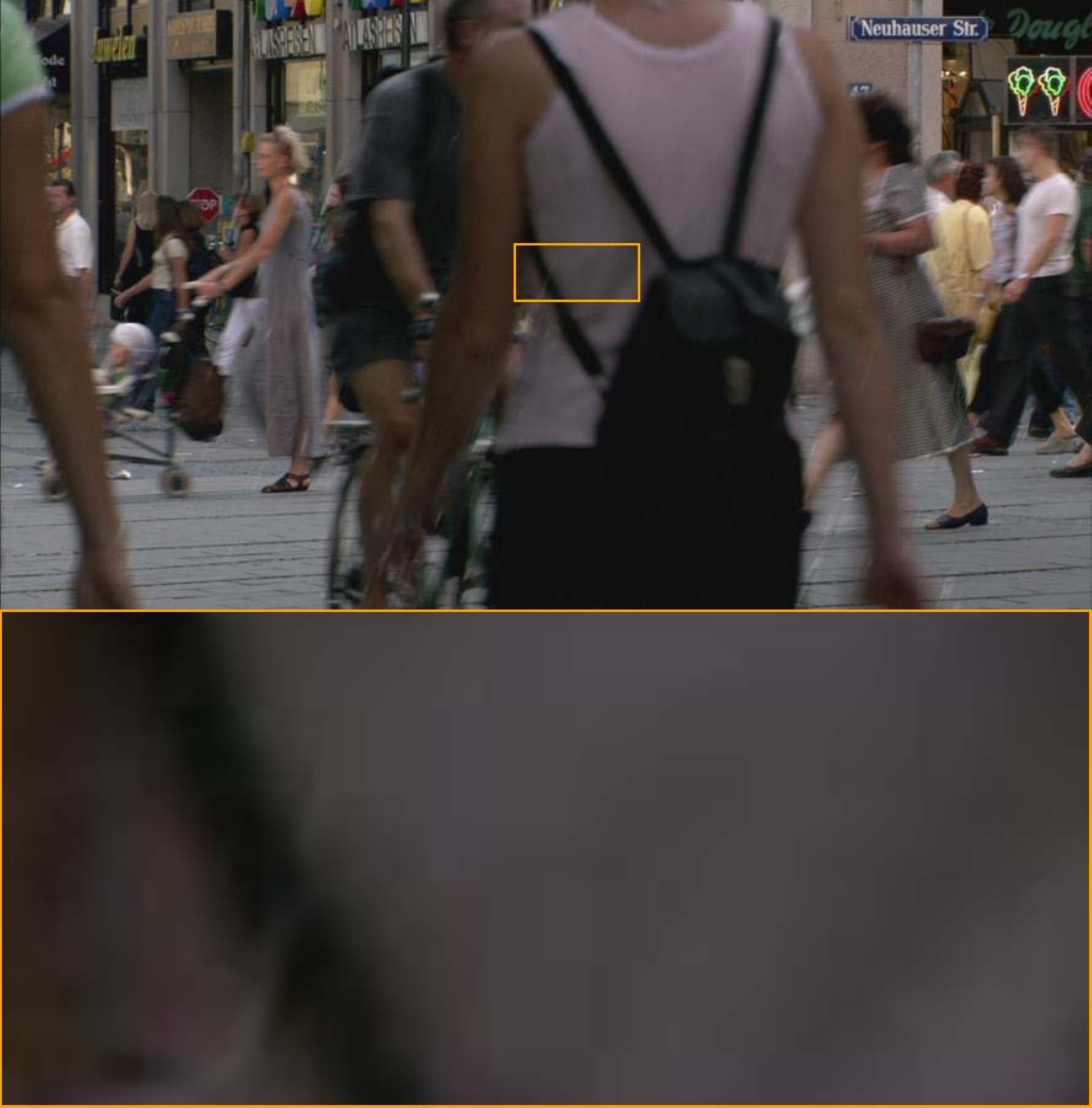}\\
		\footnotesize FastDerain &  \footnotesize J4R-Net &\footnotesize JORDER & \footnotesize SpacCNN & \footnotesize TMICS (Ours) \\
	\end{tabular}
	\caption{Video deraining performance comparison on two videos from RainSynLight25 (the top row) and RainSynComplex25 (the bottom row) respectively.} \label{fig:visual_datasets}
\end{figure*}

\begin{figure*}[t]
	\centering \begin{tabular}{c@{\extracolsep{0.2em}}c@{\extracolsep{0.2em}}c@{\extracolsep{0.2em}}c@{\extracolsep{0.2em}}c}
		\includegraphics[width=0.192\textwidth]{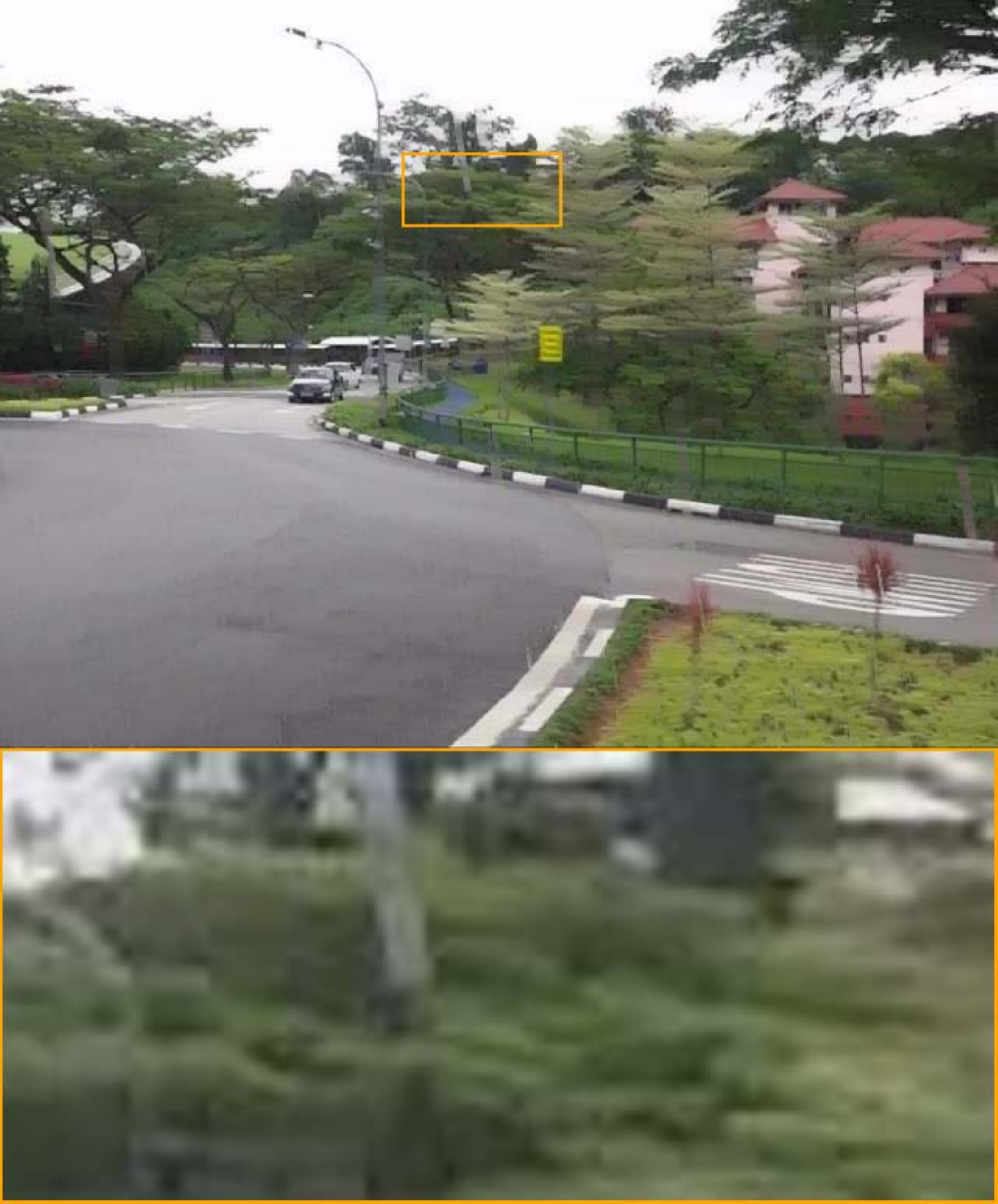}
		&\includegraphics[width=0.192\textwidth]{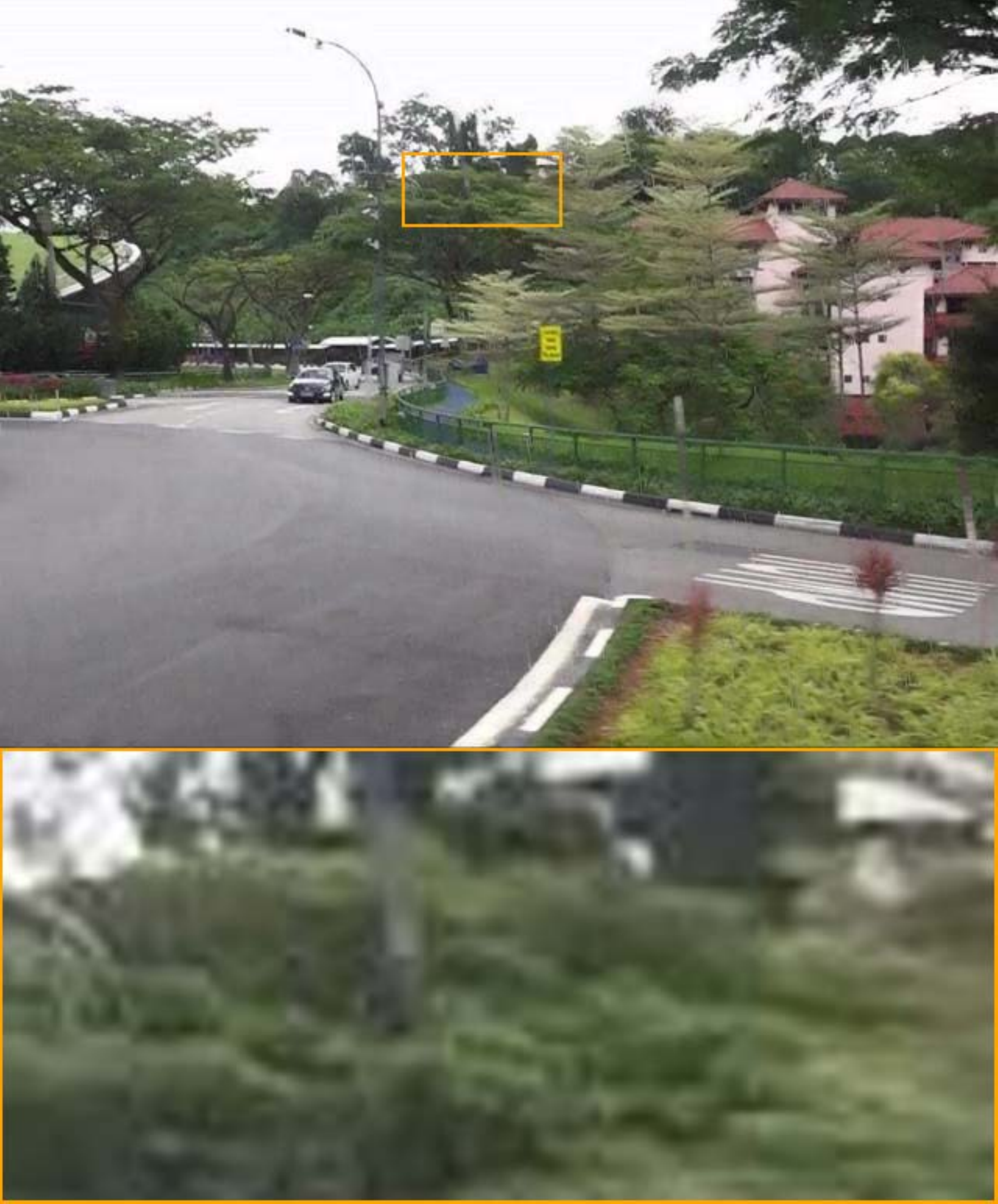}
		&\includegraphics[width=0.192\textwidth]{Figs/NTUsyn/J4RNet}	
		&\includegraphics[width=0.192\textwidth]{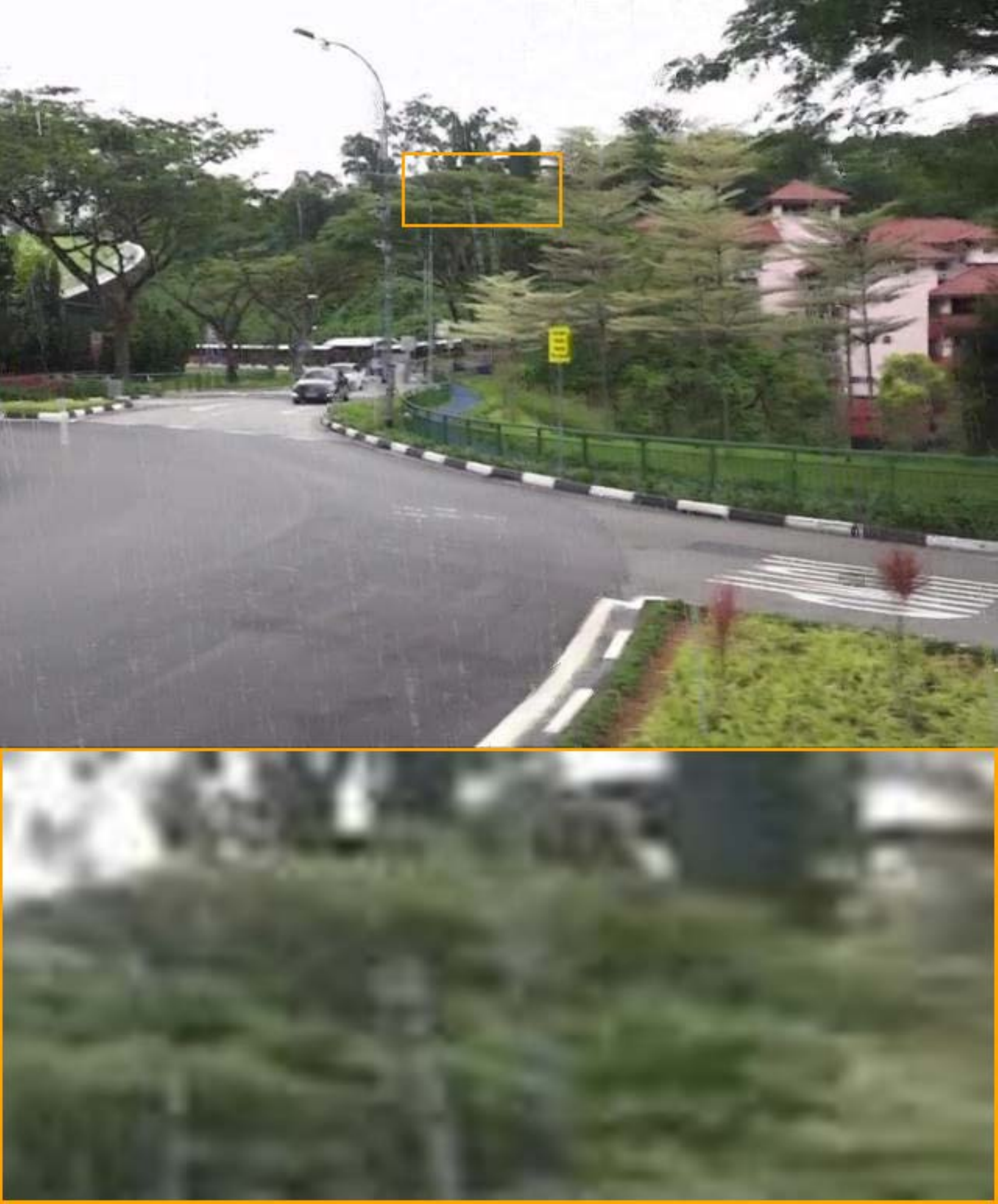}
		&\includegraphics[width=0.192\textwidth]{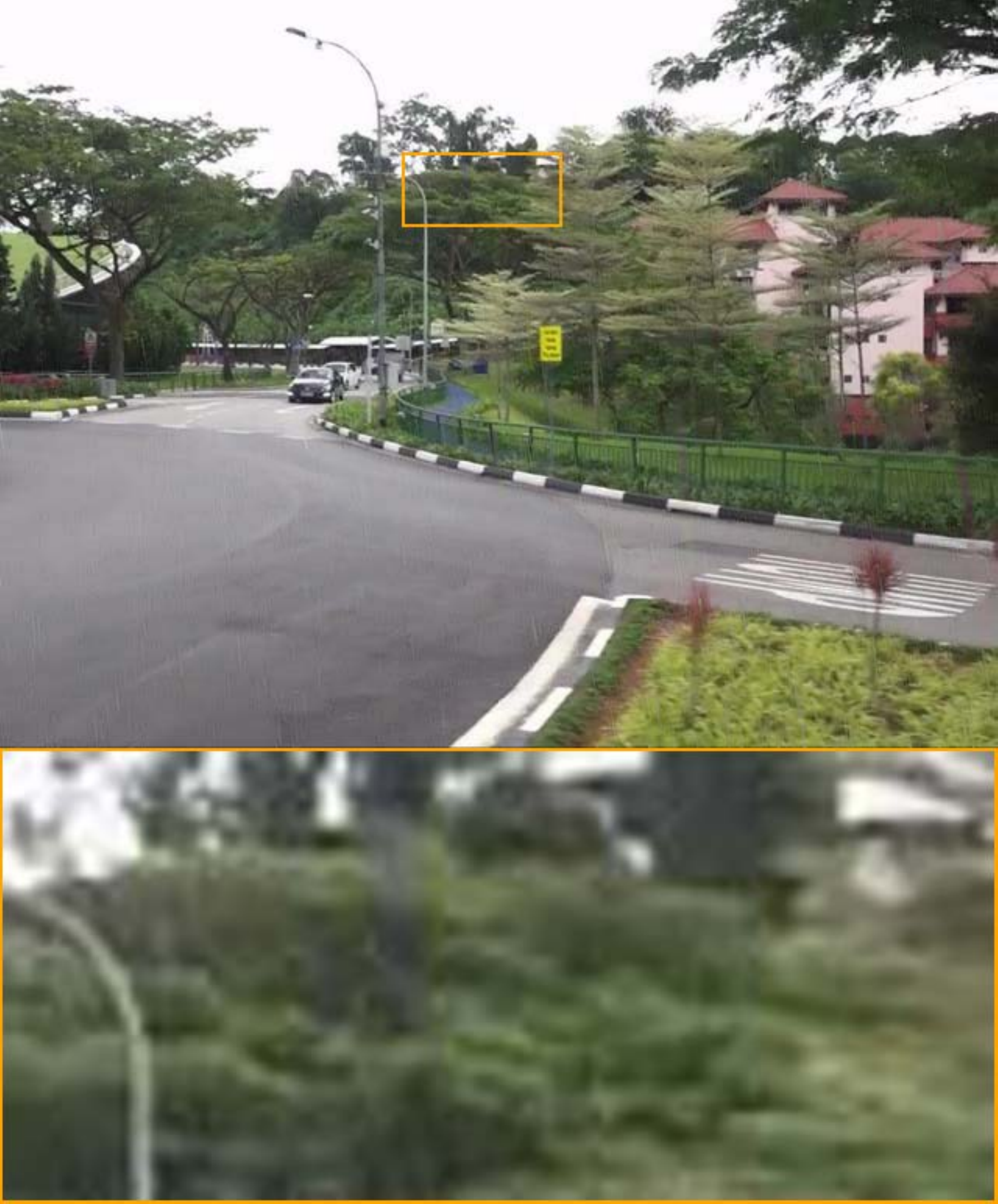}\\
		
		\includegraphics[width=0.192\textwidth]{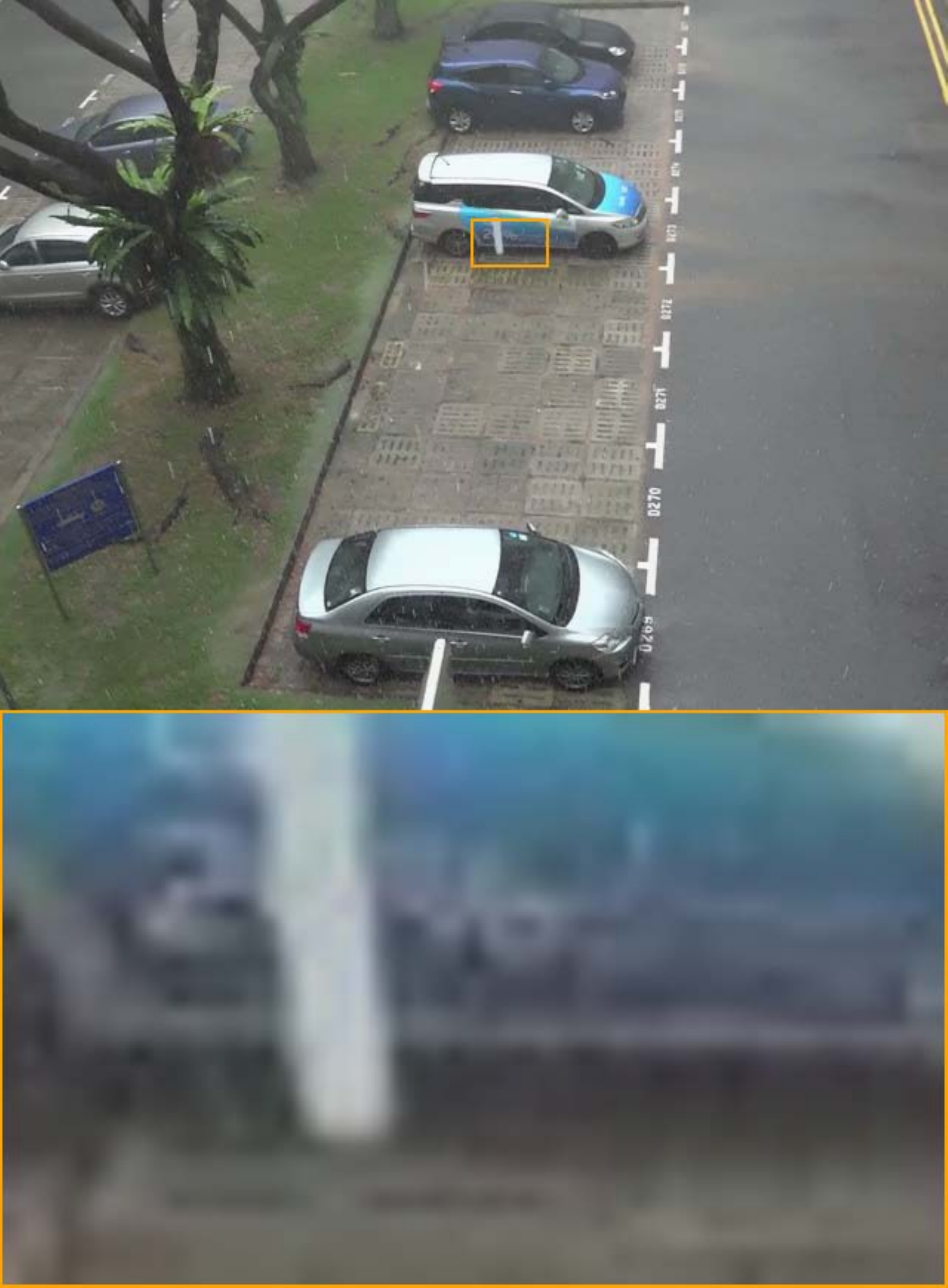}
		&\includegraphics[width=0.192\textwidth]{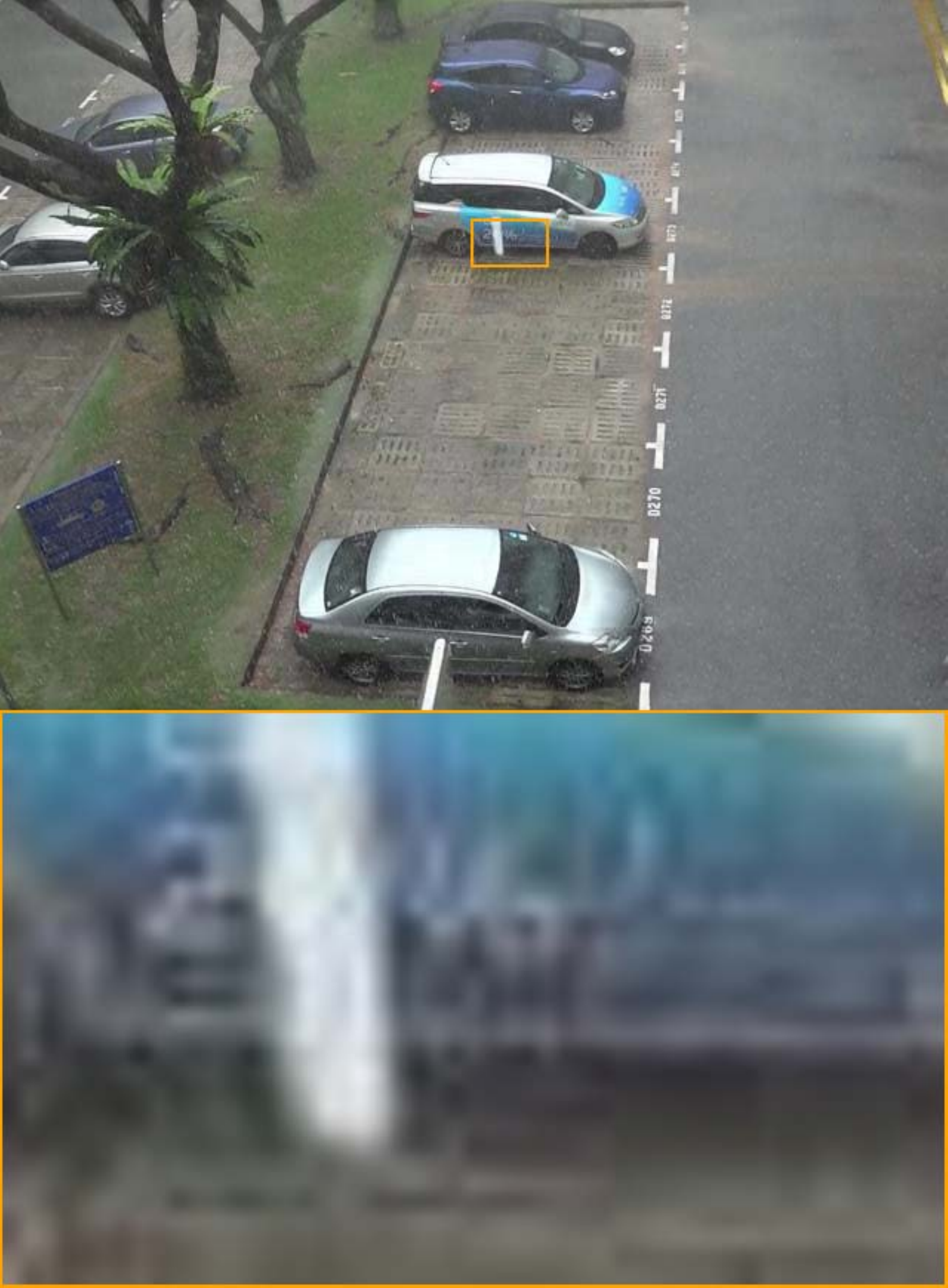}	
		&\includegraphics[width=0.192\textwidth]{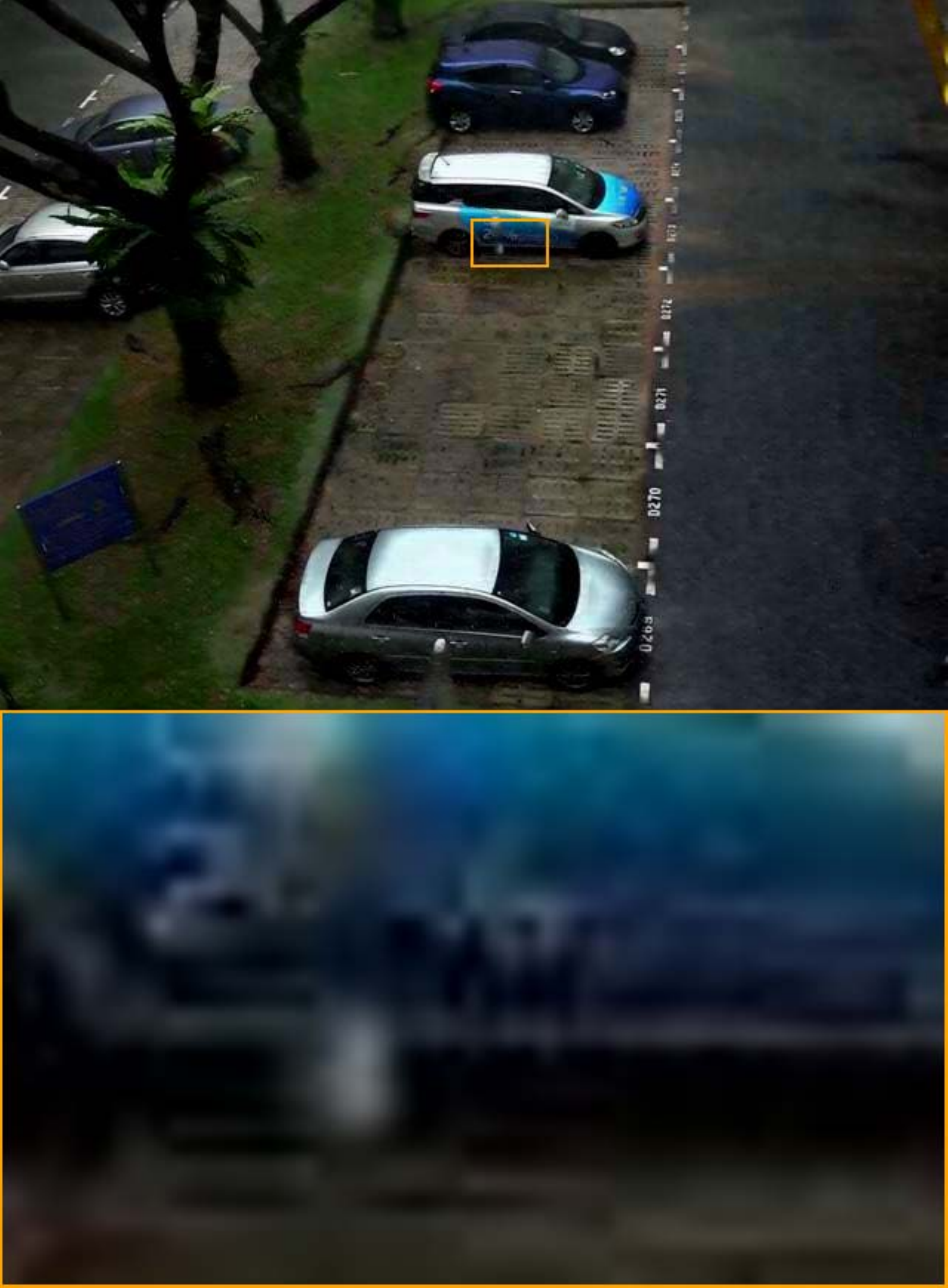}
		&\includegraphics[width=0.192\textwidth]{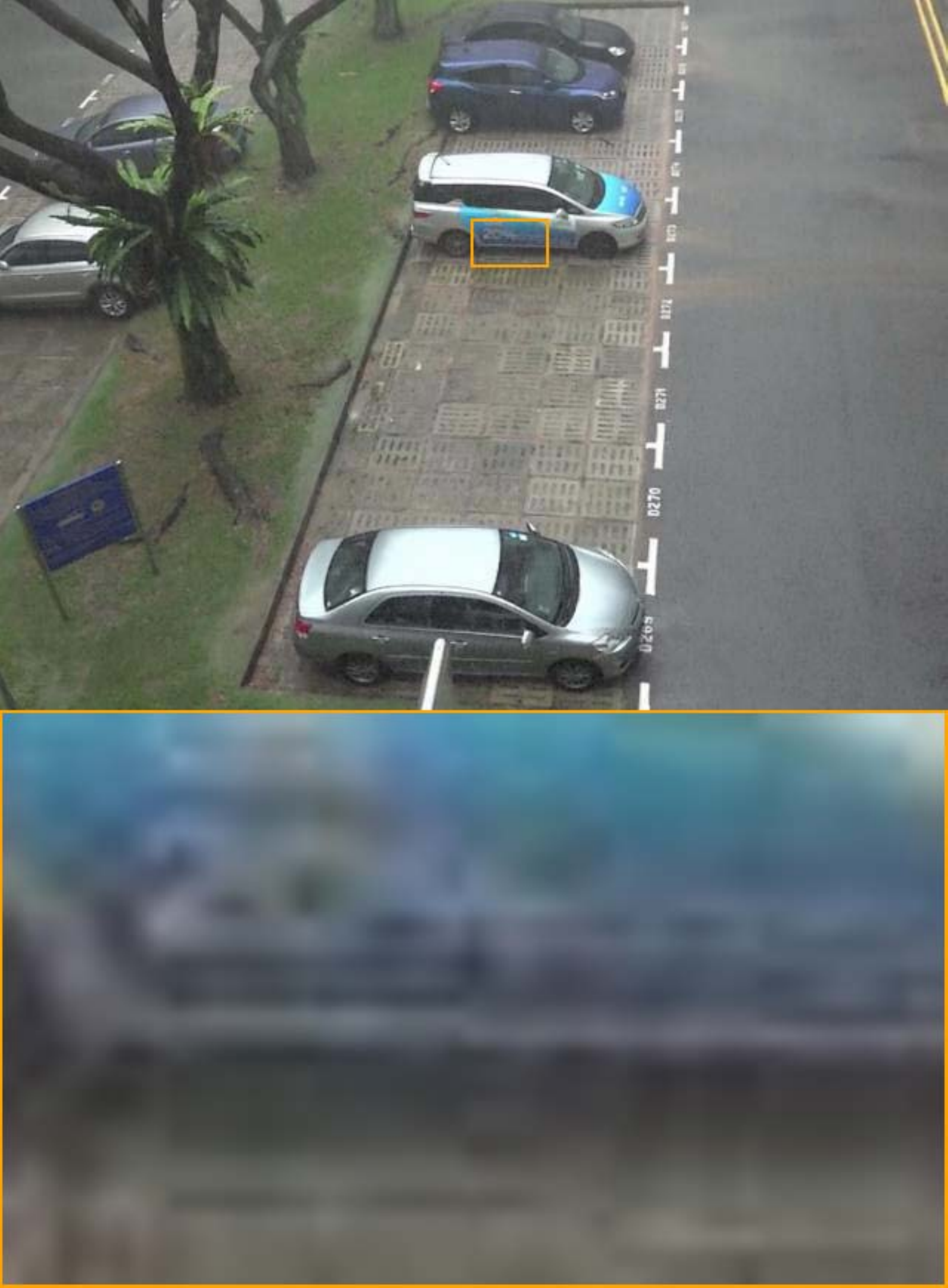}
		&\includegraphics[width=0.192\textwidth]{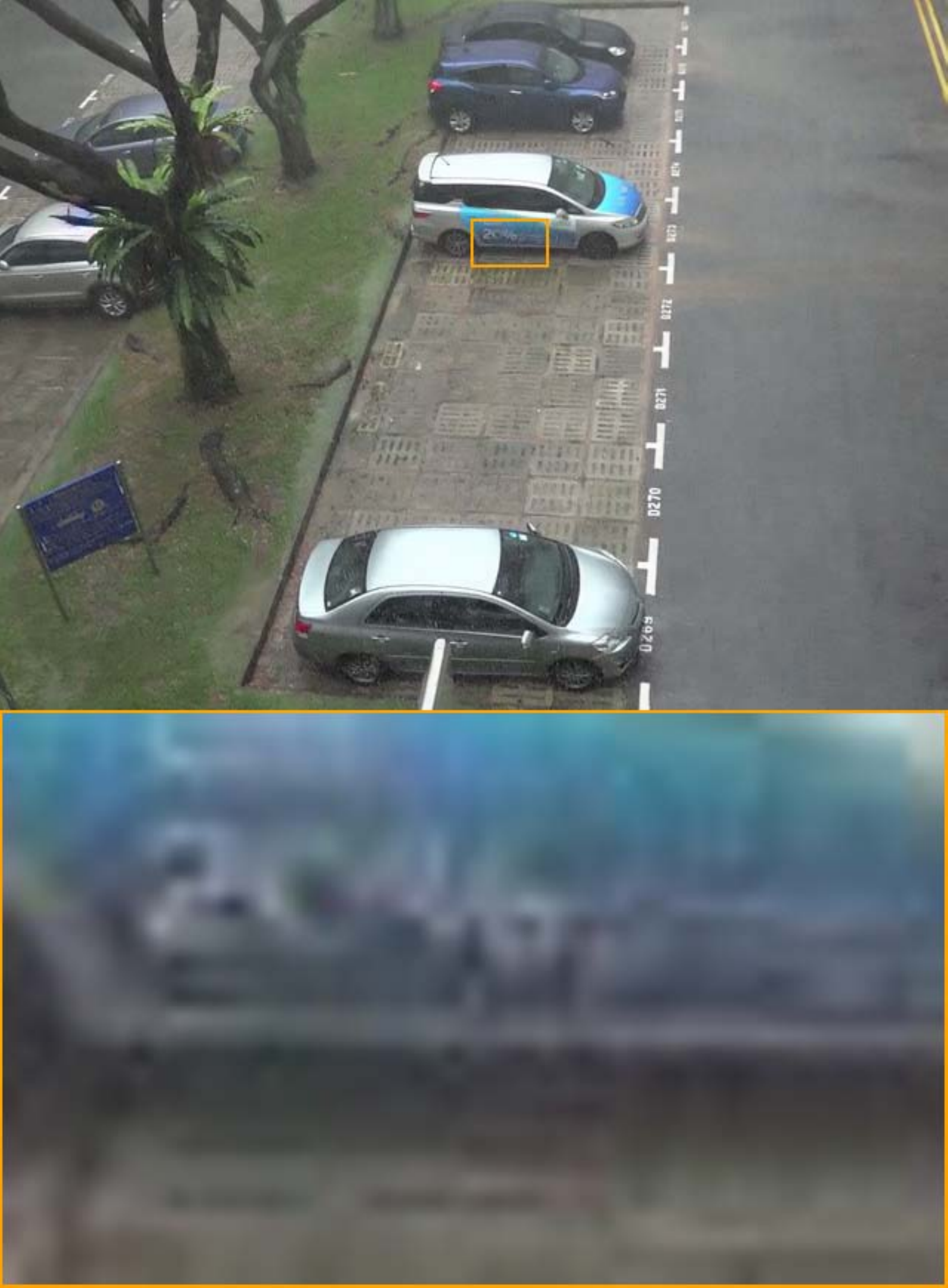}\\
		\footnotesize FastDerain &  \footnotesize J4R-Net &\footnotesize JORDER & \footnotesize SpacCNN & \footnotesize TMICS (Ours) \\
	\end{tabular}
	\caption{Video deraining results on two types of videos (i.e., synthetic and real frames in the top and bottom row respectively) from NTURain dataset. 
	} \label{fig:visual_NTU}
\end{figure*}

\section{Experimental Results}

To evaluate the proposed method, we first conduct kinds of detailed ablation studies to analyze the effectiveness of frame aligned modules, auto-searching architectures, and rain streaks generating module. Subsequently, a series of qualitative and quantitative assessments are performed comparing with existing state-of-the-art approaches. Experimental results demonstrate the effectiveness and superiority of our approach.

\subsection{Ablation Experiments}

\textbf{Datasets.} We compare the proposed method with state-of-the-arts on RainSynLight25, RainSynComplex25 and NTURain datasets. The ablation experiments are conducted on LasVR and hybrid SynComplex\&Light datasets. The detailed information (e.g., training / test numbers and rain streak types) are summarized in Table.~\ref{tab:datasets}.

\textbf{Metrics.} We use two most widely used numerical metrics, i.e., Peak Signal-to-Noise Ratio (PSNR) and Structure Similarity Index (SSIM) to evaluate the performance of different methods. Following previous works, results are evaluated on luminance channel. To further measure the performance of different methods, we also analyze the perceptual quality based on Visual Information Fidelity (VIF)~\cite{sheikh2006image}, Feature SIMilarity (FSIM)~\cite{zhang2011fsim},  Natural Image Quality Evaluator (NIQE)~\cite{mittal2012making}, Learned Perceptual Image Patch Similarity (LPIPS) \cite{zhang2018perceptual} and tLPIPS \cite{chu2018temporally}. Higher PSNR, SSIM values imply a better pixel-wise accuracy, and lower NIQE, LPIPS and tLPIPS values represent better perceptual quality. Furthermore, higher VIF and FSIM also indicate more visual-pleasant results.

\begin{figure*}[htb!]
	\centering \begin{tabular}{c@{\extracolsep{0.2em}}c@{\extracolsep{0.2em}}c@{\extracolsep{0.2em}}c}
		\includegraphics[width=0.245\textwidth]{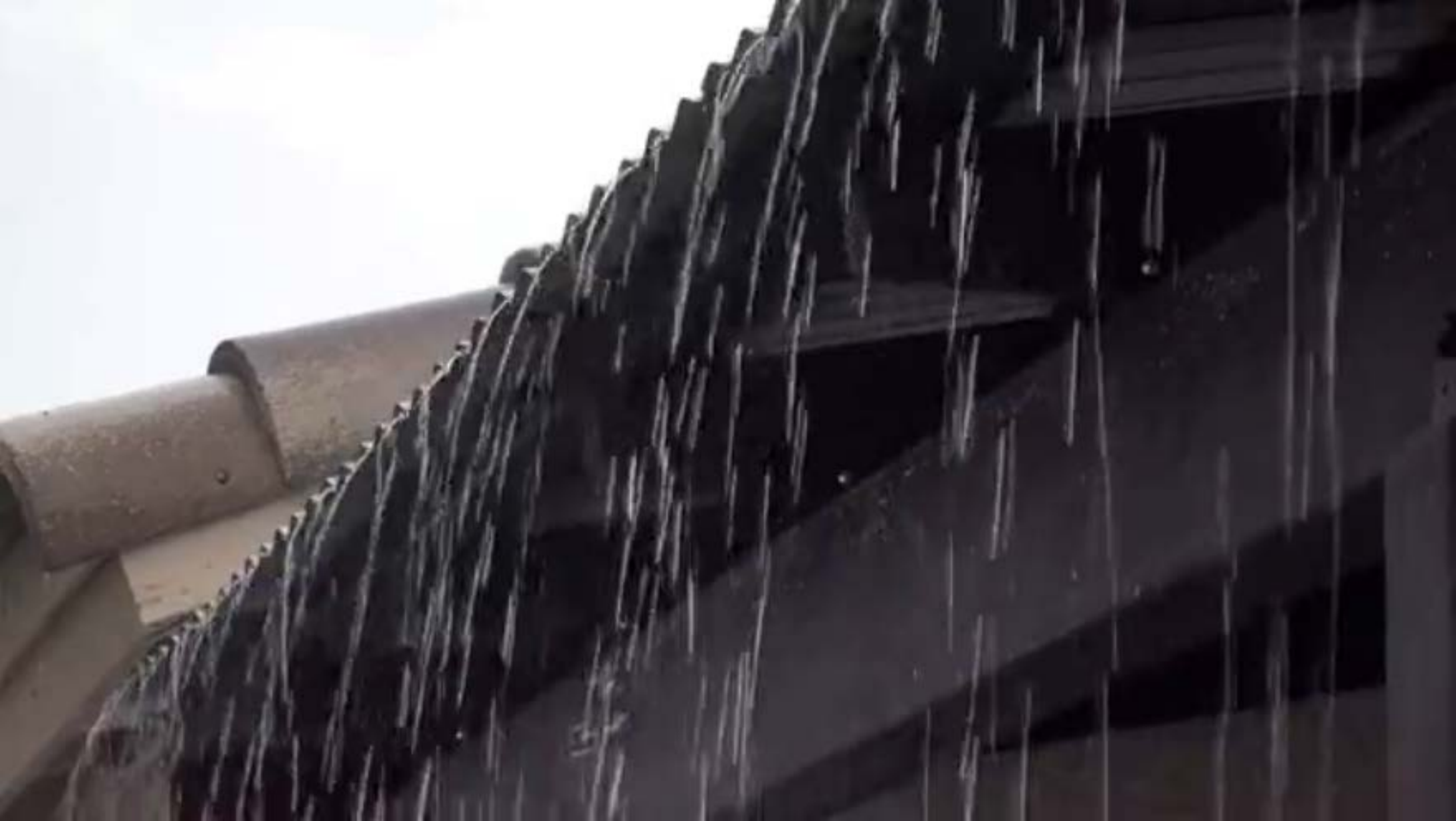}
		&\includegraphics[width=0.245\textwidth]{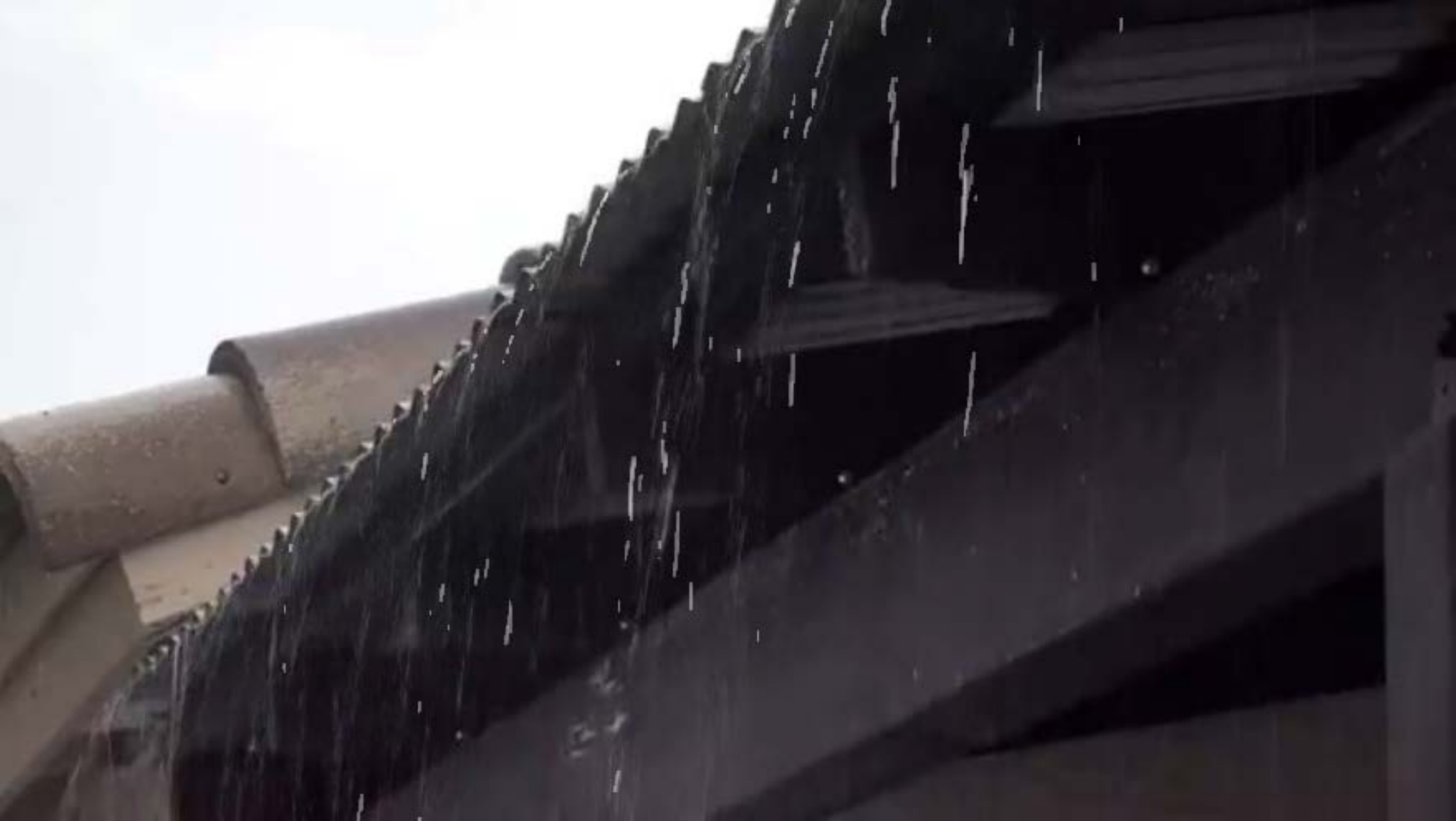}
		&\includegraphics[width=0.245\textwidth]{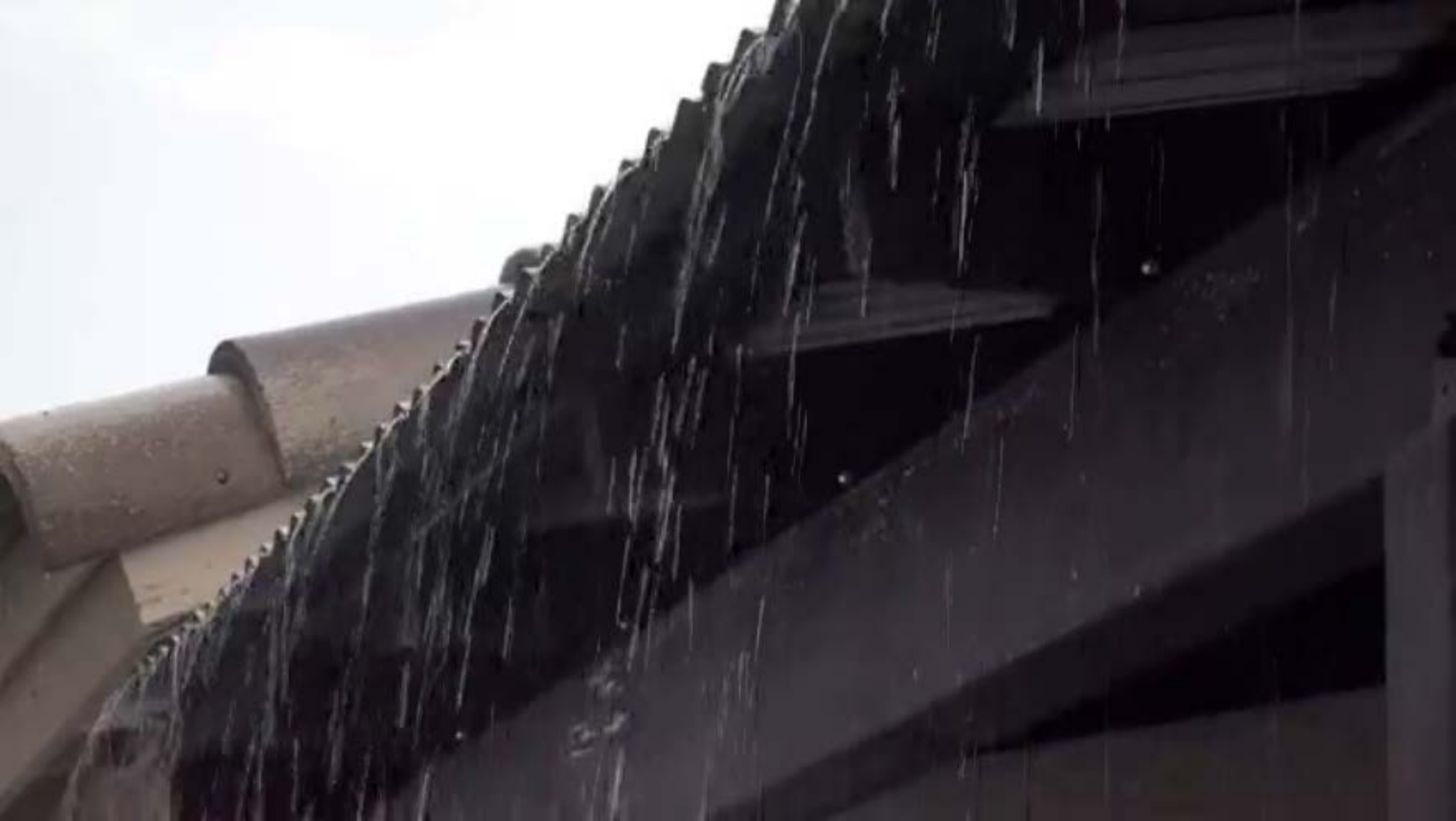}
		&\includegraphics[width=0.245\textwidth]{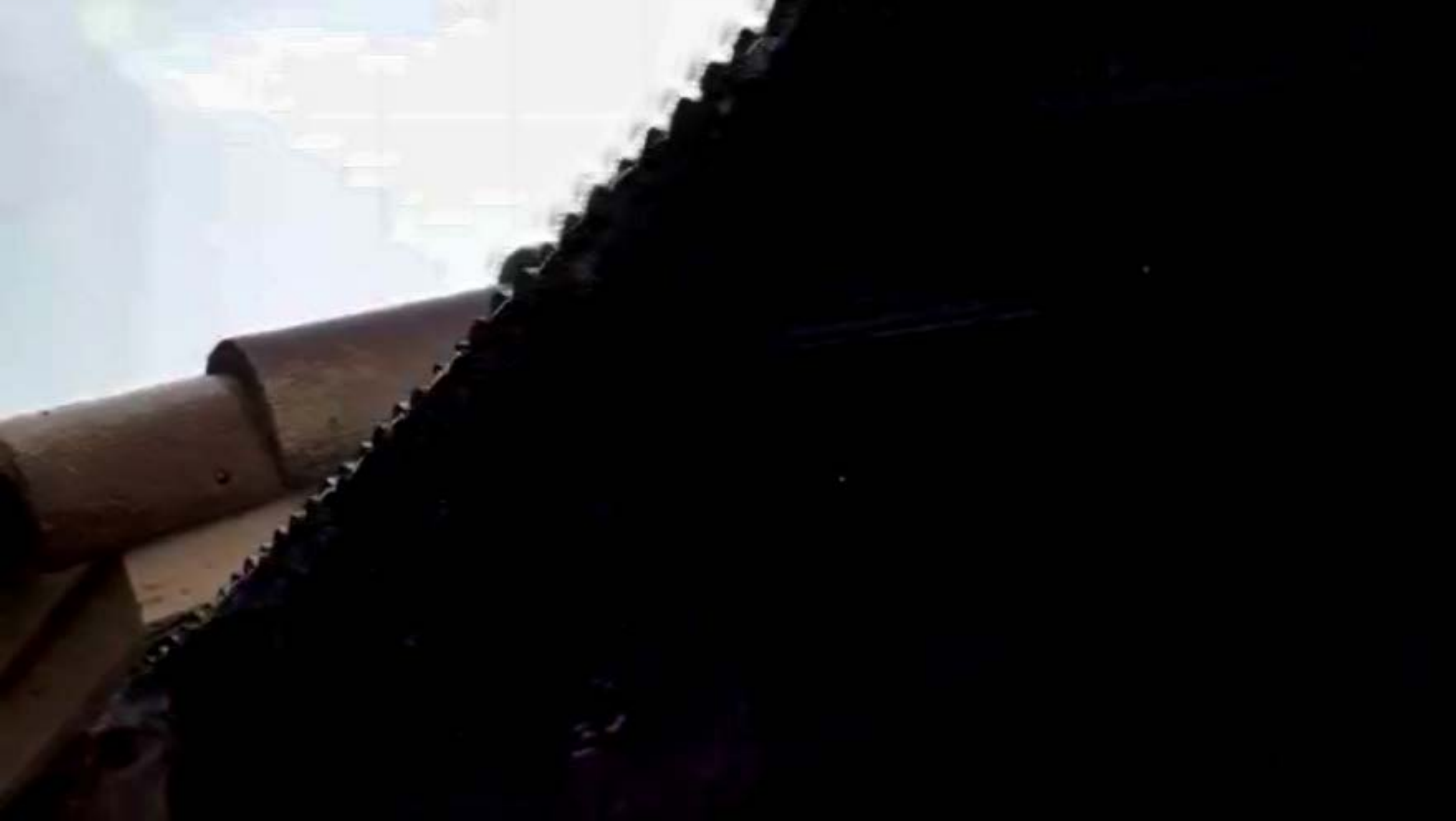}\\
		
		\includegraphics[width=0.245\textwidth]{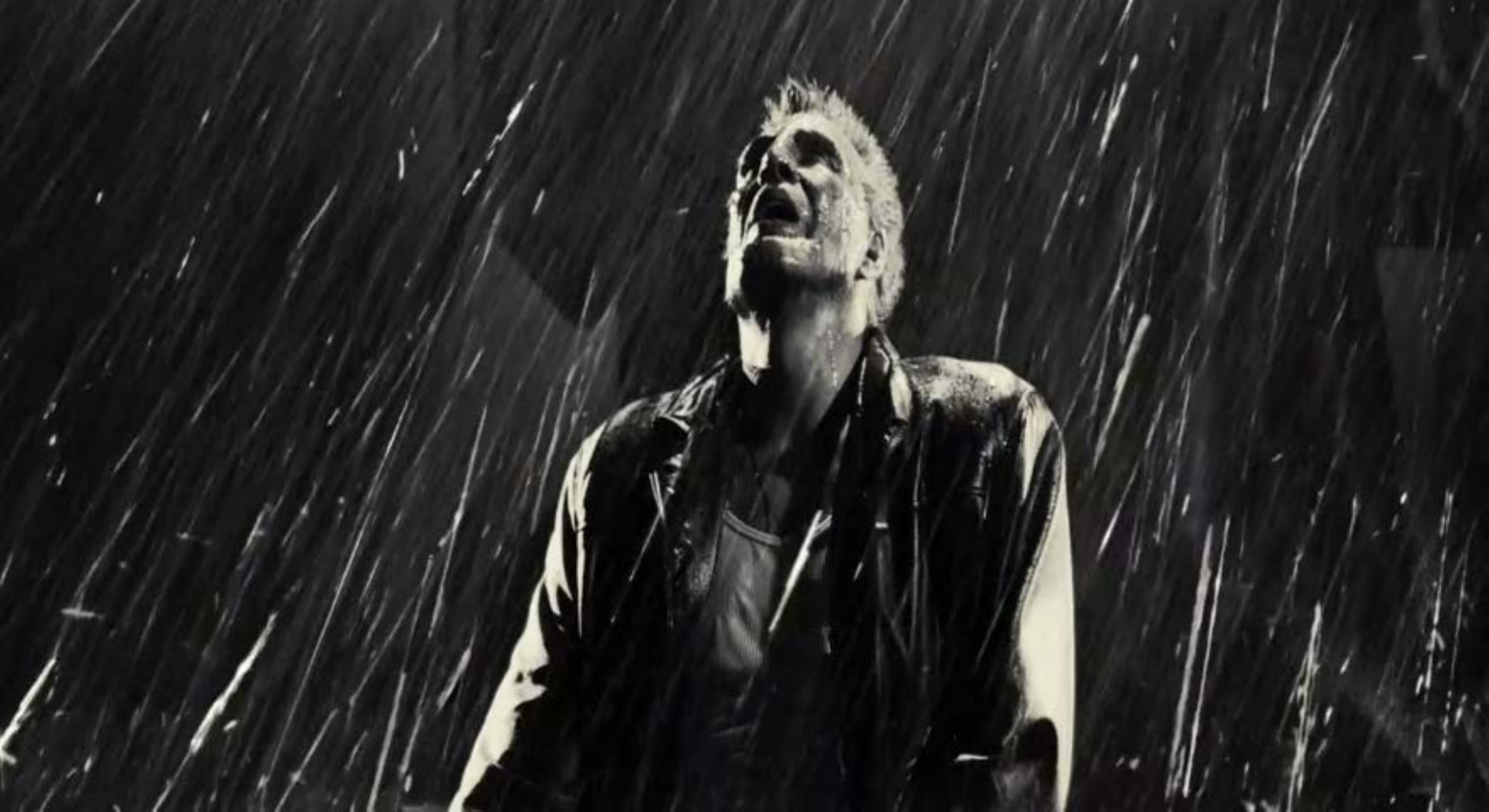}
		&\includegraphics[width=0.245\textwidth]{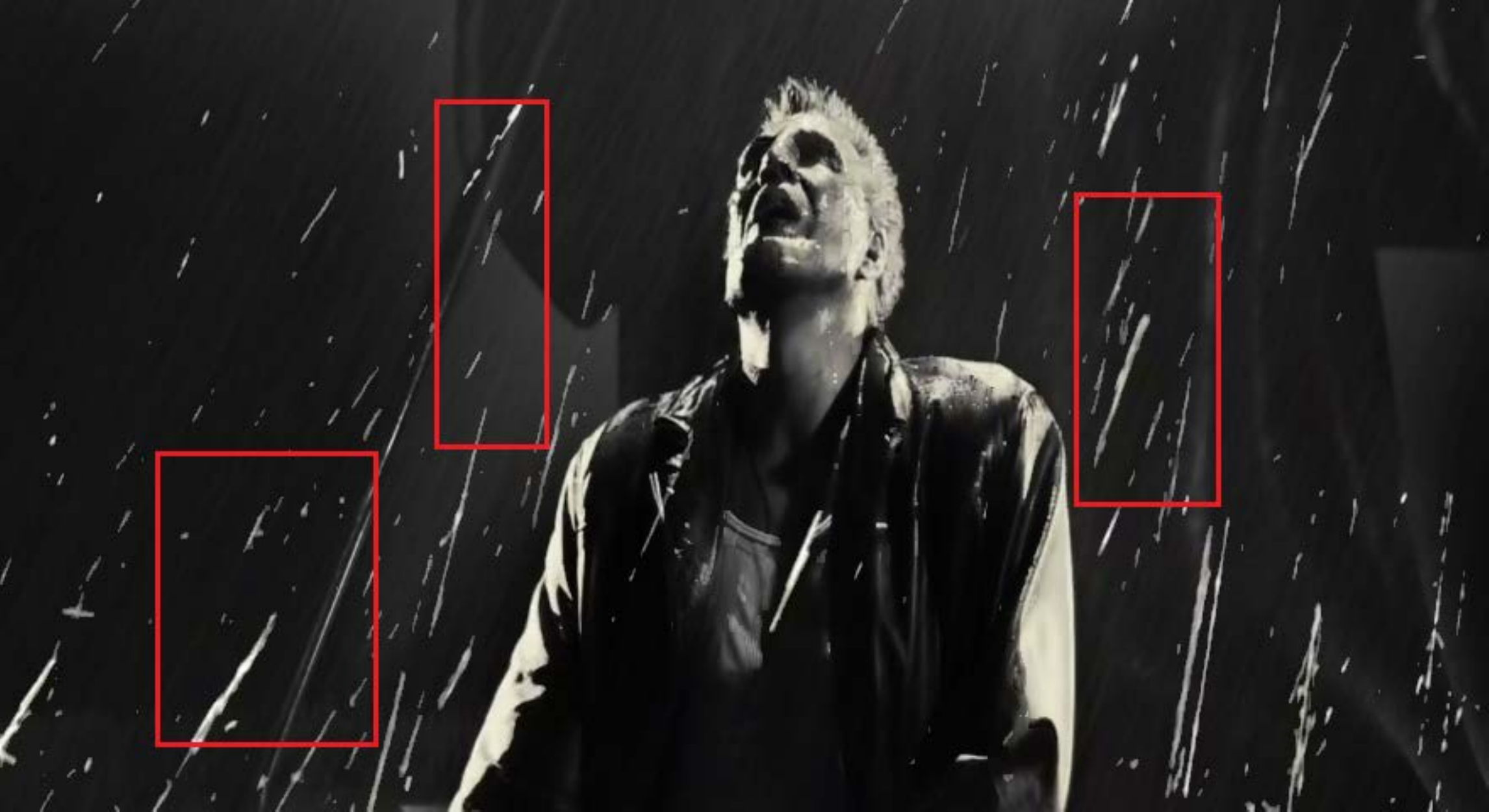}   
		&\includegraphics[width=0.245\textwidth]{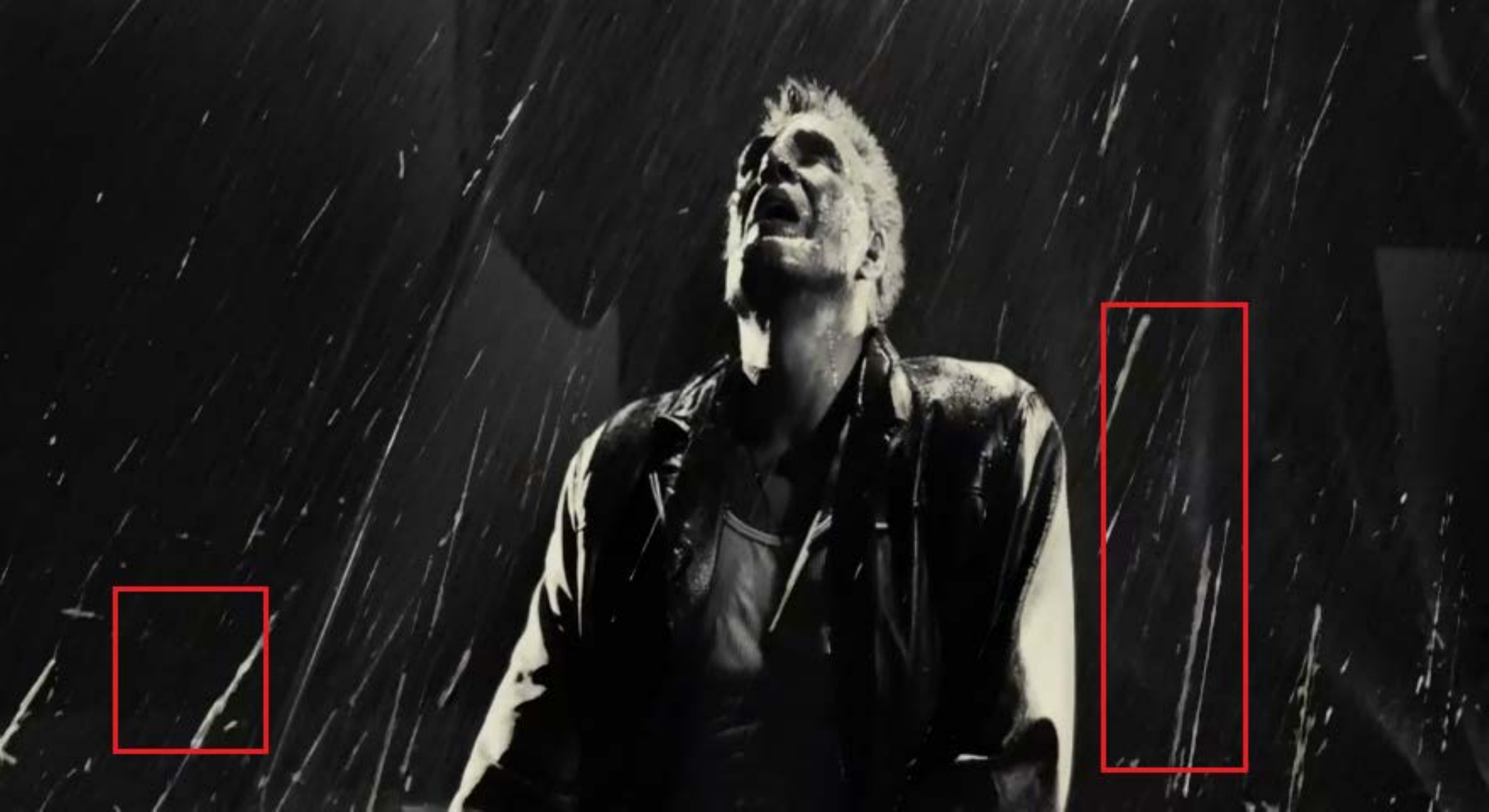}
		&\includegraphics[width=0.245\textwidth]{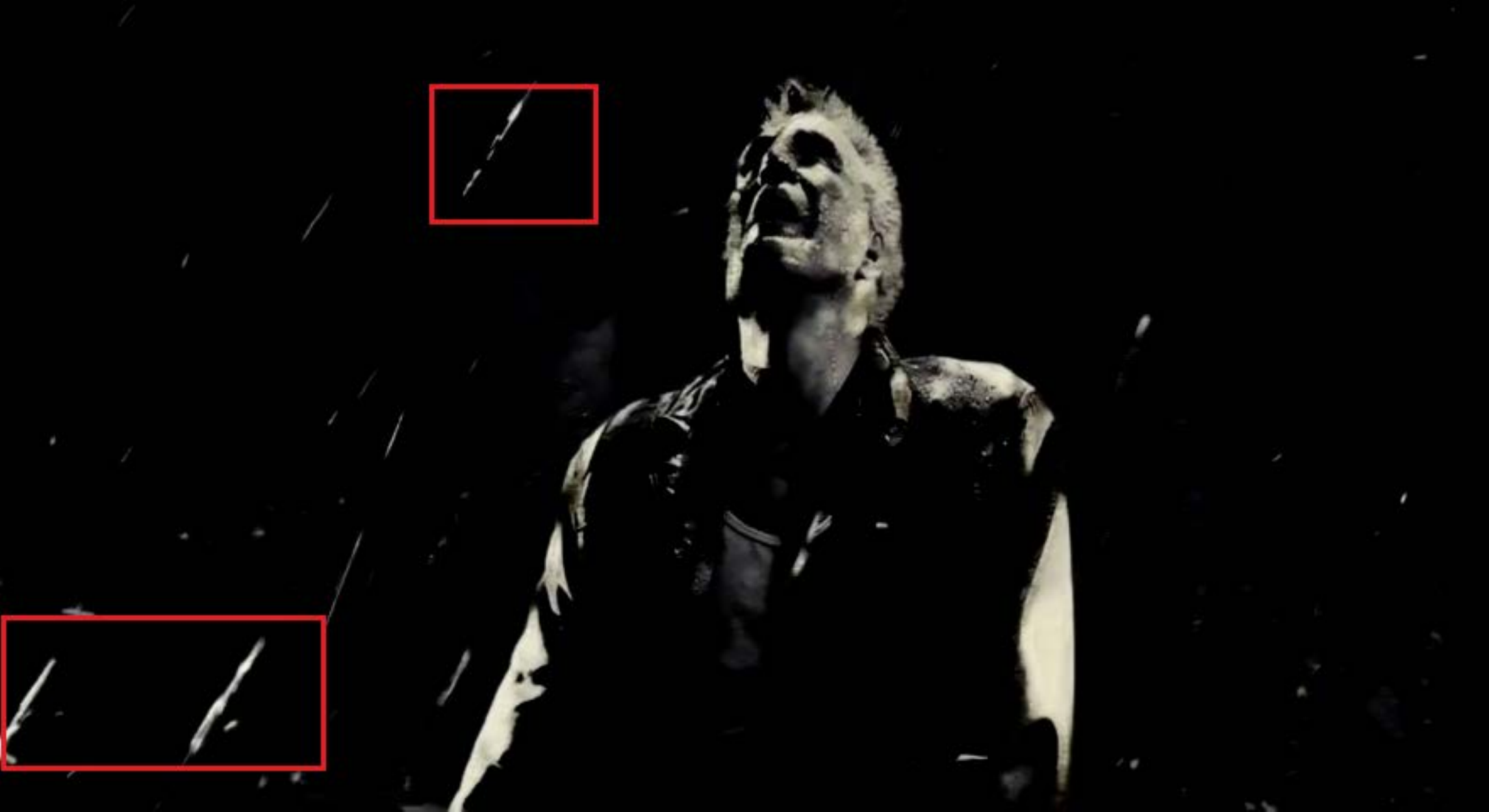}\\
		
		\includegraphics[width=0.245\textwidth]{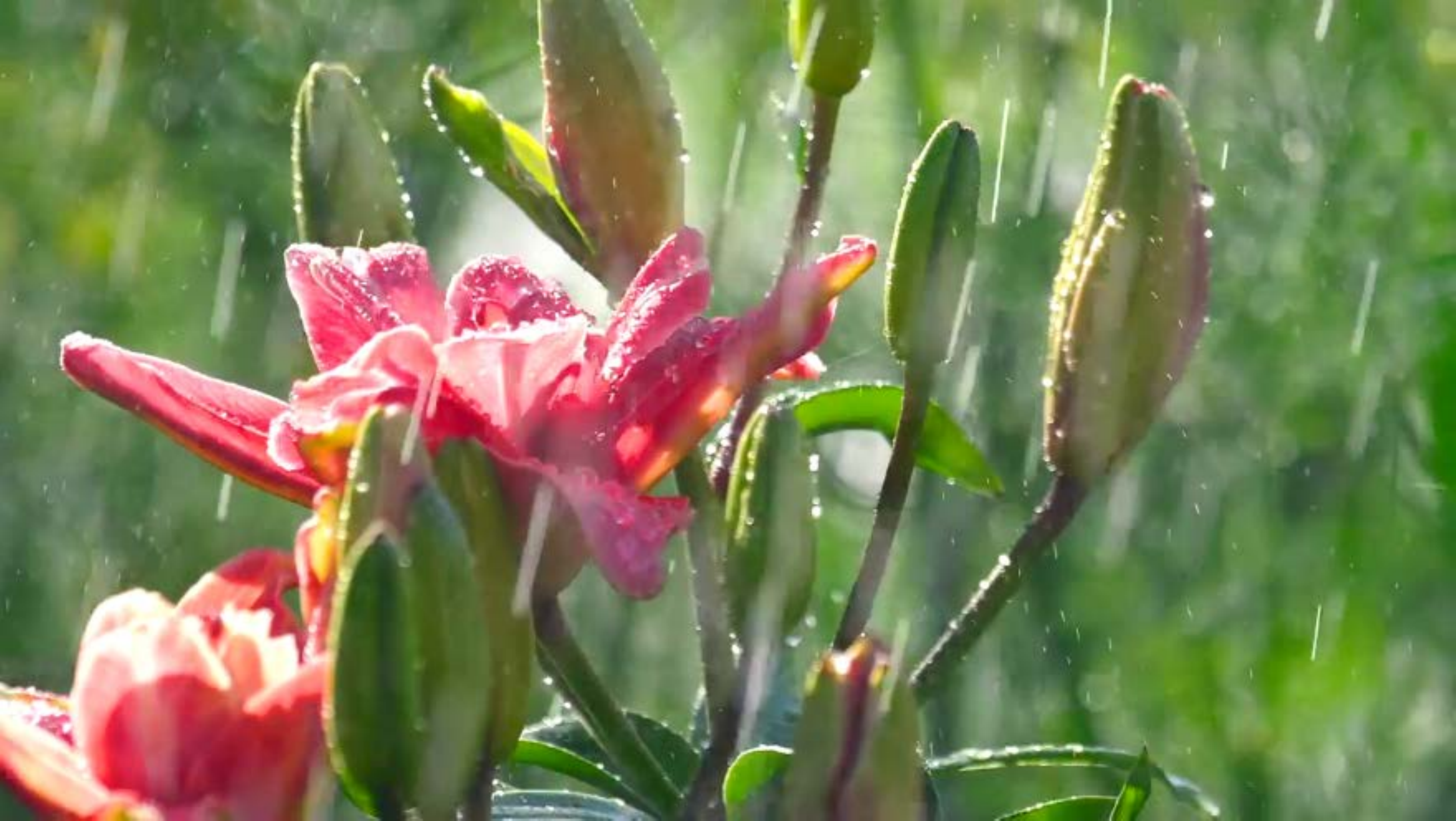}
		&\includegraphics[width=0.245\textwidth]{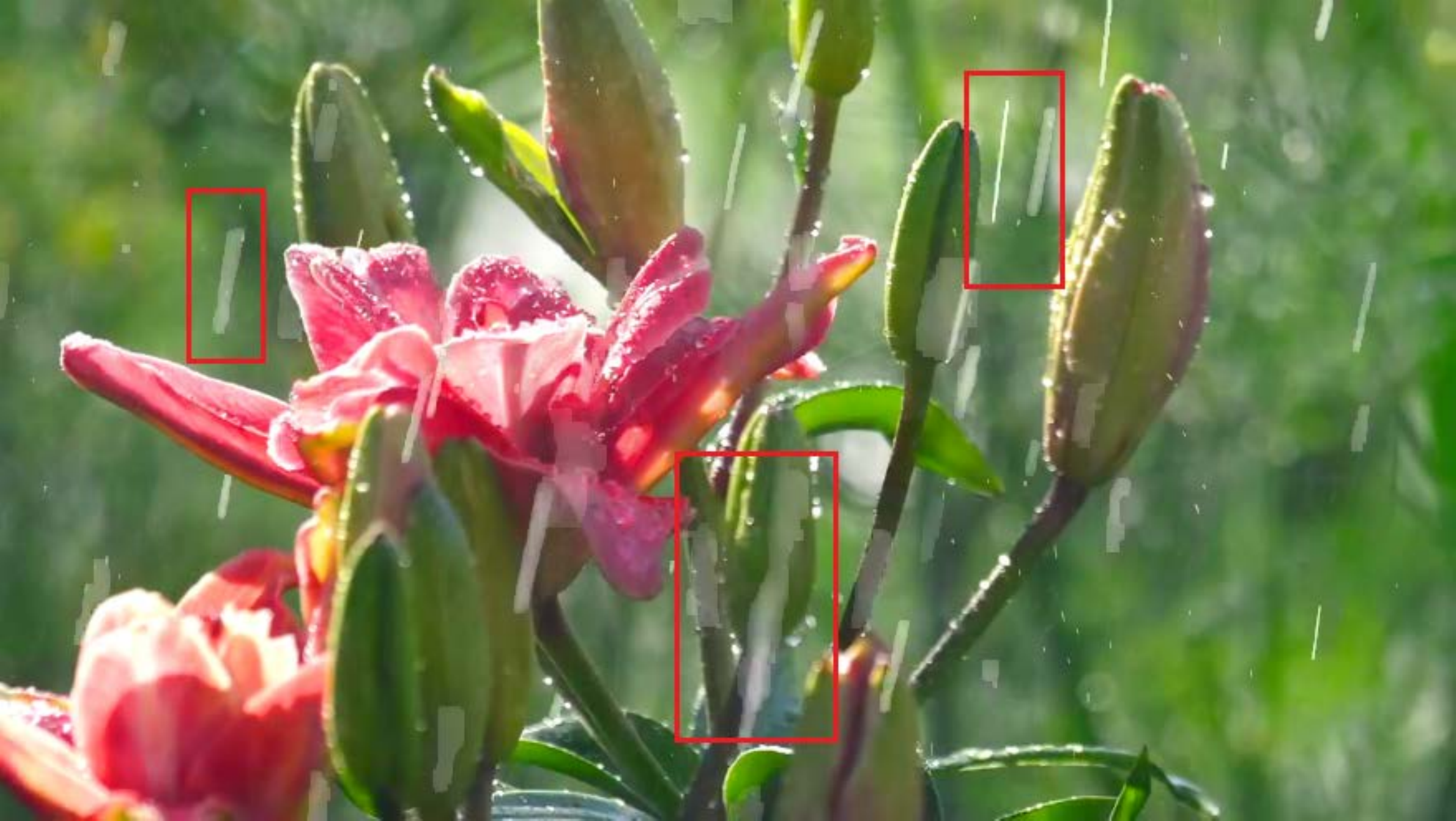}  
		&\includegraphics[width=0.245\textwidth]{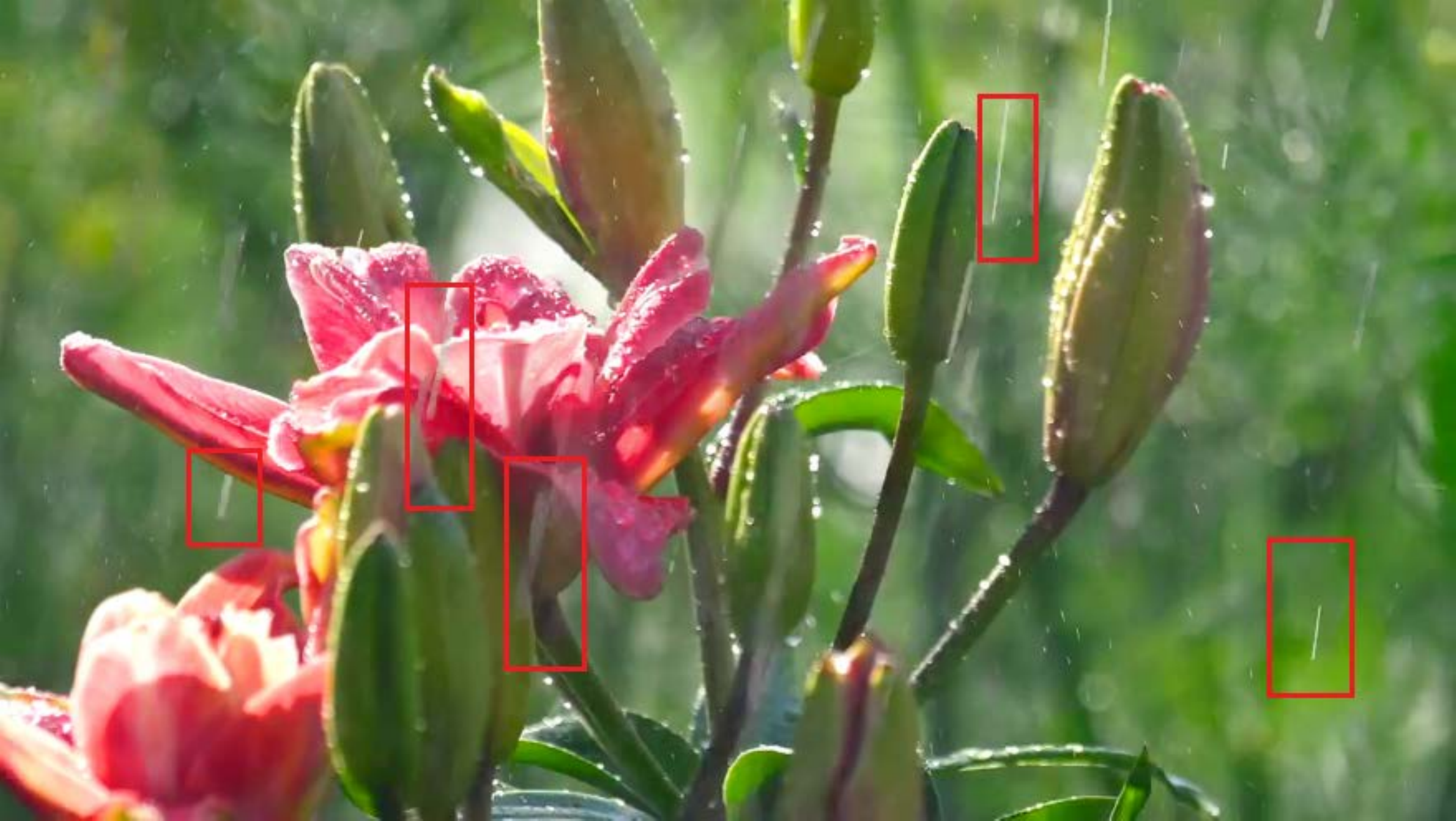}
		&\includegraphics[width=0.245\textwidth]{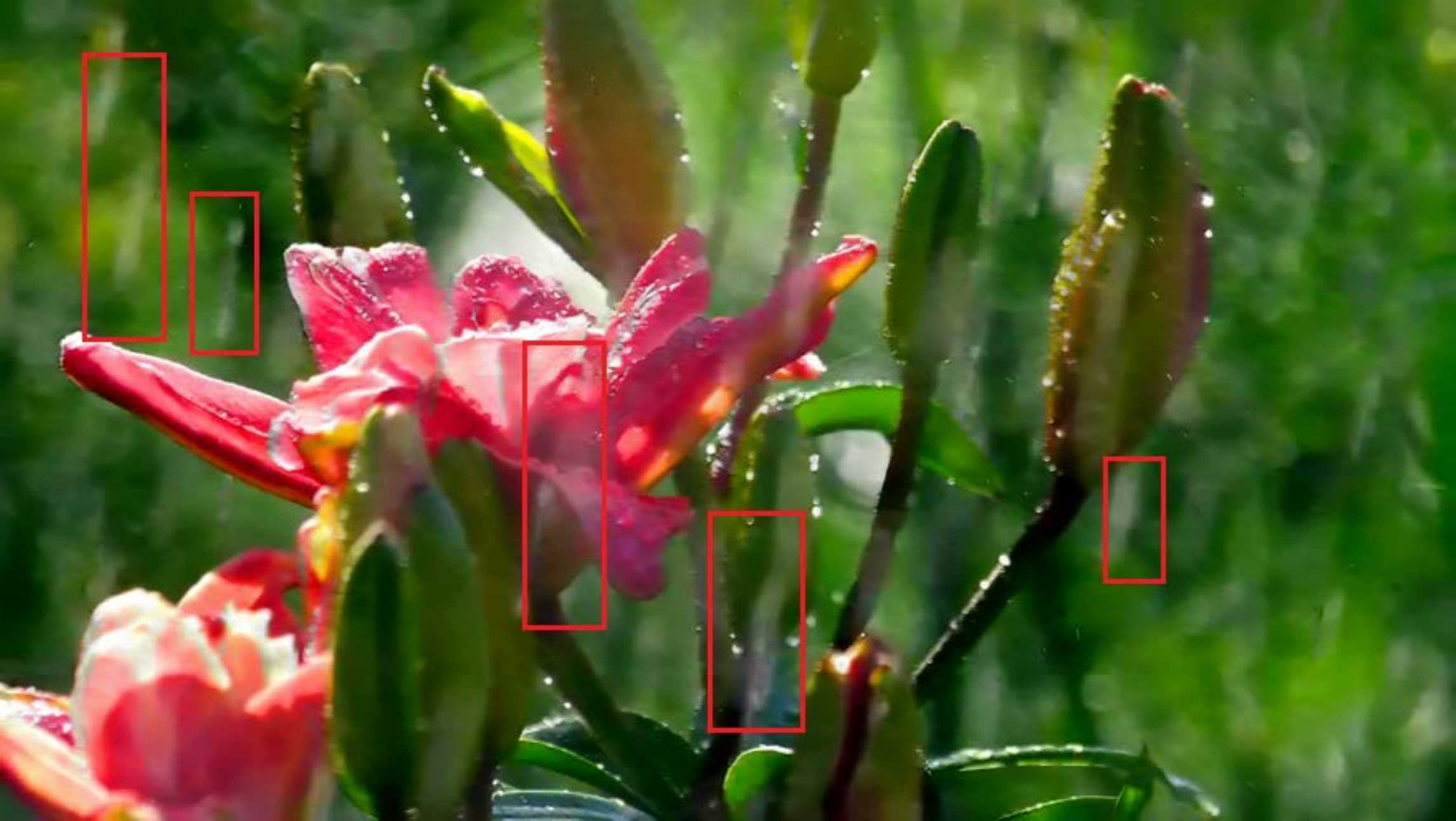}\\
		
		\footnotesize Input & \footnotesize MS-CSC & \footnotesize FastDerain & \footnotesize JORDER \\ 
		\includegraphics[width=0.245\textwidth]{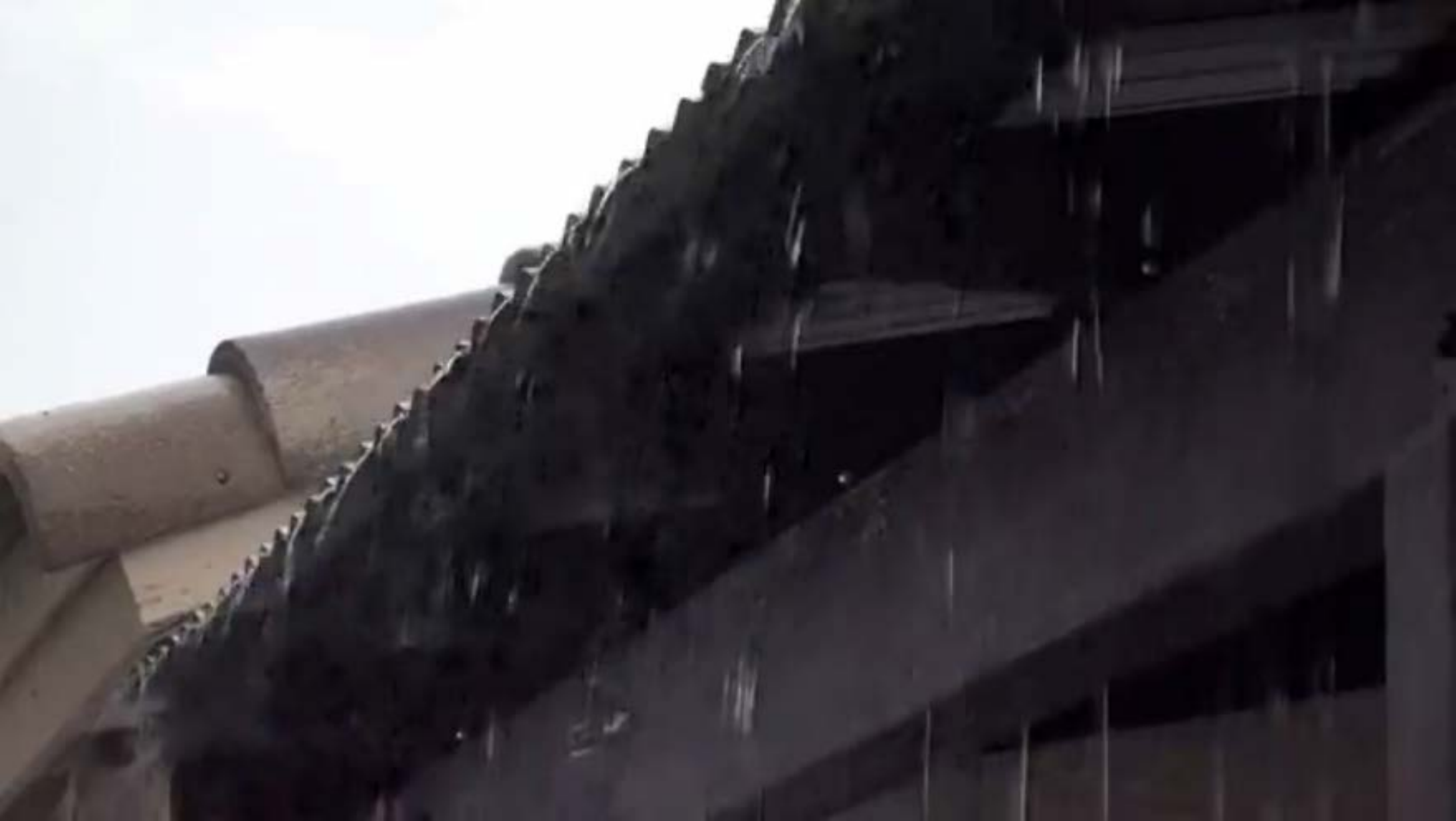}
		&\includegraphics[width=0.245\textwidth]{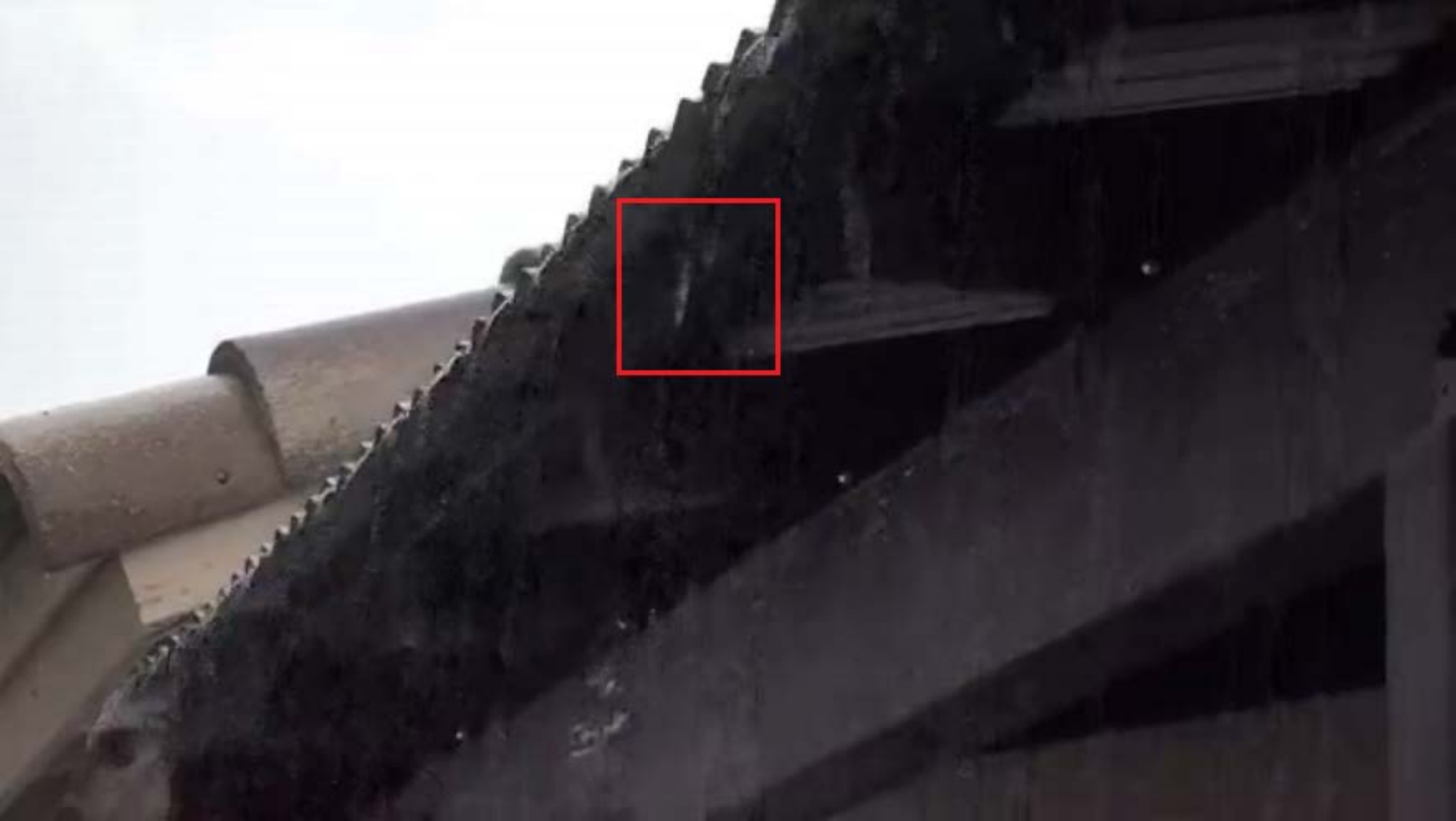}
		&\includegraphics[width=0.245\textwidth]{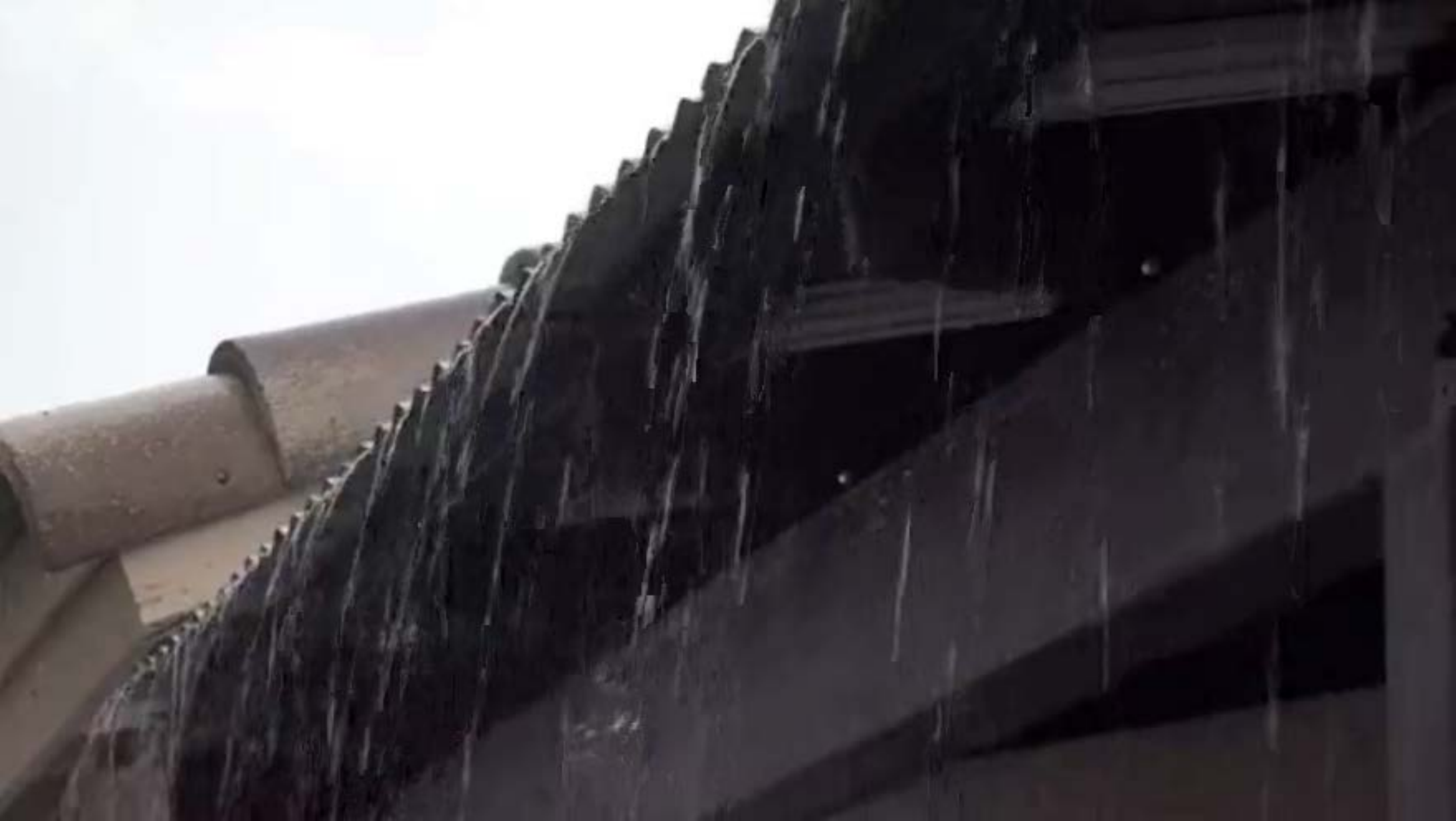}
		&\includegraphics[width=0.245\textwidth]{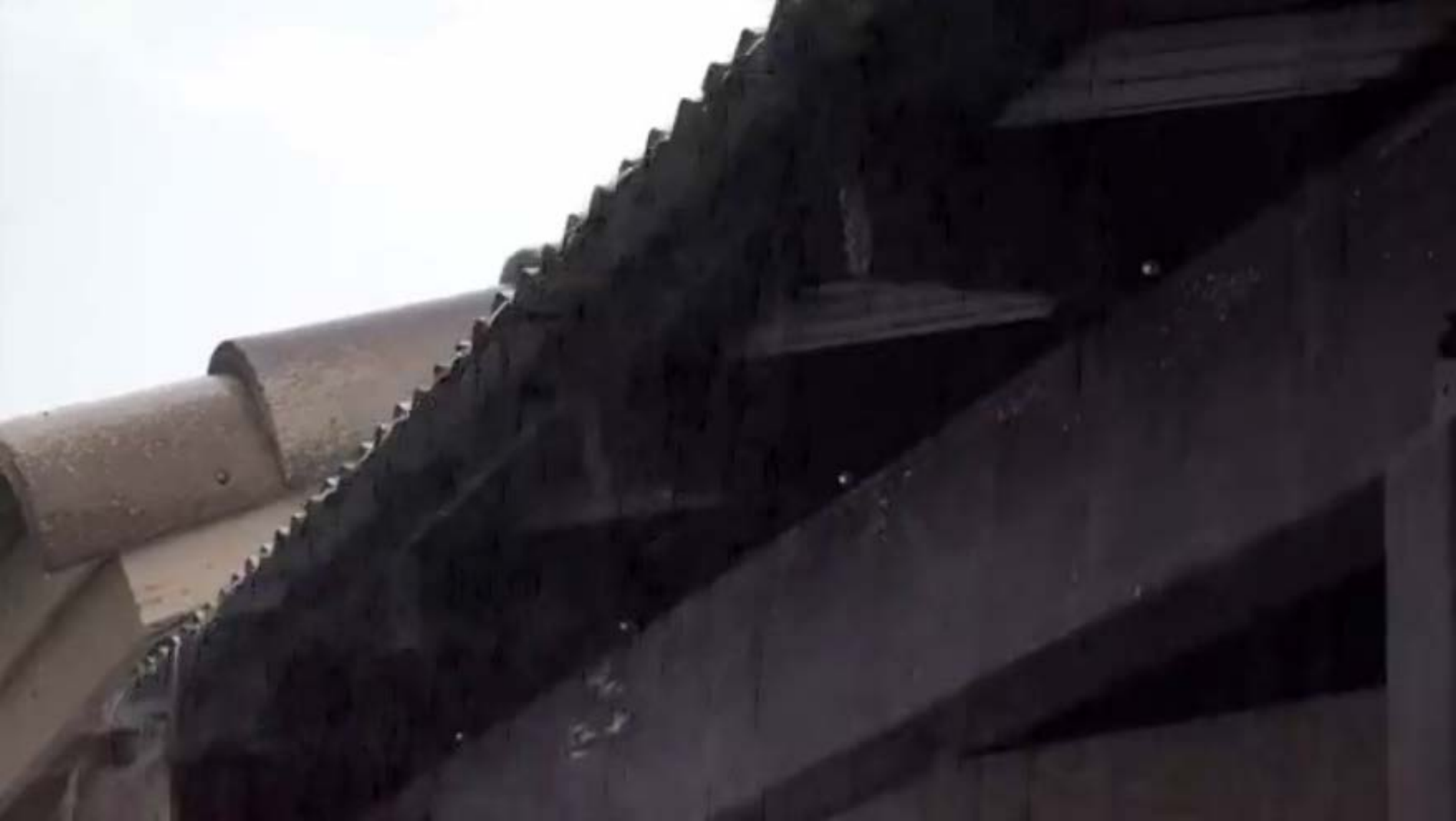}\\
		
		\includegraphics[width=0.245\textwidth]{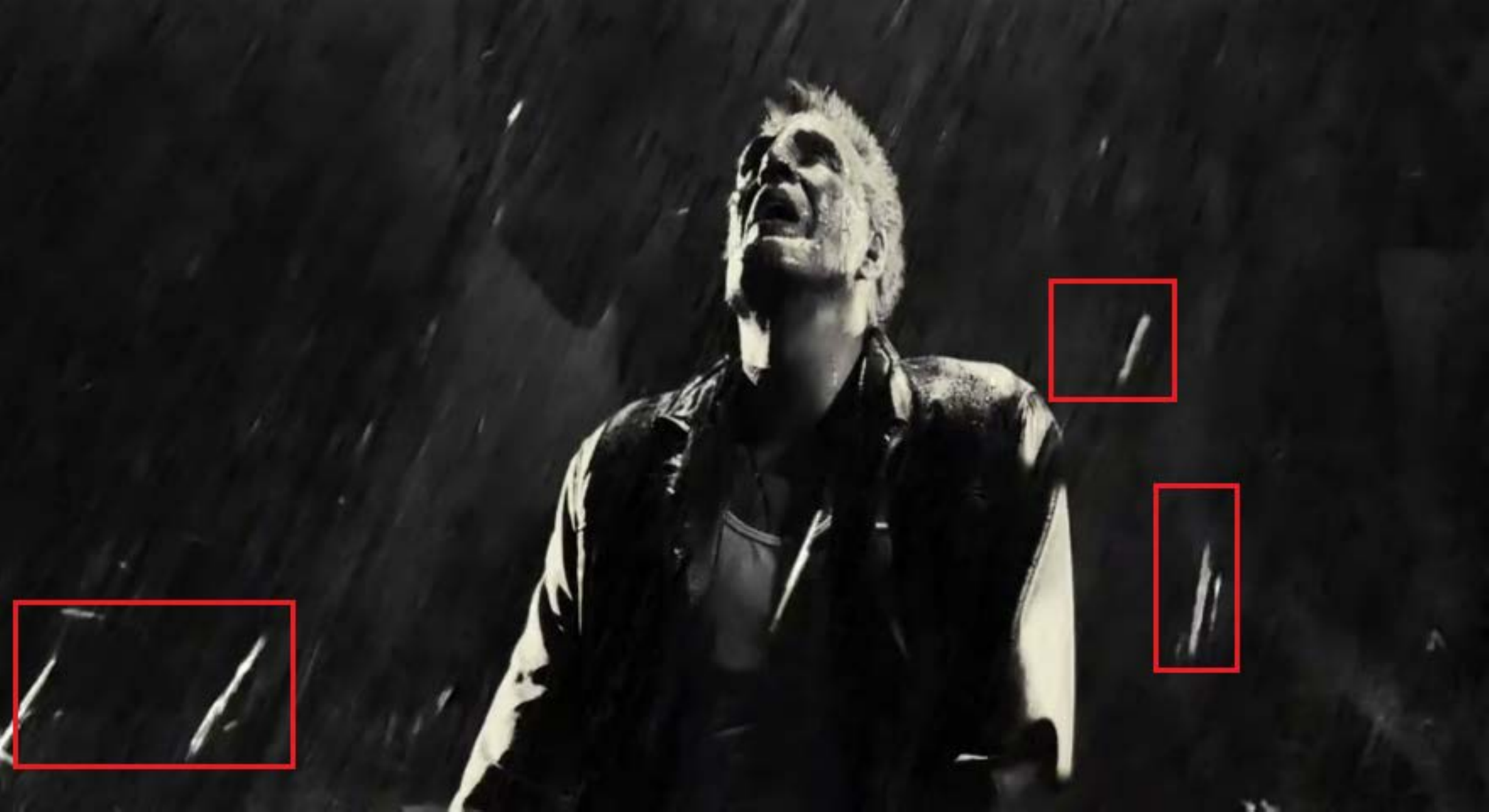}
		&\includegraphics[width=0.245\textwidth]{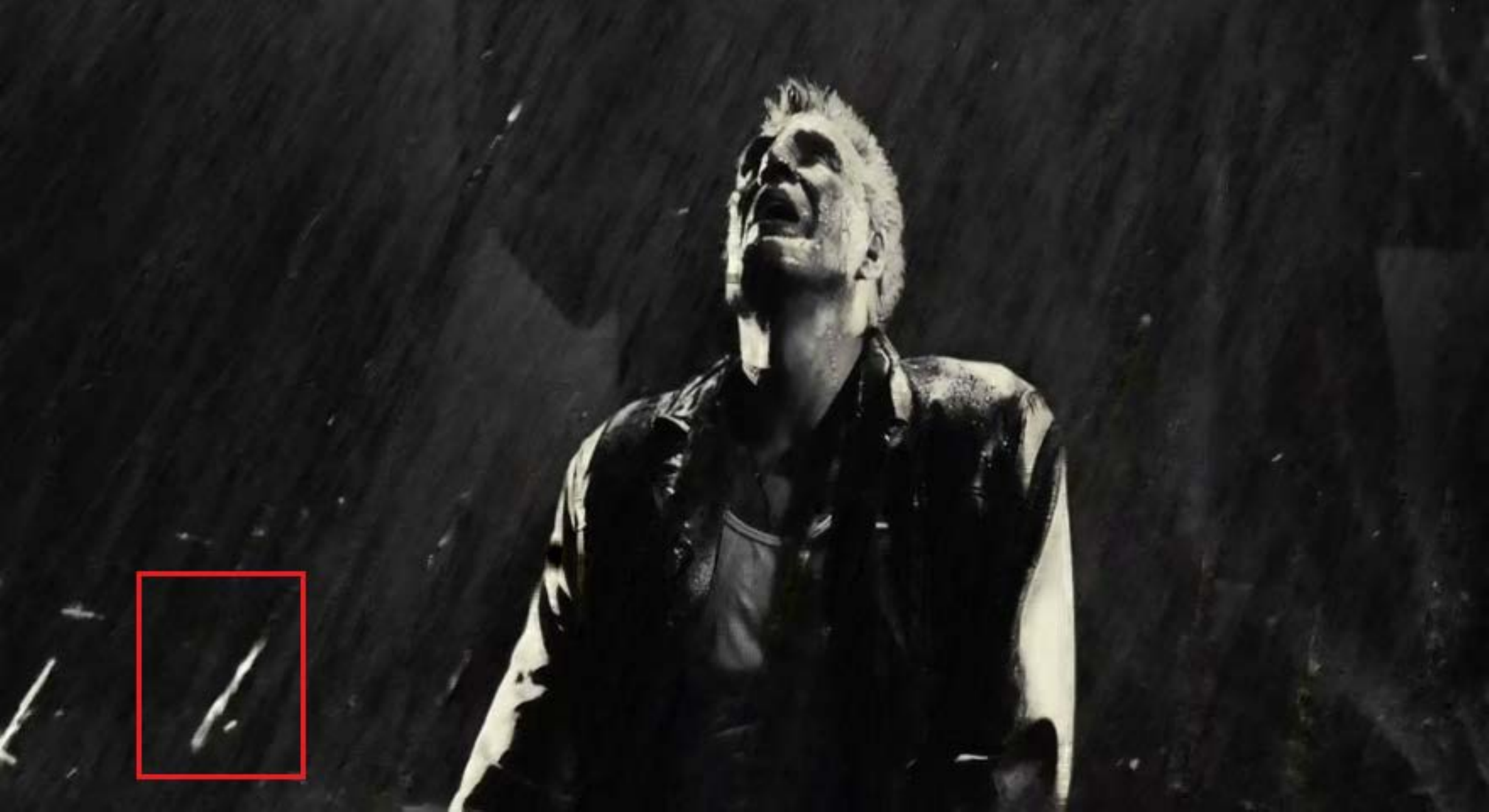}
		&\includegraphics[width=0.245\textwidth]{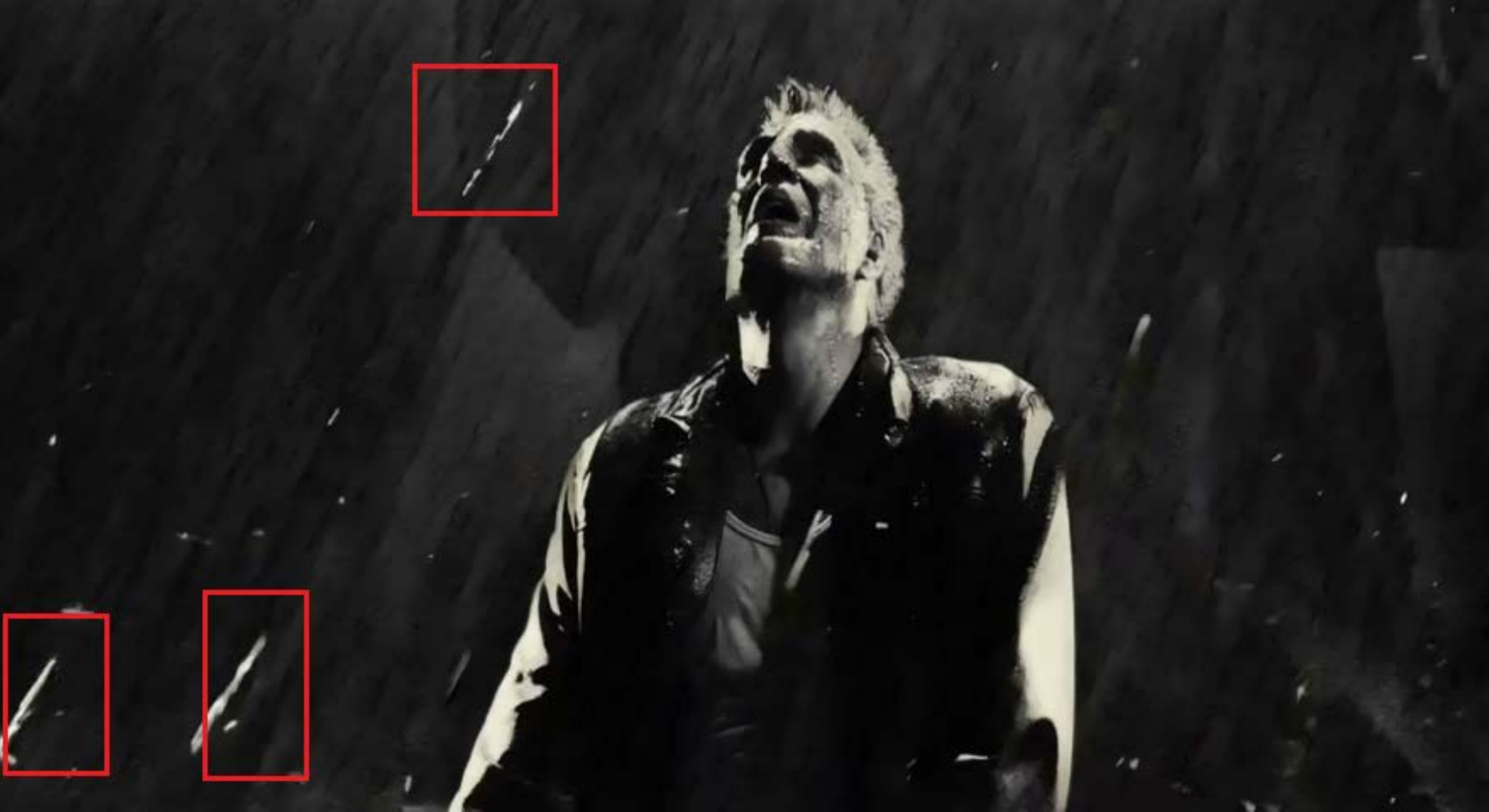}
		&\includegraphics[width=0.245\textwidth]{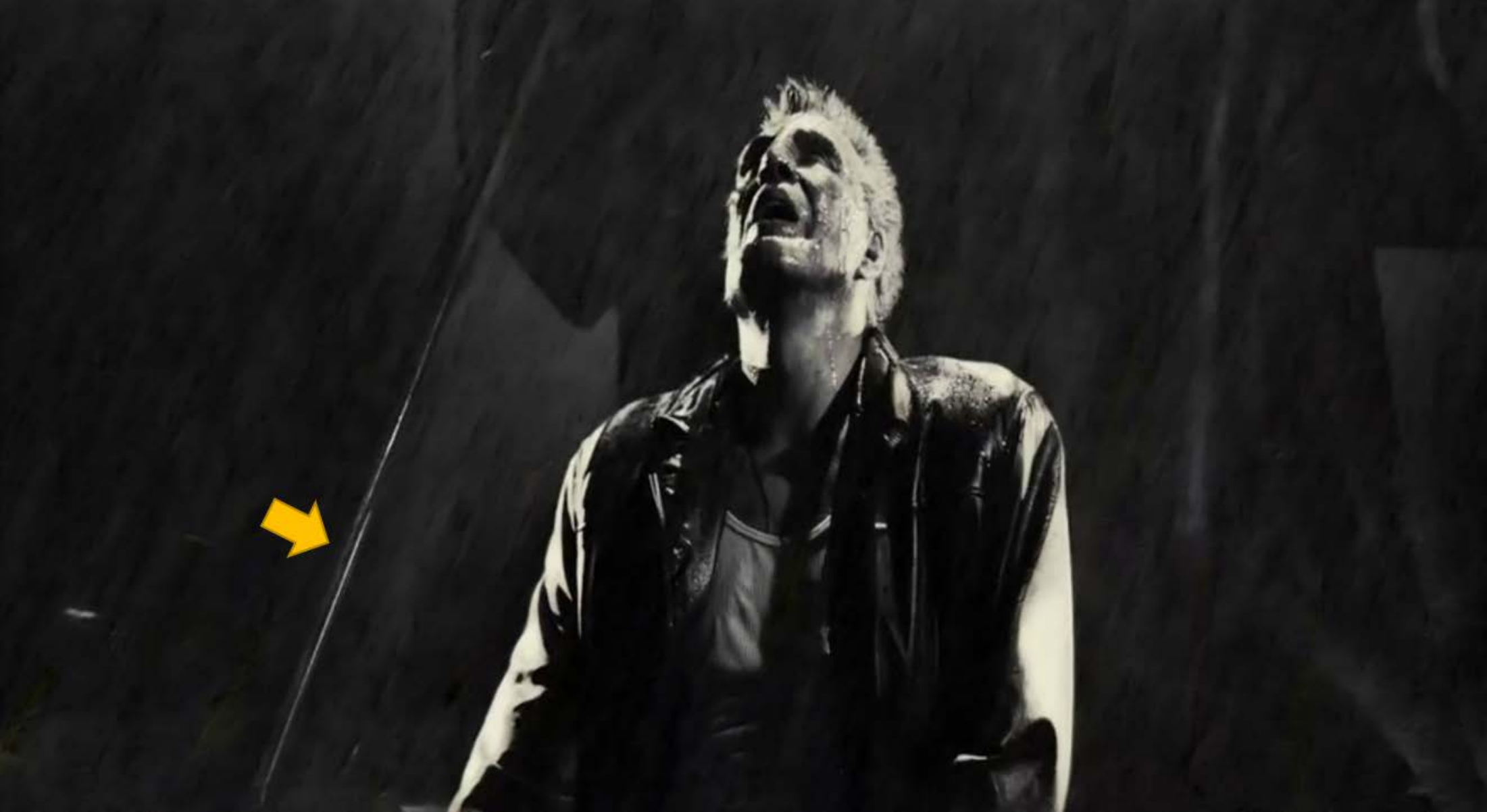}\\
		
		\includegraphics[width=0.245\textwidth]{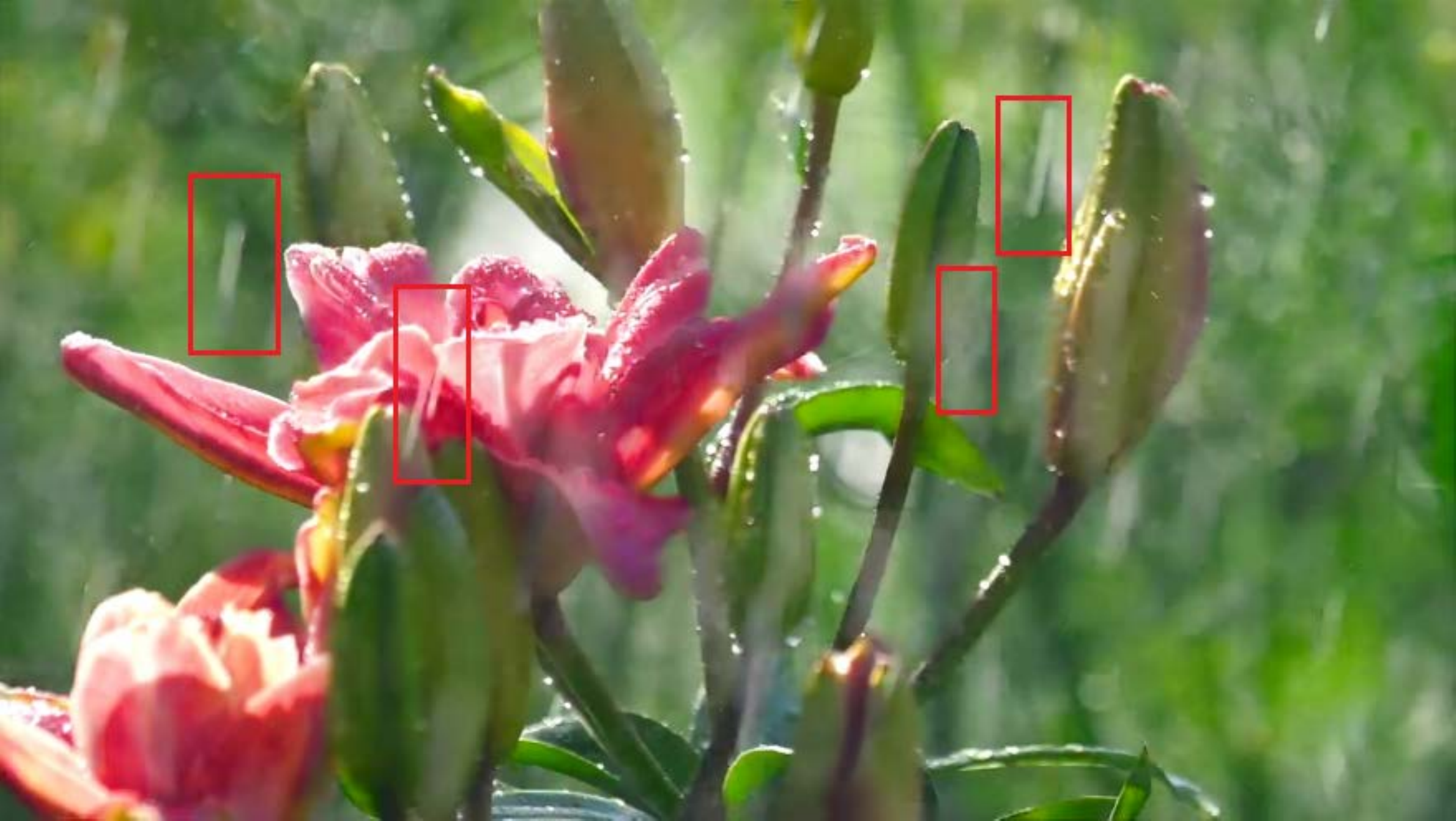}
		&\includegraphics[width=0.245\textwidth]{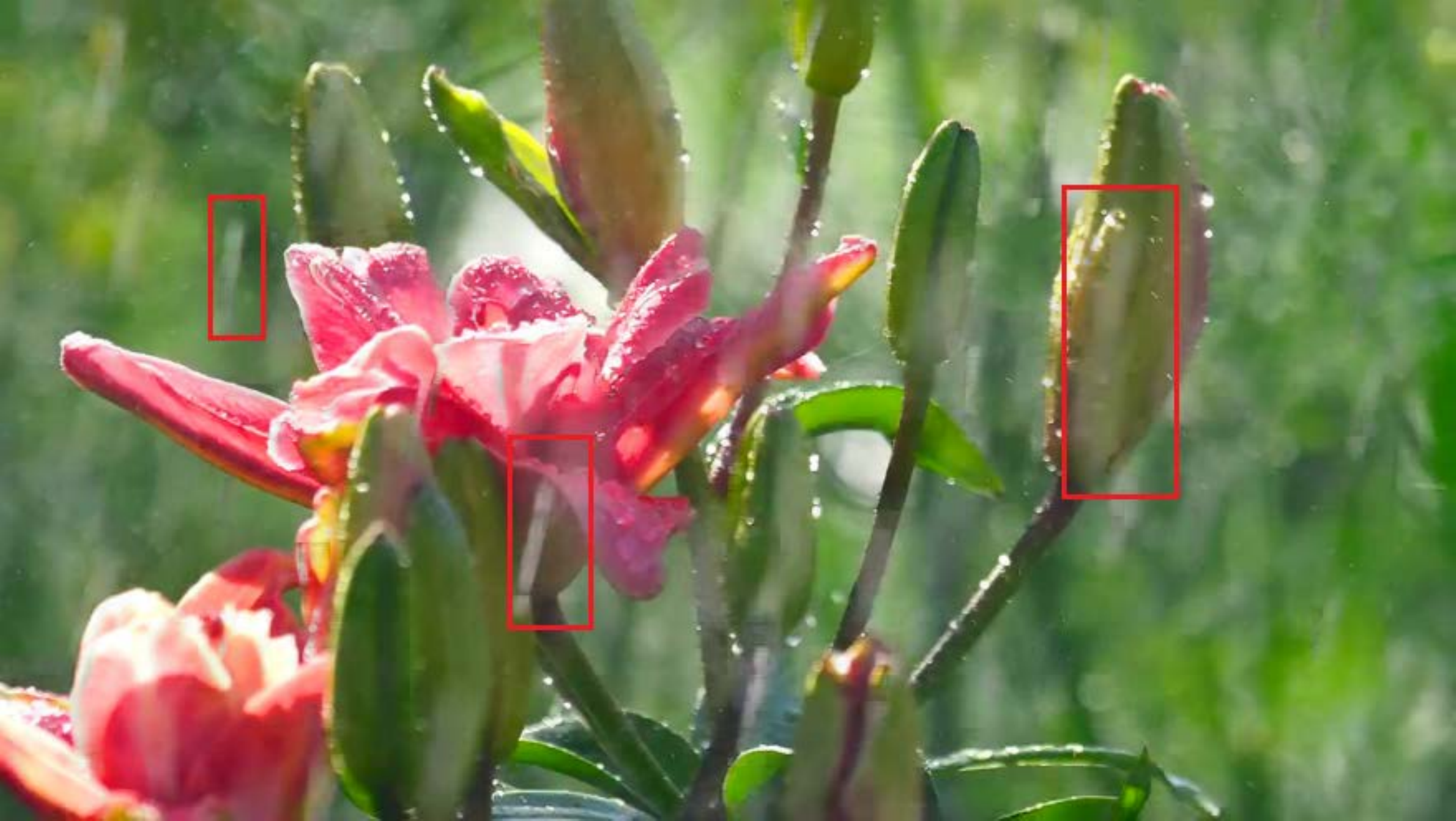}
		&\includegraphics[width=0.245\textwidth]{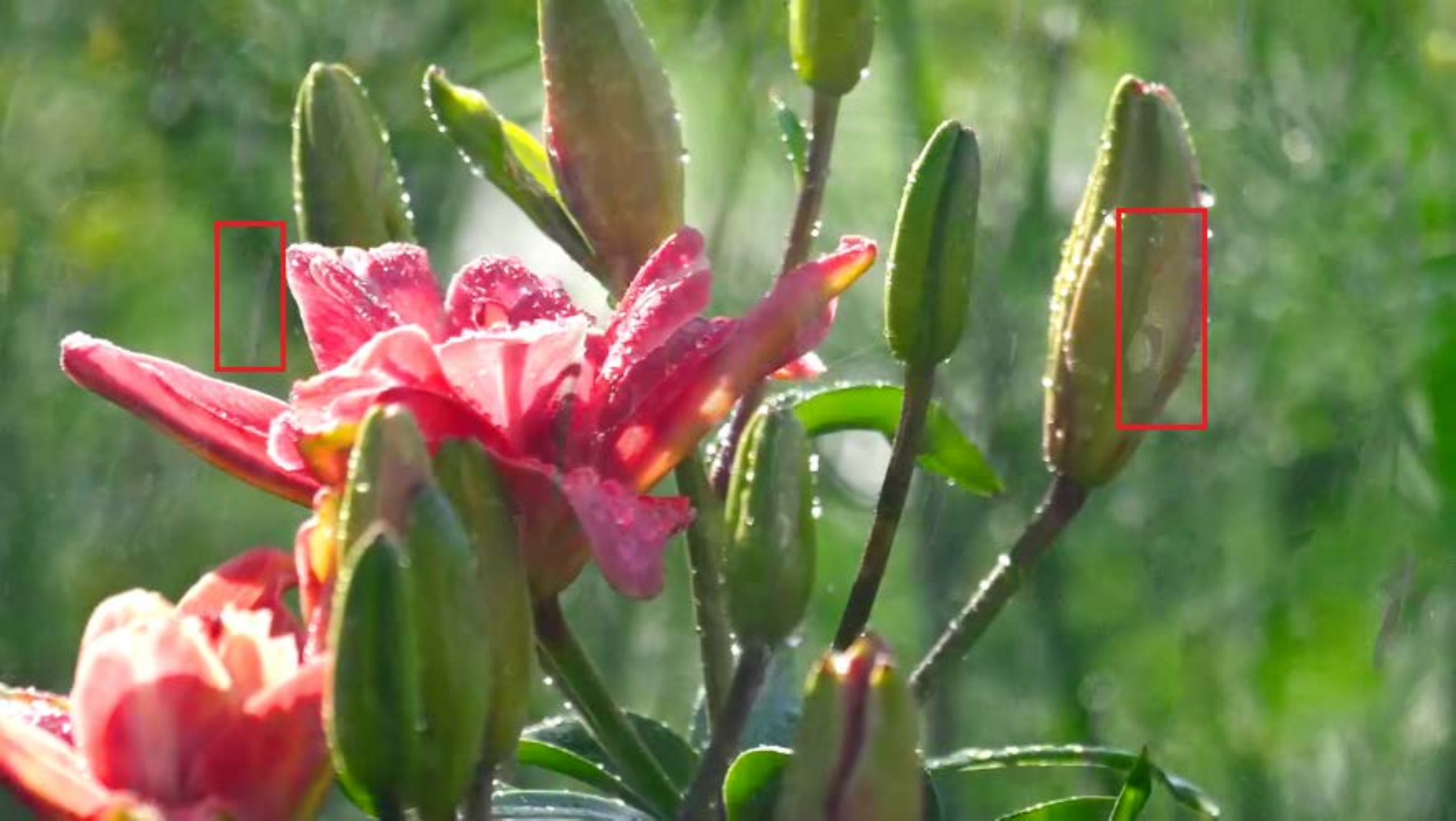}
		&\includegraphics[width=0.245\textwidth]{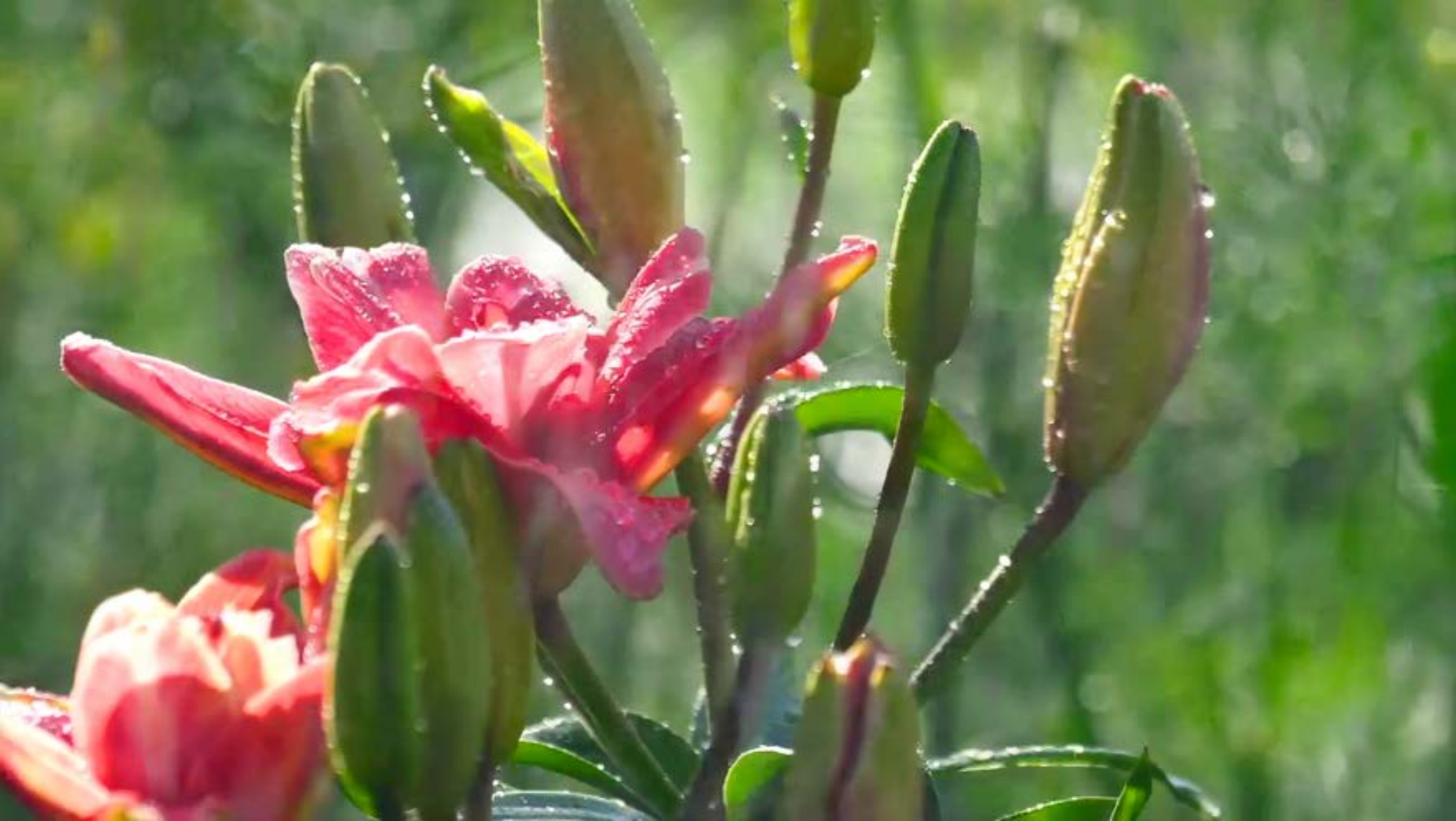}\\
		
		\footnotesize SPANet &  \footnotesize J4R-Net &	 \footnotesize SpacCNN & \footnotesize TMICS (Ours) \\
	\end{tabular}
	\caption{Video deraining performance comparison on three real-world rainy videos. The remaining rain streaks are marked with red boxes. The background details are marked with a yellow arrow.} \label{fig:real_visual}
\end{figure*}

\begin{table}[htb!]
	\centering
	\caption{Averaged VIF, FSIM, NIQE, LPIPS and tLPIPS results among different deraining methods on RainSynComplex25 dataset.}
	\setlength{\tabcolsep}{1.8mm}{
		\begin{tabular}{c|ccccc}
			\hline
			Methods & VIF & FSIM & NIQE  & LPIPS  & tLPIPS  \\
			\hline
			JORDER  & 0.2753  & 0.8243 &   5.270  & 0.407  & 0.199 \\
			FastDeRain  & 0.3347 & 0.8607   & 8.708  & 0.454  & 0.155 \\
			J4R-Net  &  0.2753 &  0.8243  & 3.804  & 0.274  & 0.137 \\
			SpacCNN   &0.1975 & 0.7592   & 4.933  & 0.386  & 0.153 \\
			\hline
			TMICS\_S   & \textcolor{blue}{\textbf{0.3927}} &  \textcolor{blue}{\textbf{0.9112}}   & \textcolor{blue}{\textbf{3.555}} & \textcolor{blue}{\textbf{0.152}} & \textcolor{blue}{\textbf{0.063}}  \\
			TMICS   & \textcolor{red}{\textbf{0.4190}} &  \textcolor{red}{\textbf{0.9243}}   & \textcolor{red}{\textbf{3.284}} & \textcolor{red}{\textbf{0.147}} & \textcolor{red}{\textbf{0.059}}  \\
			\hline
		\end{tabular}
	}
	\label{tab:datasets_NIQE}
\end{table}

\textbf{Evaluating Suitable Frames Number.} 
To evaluate the best frame number, we conduct an experiment on RainSynComplex25 and LasVR datasets about three neighborhood frame settings (i.e., 3, 5 and 7). The corresponding experimental results are listed in Table~\ref{tab:diff_frames}. Observed that unsuitable frames damage the performance in both RainSynComplex25 and LasVR datasets. Thus, in  following experiments, we utilize five continuous observations to estimate one rain-free frame.

\textbf{Impact of the Developed Triple-level Model.} 
To demonstrate the effectiveness of the propose triple-level model, we plotted one group of representative results from RainSynComplex25 in Figure~\ref{fig:diff_model}. Obviously, the result of without DNA still has some residual rain streaks. The result without CNA coupling with rain streaks generation module obtains promising visual results and remove the dominant rain streaks. Our proposed AAS (i.e., with CNA + DNA) generates the most vivid background and achieve rain-free performance. This is because the developed AAS scheme helps preserve both structure and details when removing more rain streaks.  

We then explore the hyper-parameter $\bm{\lambda}$ for AAS by comparing different settings and the experimental results are shown in Figure~\ref{fig:diff_lambda}. Specifically, by choosing a fixed $\bm{\lambda}$ (i.e., $\bm{\lambda}=0,0.5,0.7$), the performance still has space to be improved. Fortunately, the proposed attention mechanism can provide adaptive $\bm{\lambda}$ and incorporate networks leading to a better result.

\textbf{Influence of Different Settings.} 
To validate the effectiveness of each module in our developed method, we make a comprehensive ablation study about five different settings, named Optical Flow Module (OFM), Temporal Grouping Module (TGM), Manually Designed Architecture (MDA), Auto-Searching Architecture (ASA), and Generating Auxiliary Rain Streaks (GARS). The corresponding numerical results are summarized in Table~\ref{tab:diff_modules}.

\emph{Impact of Networks Architecture Search Setting:} To investigate the proposed ASA for collaborative networks, we compare this strategy with manually designed network (i.e., MDA). Specifically, for the manually designed structure, we stack $3\times3$ residual block, dense block, dilated convolution and spatial attention orderly from searching space to design the basic blocks and manually designed network. The experimental results are listed in Table~\ref{tab:diff_modules}-(a) and (b). Observed that model (b) with auto-searching strategy outperforms manually designed model (i.e., model (a)) in terms of PSNR and SSIM scores. The superiority of the proposed ASA module demonstrates that the network searching architecture can effectively exploit basic operators in search space. 

\emph{Impact of GARS Module:} We also compare the effectiveness of GARS in  Table~\ref{tab:diff_modules}-(d) and (e). The collaborative networks with GRS obtains significant improvements based on the auxiliary information of structural factors. This is mainly because the generated auxiliary rain streaks enable the capability of network to characterize a wide range of video circumstances.

\emph{Frames Estimation Modules:} To explore the proposed alternative strategy, we introduce three baseline models, i.e. (b) only with  OFM, (c) only with  TGM and (d) with alternative OFM or TGM. As can be seen in Table~\ref{tab:diff_modules}, model (d) outperforms (b) and (c) both on SynHybrid and LasVR datasets. This experiment illustrate the effectiveness of the developed alternative module.

\subsection{Comparison with State-of-the-Art}

\textbf{Comparing on Synthesized Datasets.} 
We compare the developed method with some state-of-the-art approaches on video deraining, including both single-image deraining methods (i.e., JOint Rain DEtection and Removal network (JORDER~\cite{yang2017deep}), Density-aware Single-Image Deraining using Multi-stream Dense Network (DID-MDN \cite{zhang2018density-aware}), SPatial Attentive single-image deraining network (SPANet~\cite{wang2019spatial})) and video deraining methods (i.e., Multi-Scale Convolutional Sparse Coding (MS-CSC \cite{li2018video}), FastDeRain~\cite{jiang2019fastderain}, Joint Recurrent Rain Removal and Reconstruction Network (J4R-Net~\cite{liu2018erase}), SuperPixel Alignment and Compensation CNN (SpacCNN~\cite{chen2018robust}), DualFlow~\cite{yang2019frame}, 	CLEARER~\cite{gou2020clearer} and Self-Learned Deraining Network (SLDNet~\cite{yang2020self})). The performance of video rain streaks removal is evaluated on three rainy video benchmarks: RainSynLight25, RainSynComplex25 and NTURain, which involve diversified kinds of rain streaks including direction, scale, density and intensity.

As for the quantitative comparison, we calculated two widely used metrics (i.e., PSNR and SSIM), and listed experimental results in Table~\ref{tab:datasets_results}. It can be seen that our approach shows significant superiority to previous methods on three datasets. Compared with recently proposed DualFlow, our method attains more than 0.85dB, 1.77dB and 1.33dB in PSNR on RainSynLight25, RainSynComplex25 and NTURain datasets respectively. The result of single DNA (i.e., TMICS\_S) also obtains promising performances. This corroborates the flexibility and universality of our proposed method when dealing with various video situation with different rain streaks types. Except for PSNR and SSIM metrics that measure performance by pixel-wise accuracy, we further calculate NIQE, LPIPS and tLPIPS to evaluate the quantified perceptual quality. We compare our method with four methods which have relative high PSNR and SSIM scores (i.e., JORDER, FastDerain, J4R-Net and SpacCNN), and the comparison results on challenging RainSynComplex25 are listed in Table \ref{tab:datasets_NIQE}. As the lower NIQE, LPIPS and tLPIPS values denote better perceptual quality,  our method achieves the best perceptual performance on different evaluating metrics. 

As for the qualitative performance, Figure~\ref{fig:visual_datasets} and Figure~\ref{fig:visual_NTU} show the visual comparisons between our scheme and four best methods with relative high PSNR and SSIM scores (i.e., FastDerain, J4R-Net, JORDER and SpacCNN). Observed that in Figure~\ref{fig:visual_datasets}, the proposed method performs the best while other methods (i.e., FastDeRain, JORDER, J4RNet, and SpacCNN) still have rain streaks. Besides, Figure~\ref{fig:visual_NTU} depicts the visual performance on NTURain dataset. Observed that FastDerain, J4R-Net, and JORDER still have rain streaks and mistake the background details as rain streaks. SpacCNN  generates a much blurred background. Overall, our method provides more effective performance with less remaining rain streaks, abundant details, and less blurs.

\textbf{Comparing on Real-world Video Datasets.} 
To further illustrate the performance of our method, we chose three difficult real-world rainy videos with various rain circumstances against a series of competitive methods, including MS-CSC, FastDerain, JORDER, SPANet, J4R-Net and SpacCNN. These rainy video sequences are collected from different scenes, i.e., YouTube\footnote{https://www.youtube.com/}, movie clips, and Mixkit\footnote{https://mixkit.co/free-stock-video/rain/?page=3}. As shown in Figure~\ref{fig:real_visual}, previous methods tend to leave distinct rain streaks and mistake background details as rain streaks. Fortunately, our developed method shows good capability that preserves background details as well as removes more rain streaks.

\section{Conclusion}
This work developed a model-guided auto-searching method by the formulated triple-level video deraining optimization framework for removing different video rain streaks. We first introduced a macroscopic structure searching scheme for inter-frames information extraction. Then, we designed a cooperating optimization model about task variables and hyper-parameter based on the re-constructed comprehensive model. For the task variable propagation, we designed two collaborative structures, i.e., DNA and CNA. Subsequently, we introduced an attention-based averaging scheme to effectively fuse features from collaborative structures. To obtain state-of-the-art neural network structures (i.e., DNA and CNA), we applied the microscopic network architectures searching from a compact task-specific search space to discover desirable video deraining architectures.


%

%


\ifCLASSOPTIONcaptionsoff
  \newpage
\fi



%
%
%

\bibliographystyle{IEEEtran}
\bibliography{reference}

\begin{IEEEbiography}[{\includegraphics[width=1in,height=1.25in,clip,keepaspectratio]{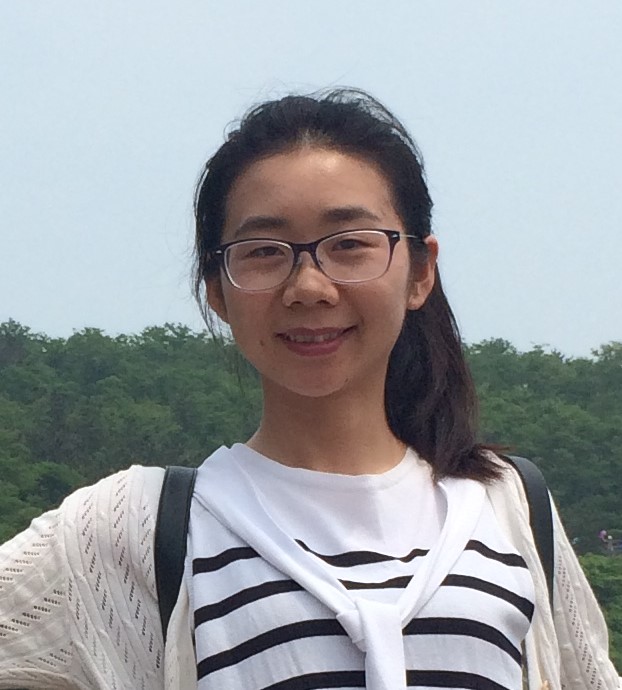}}]{Pan Mu} received the B.S. degree in Applied Mathematics from Henan University, China, in 2014, the M.S. degree in Operational Research and Cybernetics from Dalian University of Technology, China, in 2017. She received the Ph.D. degrees in mathematics from the Dalian University of Technology in 2021. She is currently a lecturer with the College of Computer Science and Technology, Zhejiang University of Technology. Her research interests include computer vision, machine learning, control and optimization. 
\end{IEEEbiography}
\vspace{-1.2cm}
\begin{IEEEbiography}[{\includegraphics[width=1in,height=1.25in,clip,keepaspectratio]{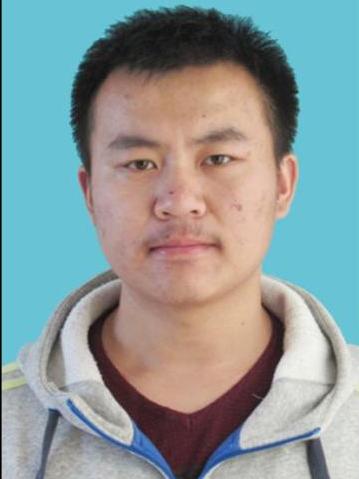}}]{Zhu Liu} received the B.E. degree in software engineering from the Dalian University of Technology, Dalian, China, in 2019, where he is currently pursuing the master’s degree in software engineering. He is with Key Laboratory for Ubiquitous Network and Service Software of Liaoning Province, Dalian University of Technology. His research interests include  computer vision and deep learning. 
\end{IEEEbiography}
\vspace{-1.2cm}
\begin{IEEEbiography}[{\includegraphics[width=1in,height=1.25in,clip,keepaspectratio]{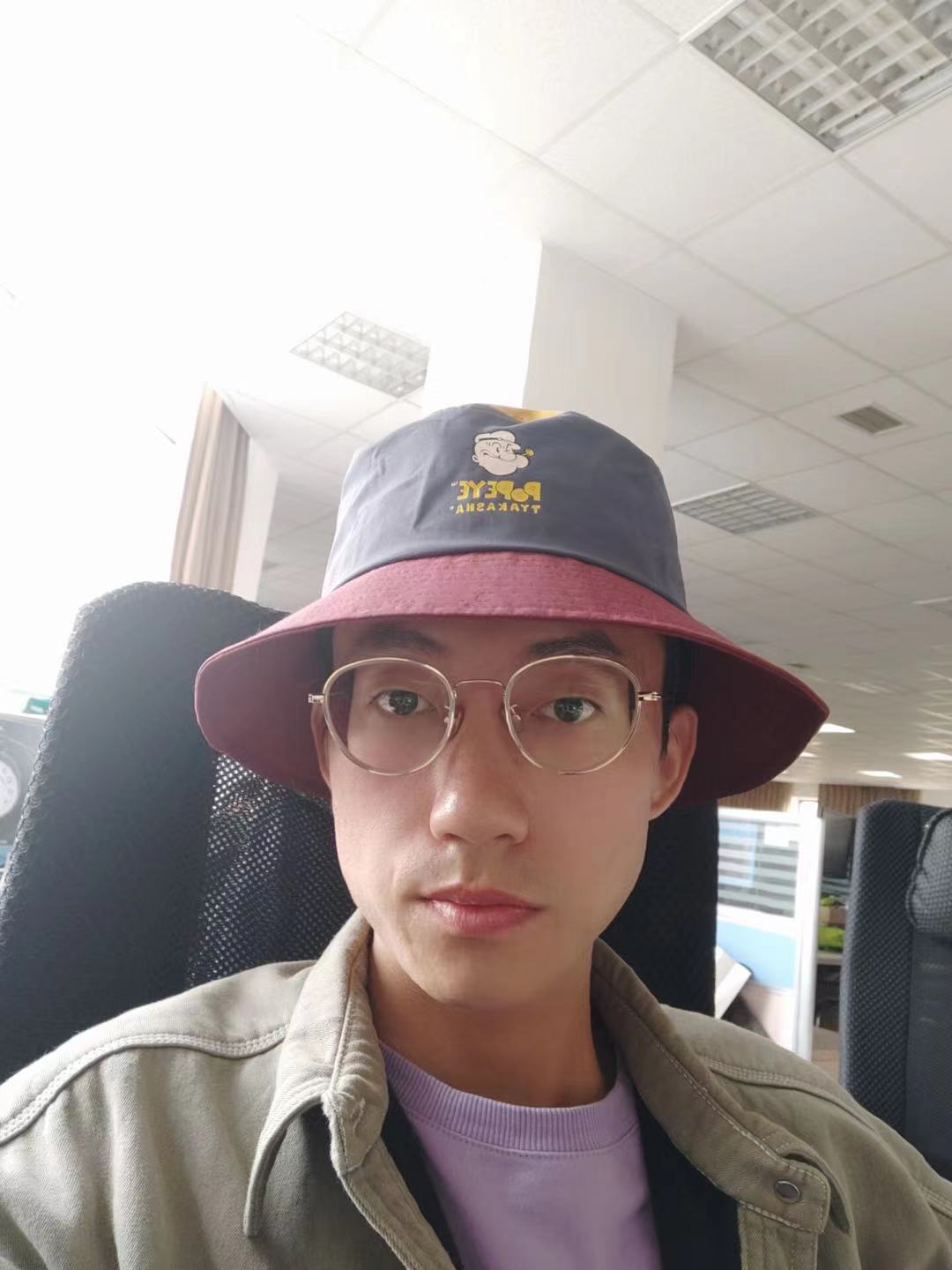}}]{Yaohua Liu} received the B.S. degree in Software Engineering from Dalian University of Technology, China in 2019. He is currently pursuing the PhD degree in software engineering at Dalian University of Technology, Dalian, China. He is with the Key Laboratory for Ubiquitous Network and Service Software of Liaoning Province, Dalian University of Technology, Dalian, China. His research interests include computer vision, machine learning and optimization. 
\end{IEEEbiography}
\vspace{-1.2cm}
\begin{IEEEbiography}[{\includegraphics[width=1in,height=1.25in,clip,keepaspectratio]{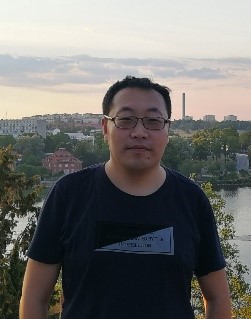}}]{Risheng Liu} received the B.S. and Ph.D. degrees both in mathematics from the Dalian University of Technology in 2007 and 2012, respectively. He was a visiting scholar in the Robotic Institute of Carnegie Mellon University from 2010 to 2012. He served as Hong Kong Scholar Research Fellow at the Hong Kong Polytechnic University from 2016 to 2017. He is currently a professor with DUT-RU International School of Information Science \& Engineering, Dalian University of Technology. He was awarded the ``Outstanding Youth Science Foundation'' of the National Natural Science Foundation of China. His research interests include machine learning, optimization, computer vision and multimedia. He was a co-recipient of the IEEE ICME Best Student Paper Award in both 2014 and 2015. His two papers were also selected as Finalist of the Best Paper Award in ICME 2017. He is a member of the IEEE and ACM.
\end{IEEEbiography}
\vspace{-1.2cm}
\begin{IEEEbiography}[{\includegraphics[width=1in,height=1.25in,clip,keepaspectratio]{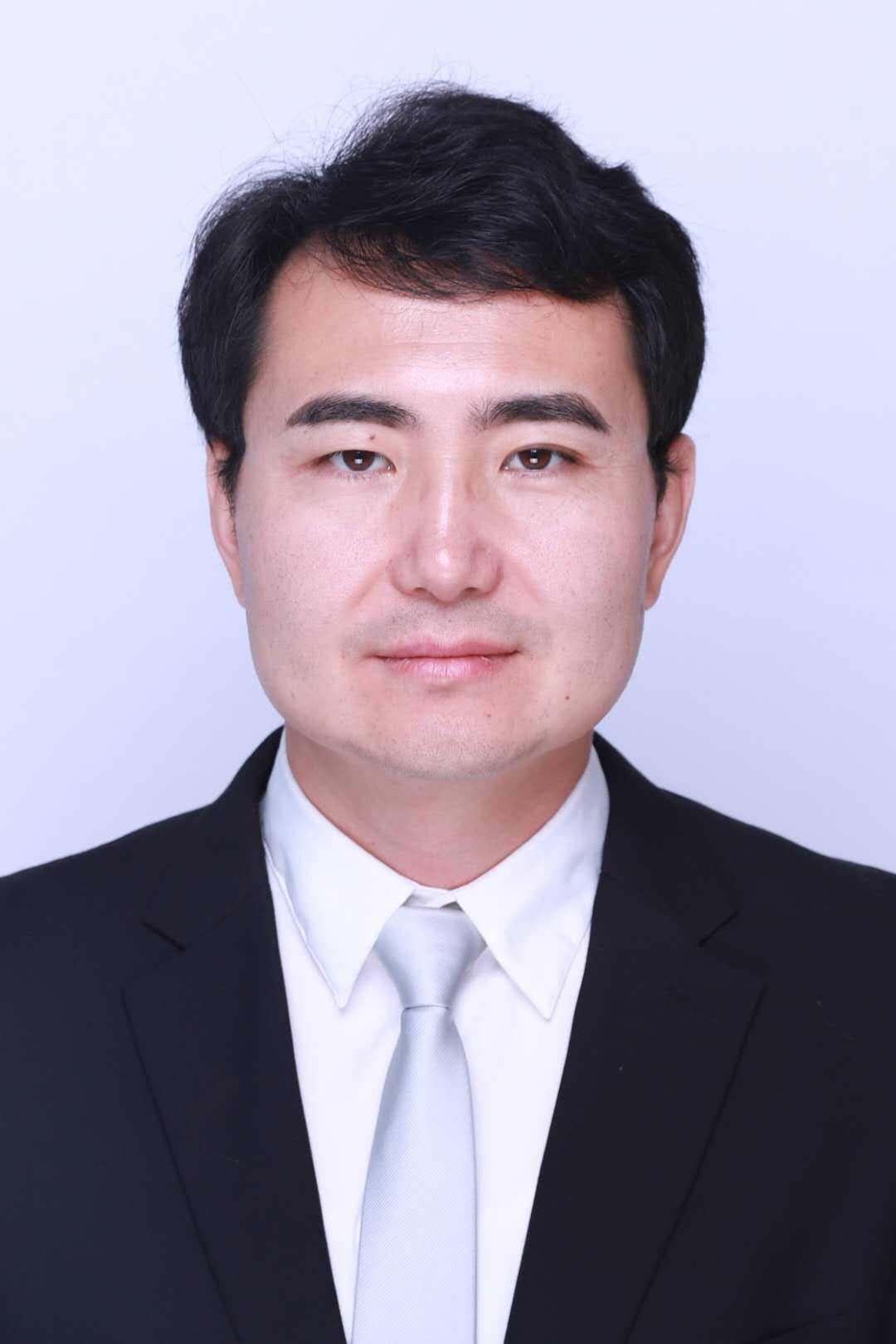}}]{Xin Fan} received the B.E. and Ph.D. degrees at Xi'an Jiaotong University, Xi'an, China, in 1998 and 2004, respectively. He was with Oklahoma State University, Stillwater, and the University of Texas Southwestern Medical Center, Dallas, from 2006 to 2009, as a post-doctoral research fellow. He joined Dalian University of Technology, Dalian, China, in 2009, where he is currently a full professor. He won the 2015 IEEE ICME Best Student Award as the corresponding author, and two papers were selected as the Finalist of the Best Paper Award at ICME 2017. His current research interests include image processing and machine vision.
\end{IEEEbiography}

%

%
%
%




\end{document}